%% file: caraten.tex
\documentstyle[11pt,epsf]{article}

\begin{document}

\begin{center}

{\bf Los Alamos Electronic Archives: physics/9909035}\\

\bigskip


\bigskip


\bigskip

{\bf CLASSICAL MECHANICS\\

\bigskip

 HARET C. ROSU\\
rosu@ifug3.ugto.mx

\bigskip
\bigskip

$\;$\\
$\;$\\

\vskip 1ex
\centerline{
\epsfxsize=190pt
\epsfbox{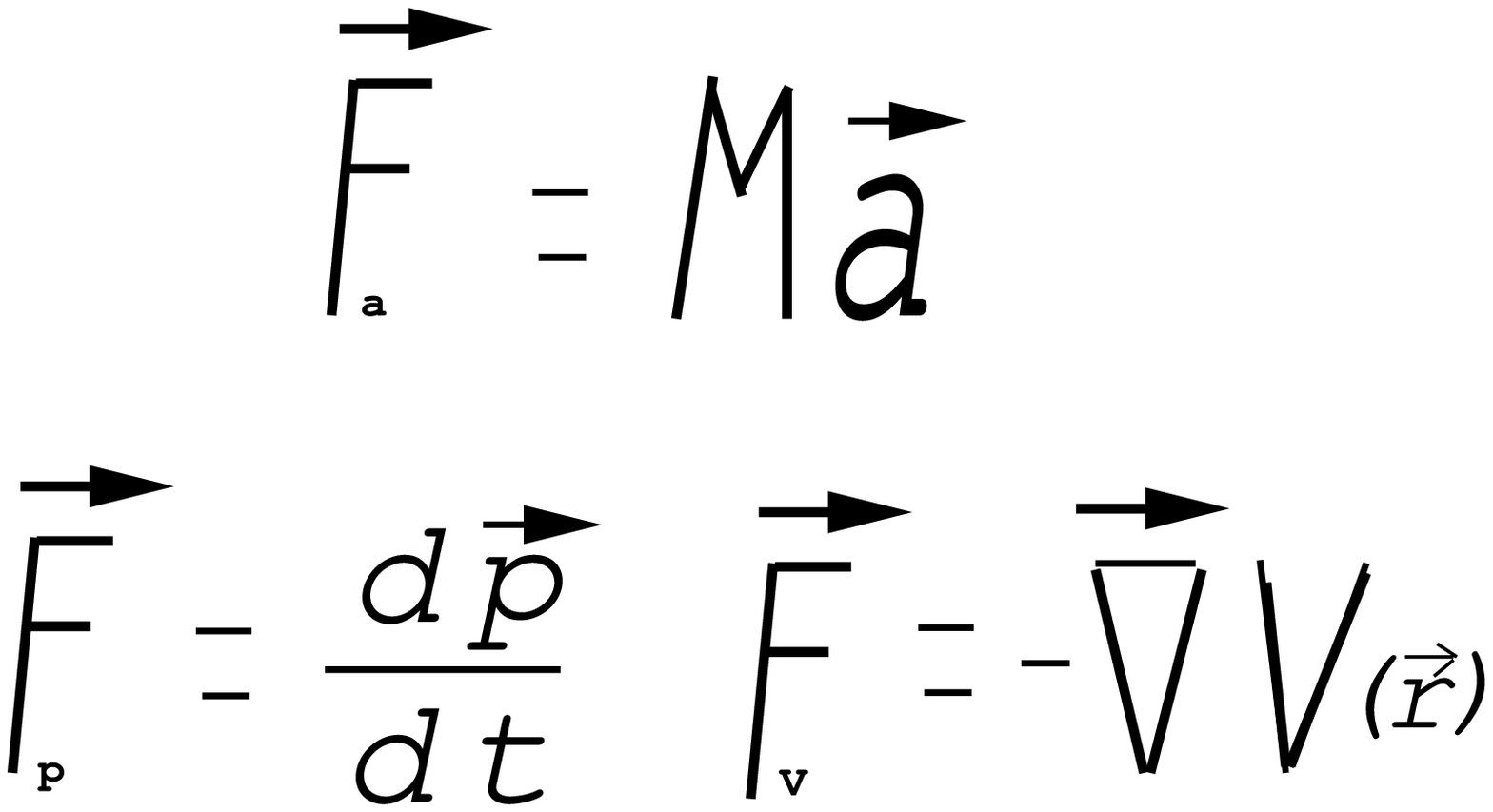}}
\vskip 2ex

\bigskip
\bigskip


{\bf graduate course}\\

\bigskip


\bigskip


$\;$\\
$\;$\\
$\;$\\
$\;$\\
$\;$\\
$\;$\\
$\;$\\
$\;$\\
$\;$\\

Copyright \copyright $\;$ 1999 H.C. Rosu\\
Le\'on, Guanajuato, Mexico\\ 
v1: September 1999.}

\end{center}

\newpage

\begin{center}

\underline{CONTENTS}

\end{center}

\bigskip

{\large

1. THE ``MINIMUM'' PRINCIPLES ... 3.

\bigskip

2. MOTION IN CENTRAL FORCES ... 19.

\bigskip

3. RIGID BODY ... 32.

\bigskip

4. SMALL OSCILLATIONS ... 52.

\bigskip

5. CANONICAL TRANSFORMATIONS ... 70.

\bigskip

6. POISSON PARENTHESES... 79.

\bigskip

7. HAMILTON-JACOBI EQUATIONS... 82.

\bigskip

8. ACTION-ANGLE VARIABLES ... 90.

\bigskip

9. PERTURBATION THEORY ... 96.

\bigskip

10. ADIABATIC INVARIANTS ... 111.

\bigskip

11. MECHANICS OF CONTINUOUS SYSTEMS ... 116.}

\bigskip
\bigskip


\bigskip









\newpage
\input erien.tex
\newpage
\input jul1en.tex
\newpage
\input jul2en.tex

\newpage
\input alben.tex
\newpage
\input c11en.tex
\newpage
\input c22en.tex
\newpage
\input c33en.tex
\newpage
\input c44en.tex
\newpage
\input peren.tex
\newpage
\input inv22en.tex

\newpage
\input clasen.tex

\end{document}

%% file: erien.tex

\newtheorem{theorem}{Theorem}
\newtheorem{acknowledgement}[theorem]{Acknowledgement}
\newtheorem{algorithm}[theorem]{Algorithm}
\newtheorem{axiom}[theorem]{Axiom}
\newtheorem{claim}[theorem]{Claim}
\newtheorem{conclusion}[theorem]{Conclusion}
\newtheorem{condition}[theorem]{Condition}
\newtheorem{conjecture}[theorem]{Conjecture}
\newtheorem{corollary}[theorem]{Corollary}
\newtheorem{criterion}[theorem]{Criterion}
\newtheorem{definition}[theorem]{Definition}
\newtheorem{example}[theorem]{Example}
\newtheorem{exercise}[theorem]{Exercise}
\newtheorem{lemma}[theorem]{Lemma}
\newtheorem{notation}[theorem]{Notation}
\newtheorem{problem}[theorem]{Problem}
\newtheorem{proposition}[theorem]{Proposition}
\newtheorem{remark}[theorem]{Remark}
\newtheorem{solution}[theorem]{Solution}
\newtheorem{summary}[theorem]{Summary}




\newlength{\defaultparindent}
\setlength{\defaultparindent}{\parindent}
\newenvironment{Default Paragraph Font}{}{}


\centerline{\Large 1. THE ``MINIMUM'' PRINCIPLES}

\bigskip
\bigskip

\noindent
{\bf Forward}:
The history of ``minimum" principles in physics is long and interesting.
The study of such principles is based on the idea that the nature acts
always in such a way that the important physical quantities are minimized
whenever a real physical process takes place. The mathematical background 
for these principles is the variational calculus.

\bigskip
\bigskip

{\bf CONTENTS}

1. Introduction

2. The principle of minimum action

3. The principle of D'Alembert

4. Phase space

5. The space of configurations

6. Constraints

7. Hamilton's equations of motion

8. Conservation laws

9. Applications of the action principle

\newpage

{\bf 1. Introduction}\\

\noindent
The empirical evidence has shown that the motion of a particle in an inertial
system
is correctly described by Newton's second law $\vec{F} = d\vec{p}/dt$,
whenever possible to neglect the relativistic effects.
When the particle happens not to be forced to a complicated motion,
the Cartesian coordinates are sufficient to describe the movement. If none
of these conditions are fulfilled, rather complicated equations of 
motion are to be expected.

\noindent
In addition, when the particle moves on a given surface,  
certain forces called constraint forces must exist
to maintain the particle in 
contact with the surface. Such forces are not so obvious from the 
phenomenological point of view; they require a separate postulate
in Newtonian mechanics, the one of action and reaction.
Moreover, other formalisms that may look more general have been 
developed. These formalisms are equivalent to Newton's laws when 
applied to simple practical problems, but they provide
a general approach for more complicated problems.
The Hamilton principle is one of these methods and its corresponding 
equations of motion are called the Euler-Lagrange equations.

\noindent
If the Euler-Lagrange equations are to be a consistent and correct
description of the dynamics of particles, they should be equivalent to 
Newton's equations. However, Hamilton's principle can also be applied to
phenomena generally not related to Newton's equations.
Thus, although HP does not give a new theory, it unifies many different
theories which appear as consequences of a simple fundamental postulate. 

\bigskip 

\bigskip 

\bigskip 




\noindent
The first ``minimum" principle was developed in the field of optics
by Heron of Alexandria about 2,000 years ago. He determined that the law of
the reflection of light on a plane mirror was such that the path taken
by a light ray to go from a given initial point to a given final point is
always the shortest one. However, Heron's minimum path principle does not
give the right law of reflection. In 1657, Fermat gave another formulation
of the principle by stating that the light ray travels on paths that
require the shortest time. Fermat's principle of minimal time led to
the right laws of reflection and refraction.
\noindent
The investigations of the minimum principles went on, and in the last half
of the XVII century, Newton, Leibniz and Bernoulli brothers initiated 
the development of the variational calculus. In the following years, 
Lagrange (1760) was able to give a solid mathematical base to this principle. 
In 1828, Gauss developed a method of studying Mechanics by means
of his principle of minimum constraint. Finally, in a sequence of works 
published during 1834-1835, Hamilton presented the dynamical principle of
minimum action.This principle has always been the base of all Mechanics and also 
of a big part of Physics. 

\noindent
Action is a quantity of dimensions of length multiplied by the momentum or 
energy multiplied by time.

\bigskip

{\bf 2.} {\bf The action principle}

\noindent
The most general formulation of the law of motion of mechanical systems is the 
{\em action  or Hamilton principle}. According to this principle every
 mechanical system is characterized by a function defined as:

\[
L\left( q_{1},q_{2},..., q_{s},\stackrel{\cdot }{q_{1}},%
\stackrel{\cdot}{q_{2}},\stackrel{\cdot }{q_{s}}, t\right) , 
\]

\noindent
or shortly $L\left(q,\stackrel{\cdot}{q},
t\right) $, and the motion of the system satisfies the following condition:
assume that at the moments $t_{1}$ and $t_{2}$ the system is in the
positions given by the set of coordinates 
$q^{\left( 1\right) }$ y $q^{\left( 2\right) };$
the system moves between these positions in such a way that the integral

\begin{equation} \label{e1}
S=\int_{t_{1}}^{t_{2}}L\left( q,\stackrel{\cdot }{q},
t\right) dt  
\end{equation}

\noindent
takes the minimum possible value. 
The function $L$ is called the {\it Lagrangian}
of the system, and the integral (1) is known as the {\it action} of the 
system. The Lagrange function
contains only $q$ and $\stackrel{\cdot}{q}$, and no other higher-order 
derivatives. This is because the mechanical state is completely defined by its
coordinates and velocities.

\noindent
Let us establish now the difererential equations that determine the minimum
of the 
integral (1). For simplicity we begin by assuming that the system has only one 
degree of freedom, therefore we are looking for only one function 
$q\left( t\right)$.
Let $q=q\left( t\right)$ be the function for which $S$ is a minimum. 
This means that $S$ grows when one $q\left(
t\right)$ is replaced by an arbitrary function

\begin{equation} \label{e2}
q\left( t\right) +\delta q\left( t\right) ,  
\end{equation}

\noindent
where $\delta q\left( t\right) $ is a small function through the 
interval from $t_{1}$ to $t_{2}$ [it is called the variation of the function
$q\left( t\right) $]. Since at $t_{1}$ and $t_{2}$
all the functions (2) should take the same values $q^{\left( 1\right) }$
and $q^{\left( 2\right) }$, one gets:

\begin{equation} \label{e3}
\delta q\left( t_{1}\right) =\delta q\left( t_{2}\right) =0. 
\end{equation}

\noindent
What makes $S$ change when $q$ is replaced by $q+\delta q$ is given by:

\[
\int_{t_{1}}^{t_{2}}L\left( q+\delta q,\stackrel{\cdot }{q}+\delta 
\stackrel{\cdot }{q},t\right) dt-\int_{t_{1}}^{t_{2}}L\left(q,
\stackrel{\cdot }{q},t\right) dt. 
\]

\noindent
An expansion in series of this difference in powers of $\delta q$ and $
\delta \stackrel{\cdot}{q}$ begins by terms of first order. The necessary
condition of minimum (or, in general, extremum) for $S$ is that the sum of 
all terms turns to
zero; Thus, the action principle can be written down as follows:

\begin{equation}\label{e4}
\delta S=\delta \int_{t_{1}}^{t_{2}}L\left( q,\stackrel{\cdot }{q},
t\right) dt=0,  
\end{equation}

\noindent
or by doing the variation:

\[
\int_{t_{2}}^{t_{1}}\left( \frac{\partial L}{\partial q}\delta q+\frac{%
\partial L}{\partial \stackrel{\cdot }{q}}\delta \stackrel{\cdot }{q}\right)
dt=0~.
\]

\noindent
Taking into account that $\delta \stackrel{\cdot}{q}=d/dt\left( \delta
q\right) $, we make an integration by parts to get:

\begin{equation}
\delta S=\left[ \frac{\partial L}{\partial \stackrel{\cdot }{q}}\delta q%
\right] _{t_{1}}^{t_{2}}+\int_{t_{2}}^{t_{1}}\left( \frac{\partial L}{%
\partial q}-\frac{d}{dt}\frac{\partial L}{\partial \stackrel{\cdot }{q}}%
\right) \delta qdt=0~.  \label{5}
\end{equation}

\noindent
Considering the conditions (3), the first term of this expresion disappears.
Only the integral remains that should be zero
for all values of $\delta q$. This is possible only if the integrand is
zero, which leads to the equation:

\[
\frac{\partial L}{\partial q}-\frac{d}{dt}\frac{\partial L}{\partial 
\stackrel{\cdot }{q}}=0~.
\]

\noindent
For more degrees of freedom, the $s$ different functions $q_{i}(t)$ should
vary independently. Thus, it is obvious that one gets $s$ equations of the form:

\begin{equation}
\frac{d}{dt}\left( \frac{\partial L}{\partial \stackrel{\cdot }{q_{i}}}%
\right) -\frac{\partial L}{\partial q_{i}}=0\qquad
\qquad\left( i=1,2,...,s\right)  \label{6}
\end{equation}

\noindent
These are the equations we were looking for; in Mechanics they are called
{\it Euler-Lagrange equations}.
If the Lagrangian of a given mechanical system is known, then the 
equations (6) form the relationship between the accelerations, the velocities 
and the coordinates; in other words, they are the equations of motion of the
system.
From the mathematical point of view, the equations 
(6) form a system of $s$ differential equations of second order
for $s$ unknown functions $q_{i}(t)$. The general solution of the system
contains $2s$ arbitrary constants. To determine them means
to completely define the movement of the mechanical system. In order to achieve
this, 
it is necessary to know the initial conditions that characterize the state
of the system at a given moment (for example, the initial values of the 
coordinates and velocities.

{\bf 3.} {\bf D'Alembert principle}

\noindent
The virtual displacement of a system is the change in its configurational
space under an arbitrary infinitesimal variation of the coordinates
$\delta {\bf r}_{i},${\it 
which is compatible with the forces and constraints imposed on the system at 
the given instant t}. It is called virtual in order to distinguish it from the 
real one, which takes place in a time interval $dt$, during which
the forces and the constraints can vary.

\noindent
The constraints introduce two types of difficulties in solving mechanics
problems:

(1) Not all the coordinates are independent.

(2) In general, the constraint forces are not known {\em a priori};
they are some unknowns of the problem and they should be obtained
from the solution looked for.

\noindent
In the case of holonomic constraints the difficulty (1) is avoided by 
introducing a set of independent coordinates ($q_{1,} q_{2,...,} q_{m}$,
where $m$ is the number of degrees of freedom involved). This means that if there
are $m$ constraint equations and $3N$ coordenates $(x_{1},...,x_{3N})$, 
we can eliminate these n
equations by introducing the independent variables $(q_{1},q_{2},..,,q_{n})$.
A transformation of the following form is used
$$
x_{1}=f_{1}(q_{1},...,q_{m},t)
$$
\begin{eqnarray}
\vdots  \nonumber
\end{eqnarray}
$$
x_{3N}=f_{3N}(q_{1},...,q_{n},t)~,%
$$

\noindent
where $\ n=3N-m$.

\noindent
To avoid the difficulty (2) Mechanics needs to be formulated in such a way
that the forces of constraint {\em do not occur} in the solution of the problem.
This is the essence of the 
``{\it principle of virtual work}".

\noindent
{\bf Virtual work:} We assume that a system of N particles is described by
$3N$ coordenates $(x_{1},x_{2},...,x_{3N})$ and let $F_{1,}F_{2,...,}F_{3N}$ be
the components of the forces acting
on each particle. If the particles of the system display infinitesimal and 
instantaneous displacements $\delta x_{1},\delta
x_{2},...,\delta x_{3N}$ under the action of the $3N$ forces, then the performed 
work is:

\begin{equation}
\delta W=\sum_{j=1}^{3N}F_{j}\delta x_{j}~.  \label{7}
\end{equation}

\noindent
Such displacements are known as {\it virtual displacements} and $\delta W$
is called {\it virtual work}; (7{\it )} can be  also written as:

\begin{equation}
\delta W=\sum_{\alpha =1}^{N}{\bf F}_{\alpha }\cdot \delta {\bf r}~. 
\label{8}
\end{equation}

\noindent
{\bf Forces of constraint}: besides the applied forces
${\bf F}_{\alpha }^{\left( e\right) }$, the particles can be acted on by forces
of constraint ${\bf F}_{\alpha }$.

\noindent
{\bf The principle of virtual work:} Let ${\bf F}_{\alpha }$ be the force acting 
on the particle $\alpha$ of the system. If we separate
${\bf F}_{\alpha }$ in a contribution from the outside
${\bf F}_{\alpha}^{\left( e\right) }$ and the constraint ${\bf R}_{\alpha}$

\begin{equation}
{\bf F}_{\alpha }={\bf F}_{\alpha }^{\left( e\right) }+{\bf R}_{\alpha}~.
\label{9}
\end{equation}
\noindent
and if the system is in equilibrium, then 

\begin{equation}
{\bf F}_{\alpha }={\bf F}_{\alpha }^{\left( e\right) }+{\bf R}_{\alpha }=0~.
\label{10}
\end{equation}

\noindent
Thus, the virtual work due to all possible forces ${\bf F}_{\alpha}$ is:

\begin{equation}
W=\sum_{\alpha =1}^{N}{\bf F}_{\alpha }\cdot \delta {\bf r}_{\alpha
}=\sum_{\alpha =1}^{N}\left( {\bf F}_{\alpha }^{\left( e\right) }+{\bf R}%
_{\alpha }\right) \cdot \delta {\bf r}_{\alpha }=0~.  \label{11}
\end{equation}

\noindent
If the system is such that the constraint forces do not make virtual work, 
then from (11) we obtain:

\begin{equation}
\sum_{\alpha =1}^{N}{\bf F}_{\alpha }^{\left( e\right) }\cdot \delta {\bf r}%
_{\alpha }=0~.  \label{12}
\end{equation}

\noindent
Taking into account the previous definition, we are now ready to introduce
the D'Alembert principle. According to Newton, the equation of motion is:

\[
{\bf F}_{\alpha }=\stackrel{\cdot }{{\bf p}}_{\alpha } 
\]

\noindent
and can be written in the form

\[
{\bf F}_{\alpha }-\stackrel{\cdot }{{\bf p}}_{\alpha }=0~,
\]

\noindent
which tells that the particles of the system would be in equilibrium under the 
action of a force equal to the real one plus an inverted force 
$-\stackrel{\cdot }{{\bf p}}_{i}$. Instead of (12) we can write

\begin{equation}
\sum_{\alpha =1}^{N}\left( {\bf F}_{\alpha }-\stackrel{\cdot }{{\bf p}}%
_{\alpha }\right) \cdot \delta {\bf r}_{\alpha }=0  \label{13}
\end{equation}

\noindent
and by doing the same decomposition in applied and constraint forces
$\left( {\bf f}_{\alpha }\right)$, we obtain:

\[
\sum_{\alpha =1}^{N}\left( {\bf F}_{\alpha }^{\left( e\right) }-\stackrel{
\cdot }{{\bf p}}_{\alpha }\right) \cdot \delta {\bf r}_{\alpha
}+\sum_{\alpha =1}^{N}{\bf f}_{\alpha }\cdot \delta {\bf r}_{\alpha }=0~.
\]

\noindent
Again, let us limit ourselves to systems for which the virtual work due to 
the forces of constraint is zero leading to

\begin{equation}
\sum_{\alpha =1}^{N}\left( {\bf F}_{\alpha }^{\left( e\right) }-\stackrel{%
\cdot }{{\bf p}}_{\alpha }\right) \cdot \delta {\bf r}_{\alpha }=0~,
\label{14}
\end{equation}

\noindent
which is the {\it D'Alembert's principle}. However, this equation does not have
a useful form yet for getting the equations of motion of the system. 
Therefore, we should change the principle to an expression entailing the virtual
displacements of the generalized coordinates, which being independent from each
other, imply zero coefficients for $\delta {\bf q}_{\alpha}$. Thus, 
the velocity in terms of the generalized coordinates reads:

\[
{\bf v}_{\alpha }=\frac{d{\bf r}_{\alpha }}{dt}=\sum_{k}\frac{\partial {\bf r
}_{\alpha }}{\partial q_{k}}\stackrel{\cdot }{q_{k}}+\frac{\partial {\bf r}
_{\alpha }}{\partial t}\ \ \ \ \ \  {\rm where}
\qquad {\bf r}_{\alpha }={\bf r}%
_{\alpha }\left( q_{1},q_{2},...,q_{n},t\right)~. 
\]

\noindent
Similarly, the arbitrary virtual displacement $\delta {\bf r}
_{\alpha }$ can be related to the virtual displacements $\delta 
{\bf q}_{j}$ through

\[
\delta {\bf r}_{\alpha }=\sum_{j}\frac{\partial {\bf r}_{\alpha }}{\partial
q_{j}}\delta q_{j}~.
\]

\noindent
Then, the virtual work ${\bf F}_{\alpha }$ expressed in terms of the 
generalized coordinates will be:

\begin{equation}
\sum_{\alpha =1}^{N}{\bf F}_{\alpha }\cdot \delta {\bf r}_{\alpha
}=\sum_{j,\alpha }{\bf F}_{\alpha }\cdot \frac{\partial {\bf r}_{\alpha }}{
\partial q_{j}}\delta q_{j}=\sum_{j}Q_{j}\delta q_{j}~,  \label{15}
\end{equation}

\noindent
where the $Q_{j}$ are the so-called components of the generalized force, defined
in the form

\[
Q_{j}=\sum_{\alpha }{\bf F}_{\alpha }\cdot \frac{\partial {\bf r}_{\alpha }}{
\partial q_{j}}~.
\]

\noindent
Now if we see eq. (14) as:

\begin{equation}
\sum_{\alpha }\stackrel{\cdot }{{\bf p}}\cdot \delta {\bf r}_{\alpha
}=\sum_{\alpha }m_{\alpha }\stackrel{\cdot \cdot }{{\bf r}_{\alpha }}\cdot 
\delta {\bf r}_{\alpha }  \label{16}
\end{equation}

\noindent
and by substituting in the previous results we can see that
(16) can be written:

\begin{equation}
\sum_{\alpha }\left\{ \frac{d}{dt}\left( m_{\alpha }{\bf v}_{\alpha }\cdot 
\frac{\partial {\bf v}_{\alpha }}{\partial \stackrel{\cdot }{q_{j}}}\right)
-m_{\alpha }{\bf v}_{\alpha }\cdot \frac{\partial {\bf v}_{\alpha }}{%
\partial q_{j}}\right\} =\sum_{j}\left[ \left\{ \frac{d}{dt}\left( \frac{
\partial T}{\partial \stackrel{\cdot }{q_{j}}}\right) -\frac{\partial T}{
\partial q_{j}}\right\} -Q_{j}\right] \delta q_{j}=0~.  \label{17}
\end{equation}

\noindent
The variables $q_{j}$ can be an arbitrary system of coordinates describing the 
motion of the system. However, if the constraints are holonomic, it is possible
to find systems of independent coordinates $q_{j}$ containing implicitly
the constraint conditions already in the equations of transformation
$x_{i}=f_{i}$ if one nullifies the coefficients by separate:

\begin{equation}
\frac{d}{dt}\left( \frac{\partial T}{\partial \stackrel{\cdot }{q}}\right) -
\frac{\partial T}{\partial q_{\alpha }}=Q_{j}~.  \label{18}
\end{equation}

\noindent
There are $m$ equations.
The equations (18) are sometimes called the Lagrange equations, although this 
terminology is usually applied to the form they take when the forces are 
conservative (derived from a scalar potential V)

\[
{\bf F}_{\alpha }=-\nabla _{i}V. 
\]

\noindent
Then $Q_{j}$ can be written as:

\[
Q_{j}=-\frac{\partial V}{\partial q_{j}}~.
\]

\noindent
The equations (18) can also be written in the form:

\begin{equation}
\frac{d}{dt}\left( \frac{\partial T}{\partial \stackrel{\cdot }{q_{j}}}%
\right) -\frac{\partial (T-V)}{\partial q_{j}}=0  \label{19}
\end{equation}

\noindent
and defining the {\it Lagrangian} $L$ in the form $L=T-V$
one gets
\begin{equation}
\frac{d}{dt}\left( \frac{\partial L}{\partial \stackrel{\cdot }{q_{j}}}%
\right) -\frac{\partial L}{\partial q_{j}}=0~.  \label{20}
\end{equation}

\noindent
These are the {\bf Lagrange equations of motion}.

\bigskip

{\bf 4. - Phase space}

\noindent
In the geometrical interpretation of mechanical phenomena the concept of
{\it phase space} is very much in use. It is a space of
$2s$ dimensions whose axes of coordinates are the $s$ generalized
coordinates and the  $s$ momenta of the given system.
Each point of this space corresponds to a definite mechanical state of the 
system. When the system is in motion, the representative point in phase space
performs a curve called {\it phase trajectory.}

{\bf 5. - Space of configurations}

\noindent
The state of a system composed of $n$ particles under the action of $m$
constraints connecting some of the $3n$ cartesian coordinates is completely
determined by $s=3n-m$ generalized coordinates.
Thus, it is possible to descibe the state of such a system by a point in the  
$s$ dimensional space usually called the {\it configuration space},
for which each of its dimensions corresponds to one $q_{j}$. 
The time evolution of the system will be represented by 
a curve in the configuration space made of points describing the instantaneous
configuration of the system.

{\bf 6. - Constraints}

\noindent
One should take into account the {\it constraints} that act on the 
motion of the system.
The constraints can be classified in various ways. In the general case in which 
the constraint equations can be written in the form:

\[
\sum_{i}c_{\alpha i}\stackrel{\cdot }{q_{i}}=0~, 
\]

\noindent
where the $c_{\alpha i}$ are functions of only the coordinates
(the index $\alpha$ counts the constraint equations).
If the first members of these equations are not total derivatives with respect
to the time they cannot be integrated.
In other words, they cannot be reduced to relationships between only 
the coordinates, that might be used to express the position by less coordinates,
corresponding to the real number of degrees of freedom. Such constraints are
called {\it non
holonomic} (in contrast to the previous ones which are
{\it holonomic} and which connect only the coordinates of the system{\it )}.

{\bf 7.} {\bf Hamilton's equations of motion}

\noindent
The formulation of the laws of Mechanics by means of the 
Lagrangian assumes that the mechanical state of the system is determined by its
generalized coordinates and velocities. However, this is not the unique possible 
method; the equivalent description in terms of its generalized coordinates and
momenta has a number of advantages.

\noindent
Turning from one set of independent variables to another one can be achieved
by what in mathematics is called {\it Legendre transformation}.
In this case the transformation takes the following form where the total 
differential of the Lagrangian as a function of coordinates and velocities is:

\[
dL=\sum_{i}\frac{\partial L}{\partial q_{i}}dq_{i}+\sum_{i}\frac{\partial L}{%
\partial \stackrel{\cdot }{q_{i}}}d\stackrel{\cdot }{q}_{i}~,
\]

\noindent
that can be written as:

\begin{equation}
dL=\sum_{i}\stackrel{\cdot }{p_{i}}dq_{i}+\sum_{i}p_{i}d\stackrel{\cdot }{q}%
_{i}~,  \label{21}
\end{equation}

\noindent
where we already know that the derivatives
$\partial L/\partial \stackrel{\cdot}{
q_{i}}$, are by definition the generalized momenta and moreover
$\partial L$ $/$ $\partial q_{i}=\stackrel{\cdot }{p_{i}}$
by Lagrange equations. The second term in eq. (21)can be written as follows

\[
\sum_{i}p_{i}d\stackrel{\cdot }{q}_{i}=d\left( \sum p_{i}\stackrel{\cdot }{q}%
_{_{i}}\right) -\sum \stackrel{\cdot }{q_{i}}dq_{i}~. 
\]

\noindent
By attaching the total diferential $d\left( \sum p_{i}\stackrel{\cdot }{q}
_{_{i}}\right)$ to the first term and changing the signs one gets from (21):

\begin{equation}
d\left( \sum p_{i}\stackrel{\cdot }{q}_{_{i}}-L\right) =-\sum \stackrel{%
\cdot }{p}_{_{i}}dq_{i}+\sum p_{i}\stackrel{\cdot }{q}_{_{i}}~.  \label{22}
\end{equation}

\noindent
The quantity under the  diferential is the energy of the system
as a function of the coordinates and momenta and is called
{\it Hamiltonian function or Hamiltonian} of the system:

\begin{equation}
H\left( p,q,t\right) =\sum_{i}p_{i}\stackrel{\cdot }{q}_{_{i}}-L~.
\label{23}
\end{equation}

\noindent
Then from ec. (22)

\[
dH=-\sum \stackrel{\cdot }{p}_{_{i}}dq_{i}+\sum p_{i}\stackrel{\cdot }{q}%
_{i} 
\]

\noindent
where the independent variables are the coordinates and the momenta, 
one gets the equations

\begin{equation}
\stackrel{\cdot }{q}_{i}=\frac{\partial H}{\partial p_{i}}\ \ \ \ \ \
\ \ \ \ \ \ \ \ \stackrel{\cdot }{p}_{_{i}}=-\frac{\partial H}{\partial
q_{i}}~.  \label{24}
\end{equation}

\noindent
These are the equations of motion in the variables $q$ y $p$ 
and they are called {\it Hamilton's equations.}

{\bf 8.} {\bf Conservation laws}

\noindent
{\bf 8.1  Energy}

\noindent
Consider first the conservation theorem resulting from the 
{\it homogeneity of time}. Because of this homogeneity, the 
Lagrangian of a closed system does not depend explicitly on time. Then, the total
time diferential of the Lagrangian (not
depending explicitly on time) can be written:

\[
\frac{dL}{dt}=\sum_{i}\frac{\partial L}{\partial q_{i}}\stackrel{\cdot }{q}%
_{i}+\sum_{i}\frac{\partial L}{\partial \stackrel{\cdot }{q}_{i}}\stackrel{%
\cdot \cdot }{q}_{i} 
\]

\noindent
and according to the Lagrange equations we can rewrite the previous equation as
follows:

\[
\frac{dL}{dt}=\sum_{i}\stackrel{\cdot }{q}_{i}\frac{d}{dt}\left( \frac{%
\partial L}{\partial \stackrel{\cdot }{q}_{i}}\right) +\sum_{i}\frac{%
\partial L}{\partial \stackrel{\cdot }{q}_{i}}\stackrel{\cdot \cdot }{q}%
_{i}=\sum_{i}\frac{d}{dt}\left( \stackrel{\cdot }{q}_{i}\frac{\partial L}{%
\partial \stackrel{\cdot }{q}_{i}}\right)~, 
\]

\noindent
or

\[
\sum_{i}\frac{d}{dt}\left( \stackrel{\cdot }{q}_{i}\frac{\partial L}{%
\partial \stackrel{\cdot }{q}_{i}}-L\right) =0~.
\]

\noindent
From this one concludes that the quantity

\begin{equation}
E\equiv \sum_{i}\stackrel{\cdot }{q}_{i}\frac{\partial L}{\partial \stackrel{%
\cdot }{q}_{i}}-L  \label{25}
\end{equation}

\noindent
remains constant during the movement of the closed system, that is it is 
an integral of motion. This constant quantity is called the {\it energy} E
of the system.

\noindent
{\bf 8.2 Momentum}

\noindent
The {\it homogeneity of space} implies another conservation theorem.
Because of this homogeneity, the mechanical properties of a cosed system
do not vary under a parallel displacement of the system as a whole through 
space. We consider an infinitesimal displacement $\epsilon$ (i.e.,
the position vectors {\bf r}$
_{\alpha }$ turned into {\bf r}$_{a}+\epsilon $) and look for the condition
for which the Lagrangian does not change. The variation of the function 
$L$ resulting from the infinitesimal change of the coordinates (maintaining 
constant the velocities of the particles) is given by:

\[
\delta L=\sum_{a}\frac{\partial L}{\partial {\bf r}_{a}}\cdot \delta {\bf r}
_{a}=\epsilon \cdot \sum_{a}\frac{\partial L}{\partial {\bf r}_{a}}~,
\]

\noindent
extending the sum over all the particles of the system. Since $\epsilon$
is arbitrary, the condition $\delta L=0$ is equivalent to

\begin{equation}
\sum_{a}\frac{\partial L}{\partial {\bf r}_{a}}=0  \label{26}
\end{equation}

\noindent
and taking into account the already mentioned equations of Lagrange

\[
\sum_{a}\frac{d}{dt}\left( \frac{\partial L}{\partial {\bf v}_{a}}\right) =%
\frac{d}{dt}\sum_{a}\frac{\partial L}{\partial {\bf v}_{a}}=0~.
\]

\noindent
Thus, we reach the conclusion that for a closed mechanical system
the vectorial quantity called {\it impetus/momentum} 

\[
{\bf P\equiv }\sum_{a}\frac{\partial L}{\partial {\bf v}_{a}} 
\]

\noindent
remains constant during the motion.

\noindent
{\bf 8.3 Angular momentum}

\noindent
Let us study now the conservation theorem coming out from 
{\it the isotropy of space}. For this we consider an infinitesimal rotation of
the system and look for the condition under which the Lagrangian does not
change.

\noindent
We shall call an infinitesimal rotation vector $\delta {\bf \phi }$ a
vector of modulus equal to the angle of rotation $\delta \phi $
and whoose direction coincide with that of the rotation axis. We shall look
first to the increment of the position vector of a particle in the system,
by taking the origin of coordinates on the axis of rotation.
The lineal displacement of the position vector as a function of angle is 

\[
\left| \delta {\bf r}\right| =r\sin \theta \delta \phi~, 
\]

\noindent
(see the figure). The direction of the vector $\delta {\bf r}$ is perpendicular
to the plane defined by ${\bf r}$ and $\delta {\bf \phi}$, and therefore,

\begin{equation}
\delta {\bf r=}\delta {\bf \phi \times r}~.  \label{27}
\end{equation}

\vskip 1ex
\centerline{
\epsfxsize=80pt
\epsfbox{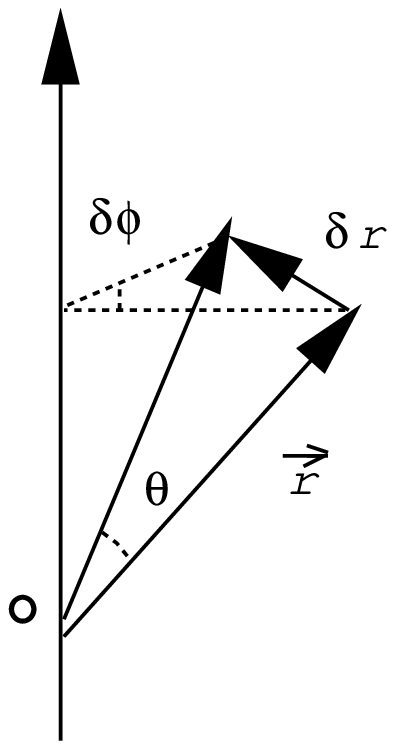}}
\vskip 2ex


\noindent
The rotation of the system changes not only the directions of the position
vectors but also the velocities of the particles that are modified by the 
same rule for all the vectors. The velocity increment 
with respect to a fixed frame system will be:

\[
\delta {\bf v}=\delta {\bf \phi \times v}~.
\]

\noindent
We apply now to these expressions the condition that the Lagrangian does not
vary under rotation:

\[
\delta L=\sum_{a}\left( \frac{\partial L}{\partial {\bf r}_{a}}\cdot \delta 
{\bf r}_{a}+\frac{\partial L}{\partial {\bf v}_{a}}\cdot \delta {\bf v}%
_{a}\right) =0 
\]

\noindent
and substituting the definitions of the derivatives $\partial L/\partial
{\bf v}_{a}$ por ${\bf p}_{a}$ and  $\partial L/\partial
{\bf r}_{a}$ from the Lagrange equations
by $\stackrel{\cdot }{{\bf p}}_{a}$; we get

\[
\sum_{a}\left( \stackrel{\cdot }{{\bf p}}_{a}\cdot \delta {\bf \phi
\times r}_{a}+{\bf p}_{a}\cdot \delta {\bf \phi \times v}_{a}\right) =0~,
\]

\noindent
or by circular permutation of the factors and getting 
$\delta {\bf \phi }$ out of the sum:

\[
\delta {\bf \phi }\sum_{a}\left( {\bf r}_{a}{\bf \times }\stackrel{\cdot }{
{\bf p}}_{a}+{\bf v}_{a}{\bf \times p}_{a}\right) =\delta {\bf \phi }\cdot 
\frac{d}{dt}\sum_{a}{\bf r}_{a}{\bf \times p}_{a}=0~,
\]

\noindent
because $\delta {\bf \phi }$ is arbitrary, one gets

\[
\frac{d}{dt}\sum_{a}{\bf r}_{a}{\bf \times p}_{a}=0 
\]

\noindent
Thus, the conclusion is that during the motion of a closed system the 
vectorial quantity called the 
{\it angular (or kinetic) momentum} is conserved.

\[
M\equiv \sum_{a}{\bf r}_{a}{\bf \times p}_{a}~.
\]

\bigskip

{\bf 9.- Applications of the action principle}

\noindent
{\bf a) Equations of motion}

\noindent
Find the eqs of motion for a pendular mass sustained by a 
resort, by directly applying Hamilton's principle.

\vskip 2ex
\centerline{
\epsfxsize=200pt
\epsfbox{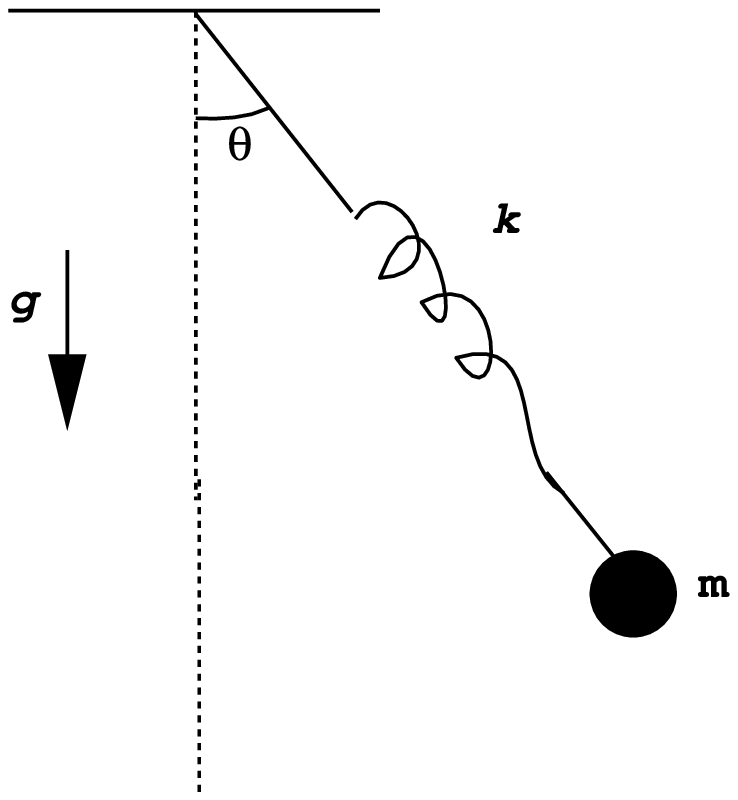}}
\vskip 4ex


\noindent
For the pendulum in the figure the Lagrangian function is 

\[
L=\frac{1}{2}m(\stackrel{\cdot}{r}^{2}+r^{2}\stackrel{\cdot}{\theta}^{2}
)+mgr\cos \theta -\frac{1}{2}k(r-r_{o})^{2}~, 
\]

\noindent
therefore

\[
\int_{t_{1}}^{t_{2}}\delta Ldt=\int_{t_{1}}^{t_{2}}\left[
 m\left( 
\stackrel{\cdot }{r}\delta \stackrel{\cdot }{r}+r\stackrel{\cdot}{\theta}
^{2}+r^{2}\stackrel{\cdot }{\theta }\delta \stackrel{\cdot }{\theta }\right)
+mg\delta r\cos \theta -mgr\delta \theta \sin \theta -k(r-r_{o})\delta r%
\right] dt 
\]

\[
m\stackrel{\cdot }{r}\delta \stackrel{\cdot }{r}dt=m\stackrel{\cdot }{r}%
d(\delta r)=d\left( m\stackrel{\cdot }{r}\delta r\right) -m\delta r\stackrel{%
\cdot \cdot }{r}dt~.
\]

\noindent
In the same way

\[
mr^{2}\theta ^{2}\delta \stackrel{\cdot }{\theta }dt=d\left( mr^{2}\stackrel{%
\cdot }{\theta }\delta \stackrel{\cdot }{\theta }\right) -\delta \theta 
\frac{d\left( mr^{2}\stackrel{\cdot }{\theta }\right) }{dt}dt 
\]

\[
=d\left( mr^{2}\stackrel{\cdot }{\theta }\delta \stackrel{\cdot }{\theta }%
\right) -\delta \theta \left( mr^{2}\stackrel{\cdot \cdot }{\theta }+2mr%
\stackrel{\cdot }{r}\stackrel{\cdot }{\theta }\right) dt~.
\]

\noindent
Therefore, the previous integral can be written 

\[
\int_{t_{1}}^{t_{2}}\left[ \left\{ m\stackrel{\cdot \cdot }{r}-mr\stackrel{%
\cdot }{\theta }^{2}-mg\cos \theta +k\left( r-r_{o}\right) \right\} +\left\{
mr^{2}\stackrel{\cdot \cdot }{\theta }+2mr\stackrel{\cdot }{r}\stackrel{%
\cdot }{\theta }+mgr\sin \theta \right\} \delta \theta \right] dt 
\]

\[
-\int_{t_{1}}^{t_{2}}\left[ d\left( m\stackrel{\cdot }{r}\delta r\right)
+d\left( mr^{2}\theta ^{2}\stackrel{\cdot }{\theta }\delta \theta \right) %
\right] =0~.
\]

\noindent
Assuming that both $\delta r$ and $\delta \theta $ are equal zero at $
t_{1}$ and $t_{2}$, the second integral is obviously nought. Since $%
\delta r$ and $\delta \theta $ are completely independent of each other, the 
first integral can be zero only if

\[
m\stackrel{\cdot \cdot }{r}-mr\stackrel{\cdot }{\theta }^{2}-mg\cos \theta
+k(r-r_{o})=0 
\]

\noindent
and

\[
mr^{2}\stackrel{\cdot \cdot }{\theta }+2mr\stackrel{\cdot }{r}\stackrel{
\cdot }{\theta }+mgr\sin \theta =0~,
\]

\noindent
These are the equations of motion of the system.

\newpage
\noindent
{\bf b) Exemple of calculating a minimum value}

\noindent
Prove that the shortest line between two given points $p_{1}$ and $p_2$
on a cilinder is a helix.

\noindent
The length $S$ of an arbitrary line on the cilinder
between $p_{1}$ and $p_{2}$ is given by

\[
S=\int_{p_{1}}^{p_{2}}\left[ 1+r^{2}\left( \frac{d\theta }{dz}%
\right) ^{2}\right] ^{1/2}dz~,
\]

\noindent
where $r$, $\theta $ and $z$ are the usual cilindrical coordinates 
for $r=const.$ A relationship between $\theta$ and $z$ can be determined
for which the last integral has an extremal value by means of

\[
\frac{d}{dz}\left( \frac{\partial \phi }{\partial \theta ^{^{\prime }}}%
\right) -\frac{\partial \phi }{\partial \theta }=0~,
\]

\noindent
where $\phi =\left[ 1+r^{2}\theta ^{\prime 2}\right] ^{1/2}$ y $\theta
^{\prime }=\frac{d\theta }{dz}$, but since $\partial \phi /\partial \theta
=0$ we have

\[
\frac{\partial \phi }{\partial \theta ^{\prime }}=\left( 1+r^{2}\theta
^{\prime 2}\right) ^{-1/2}r^{2}\theta ^{\prime }=c_{1}=const.~,
\]

\noindent
therefore $r\theta ^{\prime }=c_{2}$. Thus, $r\theta =c_{2}z+c_{3}$, 
which is the parametric equation of a helix. Assuming that in $p_{1}$ we have
$\theta =0$ and $z=0$, then $c_{3}=0$.
In $p_{2}$, make $\theta
=\theta _{2}$ and $z=z_{2}$, therefore $c_{2}=r\theta _{2}/z_{2}$, and $r\theta
=\left( r\theta _{2}/z_{2}\right) z$ is the final equation.

\begin{center}  {\bf References}  \end{center}

\noindent
L. D. Landau and E. M Lifshitz, {\it Mechanics}, Theoretical Physics,
 vol I, (Pergammon, 1976)

\noindent
H. Goldstein, {\it Classical Mechanics}, (Addison-Wesley, 1992)


%% file: jul1en.tex



\centerline{\Large 2. MOTION IN CENTRAL FORCES}

\bigskip

\noindent
{\bf Forward}:
Because of astronomical reasons, the motion under the action of central
forces has been the physical problem on which pioneer researchers focused
more, either from the observational standpoint or by trying to disentangle the
governing laws of motion. This movement is a basic example for many 
mathematical formalisms. In its relativistic version, Kepler's problem is 
yet an area of much interest. 

\bigskip
\bigskip

{\bf  CONTENTS:}

\bigskip

2.1 The two-body problem: reduction to the one-body problem

\bigskip

2.2 Equations of motion

\bigskip

2.3 Differential equation of the orbit

\bigskip

2.4 Kepler's problem

\bigskip

2.5 Dispertion by a center of forces (with example)

\bigskip

\newpage


\noindent {\bf
2.1 Two-body problem: Reduction to the one-body problem}

\noindent
Consider a system of two material points of masses $m_{1}$ 
and $m_{2}$, in which there are forces due only to an interaction potential $V$.
We suppose that $V$ is a function of any position vector
between $m_{1}$ and $m_{2}$, ${\bf r}_{2}-{\bf r}_{1}$, or of their relative 
velocities $\stackrel{\cdot }{{\bf r}}_{2}-\stackrel{\cdot }{
{\bf r}}_{1}$, or of the higher-order derivatives of ${\bf r}_{2}-{\bf r}_{1}$.
Such a system has 6 degrees of freedom and therefore 6 independent
generalized coordinates.

\noindent
We suppose that these are the vector coordinates of the center-of-mass
${\bf R}$, plus the three components of the relative difference vector
${\bf r}={\bf r}_{2}-{\bf r}_{1}$. The Lagrangian of the system
can be written in these coordinates as follows:
\setcounter{equation} {0}\\
\begin{equation}
L=T({\bf \dot{R}},{\bf \dot{r}})-V({\bf r},{\bf \dot{r}},{\bf \ddot{r}}%
,.....).  \label{eq1}
\end{equation}

\noindent
The kinetic energy $T$ is the sum of the kinetic energy of the center-of-mass
plus the kinetic energy of the motion around it, $T{\acute{}}$:
\[
T=\frac{1}{2}(m_{1}+m_{2}){\bf \dot{R}}^{2}+T{\acute{}}~,
\]
being
\[
T{\acute{}}=\frac{1}{2}m_{1}{\bf \dot{r}}_{1}^{2}{\acute{}}
+\frac{1}{2}m_{2}{\bf \dot{r}}_{2}^{2}
{\acute{}}.
\]

\noindent
Here, ${\bf r}_{1}{\acute{}}$ and ${\bf r}_{2}{\acute{}}$ are the position
vectors of the two particles with respect to the center-of-mass, and they are 
related to ${\bf r}$ by means of
\begin{equation}
{\bf r}_{1}{\acute{}}
=-\frac{m_{2}}{m_{1}+m_{2}}{\bf r},\;{\bf r}_{2}{\acute{}}
=\frac{m_{1}}{m_{1}+m_{2}}{\bf r}~.  \label{eq2}
\end{equation}
Then, $T{\acute{}}$ takes the form
\[
T{\acute{}}
=\frac{1}{2}\frac{m_{1}m_{2}}{m_{1}+m_{2}}{\bf \dot{r}}^{2} 
\]
and the total Lagrangian as given by equation (1) is:
\begin{equation}
L=\frac{1}{2}(m_{1}+m_{2}){\bf \dot{R}}^{2}+\frac{1}{2}\frac{m_{1}m_{2}}{%
m_{1}+m_{2}}{\bf \dot{r}}^{2}-V({\bf r},{\bf \dot{r}},{\bf \ddot{r}},.....)~,
\label{eq3}
\end{equation}
where from the reduced mass is defined as
\[
\mu =\frac{m_{1}m_{2}}{m_{1}+m_{2}}\;\;\;o{\acute{}}
\;\;\;\frac{1}{\mu }=\frac{1}{m_{1}}+\frac{1}{m_{2}}~. 
\]
Then, the equation (\ref{eq3}) can be written as follows
\[
L=\frac{1}{2}(m_{1}+m_{2}){\bf \dot{R}}^{2}+\frac{1}{2}\mu {\bf \dot{r}}%
^{2}-V({\bf r},{\bf \dot{r}},{\bf \ddot{r}},.....). 
\]
From this equation we see that the coordinates ${\bf \dot{R}}$ are cyclic
implying that the center-of-mass is either fixed or in uniform motion.

\noindent
Now, none of the equations of motion for ${\bf r}$ will contain a term where
${\bf R}$ or ${\bf \dot{R}}$ will occur. This term is exactly what we will have
if a center of force would have been located in the 
center of mass with an additional particle at a distance ${\bf r}$ 
away of mass $\mu$ (the reduced mass).

\noindent
Thus, the motion of two particles around their center of mass, which is due
to a central force can be always reduced to an equivalent problem of a 
single body.

\bigskip

\noindent {\bf 2.2 Equations of motion}

\noindent
Now we limit ourselves to conservative central forces for which the potential
is a function of only $r$, $V(r)$, so that the force is directed along 
{\bf \ }${\bf r}$. Since in order to solve the problem we only need to tackle
a particle of mass $m$ moving around the fixed center of force, we can put
the origin of the reference frame there. As the potential depends only on $r$,
the problem has spherical symmetry, that is any arbitrary rotation around a 
fixed axis has no effect on the solution. Therefore, an angular coordinate 
representing that rotation should be cyclic providing another considerable
simplification to the problem. Due to the spherical symmetry, the total angular
momentum 
\[
{\bf L}={\bf r}\times{\bf p} 
\]
is conserved.
Thus, it can be inferred that ${\bf r}$ is perpendicular to the fixed 
axis of ${\bf L}$. Now, if ${\bf L}=0$ the motion should be along a line 
passing through the center of force, since for ${\bf L}=0$ ${\bf r}$ and 
${\bf \dot{r}}$ are parallel. This happens only in the rectilinear
motion, and therefore central force motions proceed in one plane.

\noindent
By taking the $z$ axis as the direction of ${\bf L}$, the motion will take
place in the $(x,y)$ plane.
The spherical angular coordinate $\phi$ will have the constant value $\pi/2$ 
and we can go on as foollows. 
The conservation of the angular momentum provides
three independent constants of motion. As a matter of fact, two of them, 
expressing the constant direction of the angular momentum, are used 
to reduce the problem of three degrees of freedom 
to only two. The third coordinate corresponds to the conservation of the 
modulus of ${\bf L}$.

\noindent
In polar coordinates the Lagrangian is
\begin{equation}
L=\frac{1}{2}m(\dot{r}^{2}+r^{2}\dot{\theta}^{2})-V(r)~.  \label{eq4}
\end{equation}
As we have seen, $\theta$ is a cyclic coordinate
whose canonically conjugate momentum is the angular momentum
\[
p_{\theta }=\frac{\partial L}{\partial \dot{\theta}}=mr^{2}\dot{\theta}~, 
\]
then, one of the equations of motion will be
\begin{equation}
\dot{p}_{\theta }=\frac{d}{dt}(mr^{2}\dot{\theta})=0~.  \label{eq5}
\end{equation}
This leads us to 
\begin{equation}
mr^{2}\dot{\theta}=l=cte~,  \label{eq6}
\end{equation}
where $l$ is constant modulus of the angular momentum. From
equation (5) one also gets
\begin{equation}
\frac{d}{dt}
{r^{2}\dot{\theta} \overwithdelims() 2}
=0.  \label{eq7}
\end{equation}

\noindent
The factor $1/2$ is introduced because $(r^{2}\dot{\theta})/2$
is the {\it areolar velocity} (the area covered by the position vector
per unit of time).

\noindent
The conservation of the angular momentum is equivalent
to saying that the areolar velocity is constant.
This is nothing else than a proof of Kepler's second law of planetary motion: 
{\em the position vector of a planet covers equal areas in equal time 
intervals}. However, we stress that the constancy of the areolar velocity is
a property valid for any central force not only for inverse square ones.

\noindent
The other Lagrange equation for the $r$ coordinates reads
\begin{equation}
\frac{d}{dt}(m\dot{r})-mr\dot{\theta}^{2}+\frac{\partial V}{\partial r}=0~.
\label{eq8}
\end{equation}
Denoting the force by $f(r)$, we can write this equation as follows
\begin{equation}
m\ddot{r}-mr\dot{\theta}^{2}=f(r)~.  \label{eq9}
\end{equation}
Using the equation (6), the last equation can be rewritten as 
\begin{equation}
m\ddot{r}-\frac{l^{2}}{mr^{3}}=f(r).  \label{eq10}
\end{equation}

\noindent
Recalling now the conservation of the total energy
\begin{equation}
E=T+V=\frac{1}{2}m(\dot{r}^{2}+r^{2}\dot{\theta}^{2})+V(r)~.  \label{eq11}
\end{equation}
we say that $E$ is a constant of motion. This can be derived from the equations
of motion.
The equation (10) can be written as follows
\begin{equation}
m\ddot{r}=-\frac{d}{dr}\left[ V(r)+\frac{1}{2}\frac{l^{2}}{mr^{2}}\right]~,
\label{eq12}
\end{equation}
and by multiplying by $\dot{r}$ both sides, we get
\[
m\ddot{r}\dot{r}=\frac{d}{dt}(\frac{1}{2}m\dot{r})=-\frac{d}{dt}\left[ V(r)+%
\frac{1}{2}\frac{l^{2}}{mr^{2}}\right]~, 
\]
or
\[
\frac{d}{dt}\left[ \frac{1}{2}m\dot{r}^{2}+V(r)+\frac{1}{2}\frac{l^{2}}{%
mr^{2}}\right] =0~. 
\]
Thus
\begin{equation}
\frac{1}{2}m\dot{r}^{2}+V(r)+\frac{1}{2}\frac{l^{2}}{mr^{2}}=cte
\label{eq13}
\end{equation}
and since $(l^{2}/2mr^{2})=(mr^{2}\dot{\theta}/2)$, the equation (13) is reduced 
to (11).

\noindent
Now, let us solve the equations of motion for $r$ and $\theta$.
Taking $\dot{r}$ from equation (\ref{eq13}), we have
\begin{equation}
\dot{r}=\sqrt[2]{\frac{2}{m}(E-V-\frac{l^{2}}{2mr^{2}})}~,  \label{eq14}
\end{equation}
or
\begin{equation}
dt=\frac{dr}{\sqrt[2]{\frac{2}{m}(E-V-\frac{l^{2}}{2mr^{2}})}}~.  \label{eq15}
\end{equation}

\noindent
Let $r_{0}$ be the value of $r$ at $t=0$. The integral of the two terms of the 
equation reads
\begin{equation}
t=\int\nolimits_{r_{0}}^{r}\frac{dr}{\sqrt[2]{\frac{2}{m}(E-V-\frac{l^{2}}{%
2mr^{2}})}}.  \label{eq16}
\end{equation}

\noindent
This equation gives $t$ as a function of $r$ and of the constants of integration
$E$, $l$ and $r_{0}$. It can be inverted, at least in a formal way, 
to give $r$ as a function of $t$ and of the constants. Once we have
$r$, there is no problem  to get $\theta$ starting from equation
(6), that can be written as follows 
\begin{equation}
d\theta =\frac{ldt}{mr^{2}}~.  \label{eq17}
\end{equation}

\noindent
If $\theta _{0}$ is the initial value of $\theta$, then (17) will be
\begin{equation}
\theta =l\int\nolimits_{0}^{t}\frac{dt}{mr^{2}(t)}+\theta _{0}.  \label{eq18}
\end{equation}

\noindent
Thus, we have already get the equations of motion for the variables
$r$ and $\theta$.\bigskip

\noindent {\bf 2.3 The differential equation of the orbit}

\noindent
A change of our standpoint regarding the approach of real central force
problems prove to be convenient.
Till now, solving the problem meant seeking
$r$ and $\theta$ as functions of time and some constants of integration
such as $E$, $l$, etc.
However, quite often, what we are really looking for is the equation of the 
orbit, that is the direct dependence between $r$ and $\theta$, by 
eliminating the time parameter $t$. 
In the case of central force problems, this elimination is particularly simple
because $t$ is to be found in the equations of motion only in the form of a
variable with respect to which the derivatives are performed. Indeed, the 
equation of motion (\ref{eq6}) gives us a definite relationship between
$dt$ and $d\theta$ 
\begin{equation}
ldt=mr^{2}d\theta .  \label{eq19}
\end{equation}

\noindent
The corresponding relationship between its derivatives with respect to
$t$ and $\theta$ is 
\begin{equation}
\frac{d}{dt}=\frac{l}{mr^{2}}\frac{d}{d\theta }.  \label{eq20}
\end{equation}

\noindent
This relationship can be used to convert  (\ref{eq10})
in a differential equation for the orbit. At the same time, one can solve for 
the equations of motion and go on to get the orbit equation. For the time being,
we follow up the first route.

\noindent
From equation (\ref{eq20}) we can write the second derivative with respect to
$t$
\[
\frac{d^{2}}{dt^{2}}=\frac{d}{d\theta }\frac{l}{mr^{2}}\left( \frac{d}{%
d\theta }\frac{l}{mr^{2}}\right) 
\]
and the Lagrange equation for $r$, (\ref{eq10}), will be
\begin{equation}
\frac{l}{r^{2}}\frac{d}{d\theta }\left( \frac{l}{mr^{2}}\frac{dr}{d\theta }%
\right) -\frac{l}{mr^{3}}=f(r)~.  \label{eq21}
\end{equation}
But 
\[
\frac{1}{r^{2}}\frac{dr}{d\theta }=-\frac{d(1/r)}{d\theta }~. 
\]
Employing the change of variable $u=1/r$, we have
\begin{equation}
\frac{l^{2}u^{2}}{m}\left( \frac{d^{2}u}{d\theta ^{2}}+u\right) =-f\left( 
\frac{1}{u}\right)~.  \label{eq22}
\end{equation}
Since 
\[
\frac{d}{du}=\frac{dr}{d\theta }\frac{d}{dr}=-\frac{1}{u^{2}}\frac{d}{dr}~, 
\]
equation (\ref{eq22}) can be written as follows 
\begin{equation}
\frac{d^{2}u}{d\theta ^{2}}+u=-\frac{m}{l^{2}}\frac{d}{du}V\left( \frac{1}{u}%
\right) .  \label{eq23}
\end{equation}

\noindent
Any of the equations (\ref{eq22}) or (\ref{eq23}) is the differential equation 
of the orbit if we know the force $f$ or the potential $V$.
Vice versa, if we know the orbit equation we can get $f$ or $V$.

\noindent
For an arbitrary particular force law, the orbit equation can be obtained by
integrating the equation (\ref{eq22}). 
Since a great deal of work has been done when solving (\ref{eq10}),
we are left with the task of eliminating $t$ in the solution (\ref{eq15})
by means of (\ref{eq19}), 
\begin{equation}
d\theta =\frac{ldr}{mr^{2}\cdot \sqrt[2]{\frac{2}{m}\left[ E-V(r)-\frac{l^{2}%
}{2mr^{2}}\right] }}~,  \label{eq24}
\end{equation}
or 
\begin{equation}
\theta =\int_{r_{0}}^{r}\frac{dr}{r^{2}\cdot \sqrt[2]{\frac{2mE}{l^{2}}-%
\frac{2mU}{l^{2}}-\frac{1}{r^{2}}}}+\theta _{0}~.  \label{eq25}
\end{equation}
By the change of variable $u=1/r$, 
\begin{equation}
\theta =\theta _{0}-\int_{u_{0}}^{u}\frac{du}{\sqrt[2]{\frac{2mE}{l^{2}}-%
\frac{2mU}{l^{2}}-u^{2}}}~,  \label{eq26}
\end{equation}
which is the formal solution for the orbit equation.

\bigskip

\noindent {\bf 2.4 Kepler's problem: the case of inverse square force}

\noindent
The inverse square central force law is the most important of all and therefore
we shall pay more attention to this case.
The force and the potential are: 
\begin{equation}
f=-\frac{k}{r^{2}}\;\;\;\;\;{\rm y}\;\;\;\;\;V=-\frac{k}{r}~.  \label{eq27}
\end{equation}

\noindent
To integrate the orbit equation we put (23) in (22), 
\begin{equation}
\frac{d^{2}u}{d\theta ^{2}}+u=-\frac{mf(1/u)}{l^{2}u^{2}}=\frac{mk}{l^{2}}~.
\label{eq28}
\end{equation}
Now, we perform the change of variable $y=u-\frac{mk}{l^{2}}$~, in order that
the differential equation be written as follows 
\[
\frac{d^{2}y}{d\theta ^{2}}+y=0~, 
\]
possessing the solution
\[
y=B\cos (\theta -\theta {\acute{}})~, 
\]
where $B$ and $\theta {\acute{}}$ are the corresponding integration constants.
The solution in terms of $r$ is 
\begin{equation}
\frac{1}{r}=\frac{mk}{l^{2}}\left[ 1+e\cos (\theta -\theta {\acute{}}
)\right] ,  \label{eq29}
\end{equation}
where
\[
e=B\frac{l^{2}}{mk}~. 
\]

\noindent
We can get the orbit equation from the formal solution
(\ref{eq26}). Although the procedure is longer than solving the equation
(\ref{eq28}), it is nevertheless to do it since the integration constant
$e$ is directly obtained as a function of $E$ and $l$.

\noindent
We write equation (\ref{eq26}) as follows
\begin{equation}
\theta =\theta {\acute{}}
-\int \frac{du}{\sqrt[2]{\frac{2mE}{l^{2}}-\frac{2mU}{l^{2}}-u^{2}}}~,
\label{eq30}
\end{equation}
where now one deals with a definite integral. Then $\theta {\acute{}}$
of (\ref{eq30}) is an integration constant determined  through the initial
conditions
and is not necessarily the initial angle $\theta _{0}$ at $t=0$.
The solution for this type of integrals is
\begin{equation}
\int \frac{dx}{\sqrt[2]{\alpha +\beta x+\gamma x^{2}}}=\frac{1}{\sqrt[2]{%
-\gamma }}\arccos \left[ -\frac{\beta +2\gamma x}{\sqrt[2]{q}}\right]~,
\label{eq31}
\end{equation}
where
\[
q=\beta ^{2}-4\alpha \gamma . 
\]

\noindent
In order to apply this type of solutions to the equation (\ref{eq30}) we should
make 
\[
\alpha =\frac{2mE}{l^{2}},\;\;\;\;\;\beta =\frac{2mk}{l^{2}}%
,\;\;\;\;\;\gamma =-1, 
\]
and the discriminant $q$ will be
\[
q=\left( \frac{2mk}{l^{2}}\right) ^{2}\left( 1+\frac{2El^{2}}{mk^{2}}\right)
. 
\]

\noindent
With these substitutions, (\ref{eq30}) is
\[
\theta =\theta {\acute{}}
-\arccos \left[ \frac{\frac{l^{2}u}{mk}-1}{\sqrt[2]{1+\frac{2El^{2}}{mk^{2}}}
}\right]~. 
\]
For $u\equiv 1/r$, the resulting orbit equation is
\begin{equation}
\frac{1}{r}=\frac{mk}{l^{2}}\left[ 1+\sqrt[2]{1+\frac{2El^{2}}{mk^{2}}}\cos
(\theta -\theta {\acute{}})\right] .  \label{eq32}
\end{equation}

\noindent
Comparing (\ref{eq32}) with the equation (\ref{eq29}) we notice that the value
of $e$ is: 
\begin{equation}
e=\sqrt[2]{1+\frac{2El^{2}}{mk^{2}}}~.  \label{eq33}
\end{equation}

\noindent
The type of orbit depends on the value of $e$ according to the following table:

\begin{center}
\begin{tabular}{lll}
$e>1,$ & $E>0:$ & hyperbola, \\ 
$e=1,$ & $E=0:$ & parabola, \\ 
$e<1,$ & $E<0:$ & elipse, \\ 
$e=0$ & $E=-\frac{mk^{2}}{2l^{2}}:$ & circumference.
\end{tabular}

\bigskip
\end{center}

\noindent {\bf 2.5 Dispersion by a center of force}

\noindent
From a historical point of view, the interest on central forces was related
to the astronomical problem of planetary motions. However, there is no reason
to consider them only under these circumstances.
Another important issue that one can study within Classical Mechanics is the 
dispersion of particles by central forces. Of course, if the particles are
of atomic size, we should keep in mind that the classical formalism may not
give the right result because of quantum effects that begin to be important 
at those scales. Despite this, there are classical predictions that continue 
to be correct. Moreover, the main concepts of the dispersion phenomena
are the same in both Classical Mechanics and Quantum Mechanics; 
thus, one can learn this scientific idiom in the classical picture, usually
considered more convenient.

\noindent
In its one-body formulation, the dispersion problem refers to the action
of the center of force on the trajectories of the coming particles.
Let us consider a uniform beam of particles, (say electrons, protons, or
planets and comets), but all of the same mass and energy
impinging on a center of force.
We can assume that the force diminishes to zero at large distances.
The incident beam is characterized by its {\it intensity} $I$ 
(also called flux density), which is the number of particles that
pass through per units of time and normal surface.
When one particle comes closer and closer to the center of force will be
attracted or repelled, and its orbit will deviate from the initial rectilinear 
path.
Once it passed the center of force, the perturbative effects will diminish
such that the orbit tends again to a streight line. 
In general, the final direction of the motion does not coincide 
with the incident one. One says that the particle has been dispersed.
By definition, the 
{\it differential cross section} $\sigma (\Omega )$ is 
\begin{equation}
\sigma (\Omega )d\Omega =\frac{dN}{I},
\label{eq34}
\end{equation}
where $dN$ is the number of particles dispersed per unit of
time in the element of solid angle $d\Omega$ around the
$\Omega$ direction.
In the case of central forces there is a high degree of symmetry around the 
incident beam axis. Therefore, the element of solid angle can be written 
\begin{equation}
d\Omega =2\pi \sin \Theta d\Theta ,  \label{eq35}
\end{equation}
where $\Theta$ is the angle between two 
incident dispersed directions, and is called {\it the dispersion angle}.

\noindent
For a given arbitrary particle the constants of the orbit
and therefore the degree of dispersion are determined by its energy and angular
momentum. It is convenient to express the latter in terms of a function of 
energy and the so-called impact parameter $s$, which by definition 
is the distance from the center of force to the straight suport line
of the incident velocity. If $u_{0}$ is the incident velocity
of the particle, we have 
\begin{equation}
l=mu_{0}s=s\cdot \sqrt[2]{2mE}.  \label{eq36}
\end{equation}

\noindent
Once $E$ and $s$ are fixed, the angle of dispersion
$\Theta$ is uniquely determined. For the time being, we suppose that different
values of $s$ cannot lead to the same dispersion angle. Therefore, the number of
dispersed particles in the element of solid angle $d\Omega$ 
between $\Theta$ and $\Theta +d\Theta$ should be equal to the number of 
incident particles whose impact parameter ranges within
the corresponding $s$ and $s+ds$: 
\begin{equation}
2\pi Is\left| ds\right| =2\pi \sigma (\Theta )I\sin \Theta \left| d\Theta
\right| .  \label{eq37}
\end{equation}

\noindent
In the equation (\ref{eq37}) we have introduced absolute values
because while the number of particles is always positive,
$s$ and $\Theta$ can vary in opposite directions. If we consider 
$s$ as a function of the energy and the corresponding dispersion angle, 
\[
s=s(\Theta ,E), 
\]
the dependence of the cross section of $\Theta$ will be given by
\begin{equation}
\sigma (\Theta )=\frac{s}{\sin \Theta }\left| \frac{ds}{d\Theta }\right| .
\label{eq38}
\end{equation}

\noindent
From the orbit equation (\ref{eq25}), one can obtain directly
a formal expression for the dispersion angle.
In addition, for the sake of simplicity, we tackle the case of a pure repulsive
dispersion. Since the orbit should be symmetric with respect to the 
direction of the periapsis, the dispersion angle is 
\begin{equation}
\Theta =\pi -2\Psi~,  \label{eq39}
\end{equation}
where $\Psi$ is the angle between the direction of the incident asymptote
and the direction of the periapsis. In turn, $\Psi$ can be obtained 
from the equation (\ref{eq25}) by making $r_{0}=\infty$ when 
$\theta _{0}=\pi$ (incident direction). Thus,
$\theta =\pi -\Psi $ when $r=r_{m}$, the closest distance of the particle to the 
center of force. Then, one can easily obtain 
\begin{equation}
\Psi =\int\nolimits_{r_{m}}^{\infty }\frac{dr}{r^{2}\cdot \sqrt[2]{\frac{2mE%
}{l^{2}}-\frac{2mV}{l^{2}}-\frac{1}{r^{2}}}}~.  \label{eq40}
\end{equation}

\noindent
Expressing $l$ as a function of the impact parameter $s$ (eq. (\ref{eq36})), 
the result is
\begin{equation}
\Theta =\pi -2\int\nolimits_{r_{m}}^{\infty }\frac{sdr}{r\cdot \sqrt[2]{r^{2}%
\left[ 1-\frac{V(r)}{E}\right] -s^{2}}}~,  \label{eq41}
\end{equation}
or 
\begin{equation}
\Theta =\pi -2\int\nolimits_{0}^{u_{m}}\frac{sdu}{\sqrt[2]{1-\frac{v(u)}{E}%
-s^{2}u^{2}}}~.  \label{eq42}
\end{equation}

\noindent
The equations (\ref{eq41}) and (\ref{eq42}) are used rarely, as they do not 
enter in a direct way in the numerical calculation of the dispersion angle.
However, when an analytic expression for the orbits is available, one can often
get, merely by simple inspection, a relationship between $\Theta$ and $s$.

\bigskip

\noindent \underline{{\it EXAMPLE}}:

\noindent
This example is very important from the historical point of view.
It refers to the repulsive dispersion  of charged particles in a Coulomb field.
The field is produced by a fixed charge $-Ze$
and acts on incident particles of charge $-Z{\acute{}}e$;
therefore, the force can be written as follows 
\[
f=\frac{ZZ{\acute{}}e^{2}}{r^{2}}~,
\]
that is, one deals with a repulsive inverse square force.
The constant is 
\begin{equation}
k=-ZZ{\acute{}}e^{2}.  \label{eq43}
\end{equation}

\noindent
The energy $E$ is positive implying a hyperbolic orbit of eccentricity 
\begin{equation}
\epsilon =\sqrt[2]{1+\frac{2El^{2}}{m(ZZ{\acute{}}
e^{2})^{2}}}=\sqrt[2]{1+\left( \frac{2Es}{ZZ{\acute{}}
e^{2}}\right) ^{2}},  \label{eq44}
\end{equation}
where we have taken into account the equation (\ref{eq36}). If the 
angle $\theta {\acute{}}$ is taken to be $\pi$, then from the 
equation (\ref{eq29}) we come to the conclusion that
the periapse corresponds to $\theta =0$ and the orbit equation reads
\begin{equation}
\frac{1}{r}=\frac{mZZ{\acute{}}
e^{2}}{l^{2}}\left[ \epsilon \cos \theta -1\right] .  \label{eq45}
\end{equation}

\noindent
The direction $\Psi$ of the incident asymptote is thus determined by the 
condition $r\rightarrow \infty$: 
\[
\cos \Psi =\frac{1}{\epsilon}~, 
\]
that is, according to equation (\ref{eq39}), 
\[
\sin \frac{\Theta }{2}=\frac{1}{\epsilon }~. 
\]

\noindent
Then,
\[
\cot ^{2}\frac{\Theta }{2}=\epsilon ^{2}-1, 
\]
and by means of equation (\ref{eq44}) 
\[
\cot \frac{\Theta }{2}=\frac{2Es}{ZZ{\acute{}}e^{2}}~.
\]

\noindent
The functional relationship between the impact parameter and the dispersion
angle will be 
\begin{equation}
s=\frac{ZZ{\acute{}}
e^{2}}{2E}\cot \frac{\Theta }{2},  \label{eq46}
\end{equation}
and by effecting the transformation required by the equation
(\ref{eq38}) we find that $\sigma (\Theta )$ is given by
\begin{equation}
\sigma (\Theta )=\frac{1}{4}\left( \frac{ZZ{\acute{}}
e^{2}}{2E}\right) ^{2}\csc ^{4}\frac{\Theta }{2}.  \label{eq47}
\end{equation}

\noindent
The equation (\ref{eq47}) gives the famous Rutherford scattering cross section
derived by him for the dispersion of
$\alpha$ particles on atomic nuclei. In the nonrelativistic limit,
the same result is provided by the quantum mechanical calculations.

\noindent
The concept of {\it total cross section} $\sigma _{T}$ is very important in 
atomic physics. Its definition is
\[
\sigma _{T}=\int\nolimits_{4\pi }\sigma (\Omega )d\Omega =2\pi
\int\nolimits_{0}^{\pi }\sigma (\Theta )d\Theta~. 
\]
However, if we calculate the total cross section for the Coulombian dispersion
by substituting the equation (\ref{eq47}) in the definition above we get an 
infinite result. The physical reason is easy to see.
According to the definition, the total cross section is the number of particles
per unit of incident intensity that are dispersed in all directions.
The Coulombian field is an example of long-range force; its effects are still
present in the infinite distance limit.
The small deviation limit is valid only for particles of large impact parameter.
Therefore, for an incident beam of infinite lateral extension all the particles
will be dispersed and should be included in the total cross section.
It is clear that the infinite value of $\sigma _{T}$ is not a special 
property of the Coulombian field and occurs for any type of long-range field.

\bigskip
\bigskip

\begin{center}  Further reading  \end{center}

\bigskip

\noindent
L.S. Brown, {\it Forces giving no orbit precession}, Am. J. Phys. 46, 930
(1978)

\bigskip

\noindent
H. Goldstein, {\it More on the prehistory of the Laplace-Runge-Lenz vector},
Am. J. Phys. 44, 1123 (1976)



%% file: jul2en.tex



\centerline{{\Large 3. THE RIGID BODY}}

\bigskip
\bigskip

\noindent
{\bf Forward}:
Due to its particular features, 
the study of the motion of the rigid
body has generated several interesting mathematical techniques and methods.
In this chapter, we briefly present the basic rigid body concepts.

\bigskip
\bigskip

{\bf CONTENTS:}

\bigskip

3.1 Definition

\bigskip

3.2 Degrees of freedom

\bigskip

3.3 Tensor of inertia (with example)

\bigskip

3.4 Angular momentum

\bigskip

3.5 Principal axes of inertia (with example)

\bigskip

3.6 The theorem of parallel axes (with 2 examples)

\bigskip

3.7 Dynamics of the rigid body (with example)

\bigskip

3.8 Symmetrical top free of torques

\bigskip

3.9 Euler angles

\bigskip

3.10 Symmetrical top with a fixed point

\bigskip

\newpage

\noindent {\bf 3.1 Definition}

\noindent
A rigid body (RB) is defined as a system of particles whose relative distances 
are forced to stay constant during the motion.

\bigskip

\noindent{\bf 3.2 Degrees of freedom}

\noindent
In order to describe the general motion of a RB in the 
three-dimensional space one needs six variables, for example the three
coordinates of the center of mass measured with respect to an inertial
frame and three angles for labeling the orientation of the body in space (or
of a fixed system within the body with the origin in the center of mass).
In other words, in the three-dimensional space
the RB can be described by at most six degrees of freedom.

\noindent
The number of degrees of freedom may be less when the rigid body is subjected
to various conditions as follows:

\begin{itemize}
\item  If the RB rotates around a single axis
       there is only one degree of freedom (one angle).

\item  If the RB moves in a plane, its motion can be described 
       by five degrees of freedom (two coordinates and three angles).
\end{itemize}

\bigskip

\noindent{\bf 3.3 Tensor of inertia}.

\noindent
We consider a body made of $N$ particles of masses
$m_{\alpha }$, $\alpha =1,2,3...,N$. If the body rotates at angular velocity
$
{\bf \omega }$ around a fixed point in the body and this point, in turn, moves
at velocity ${\bf v}$ with respect to a fixed inertial system, then the 
velocity of the $\alpha$th particle w.r.t. the inertial system is given by
\setcounter{equation} {0}\\
\begin{equation}
{\bf v}_{\alpha }={\bf v+\omega \times r}_{\alpha }.  \label{eq1a}
\end{equation}

\noindent
The kinetic energy of the $\alpha$th particle is
\begin{equation}
T_{\alpha }=\frac{1}{2}m_{\alpha }{\bf v}_{\alpha }^{2}~,  \label{eq2a}
\end{equation}
where
\begin{equation}
{\bf v}_{\alpha }^{2}={\bf v}_{\alpha }\cdot {\bf v}_{\alpha }=({\bf %
v+\omega \times r}_{\alpha })\cdot ({\bf v+\omega \times r}_{\alpha }) 
\nonumber
\end{equation}
\[
={\bf v}\cdot {\bf v}+2{\bf v}\cdot ({\bf \omega \times r}_{\alpha })+({\bf %
\omega \times r}_{\alpha })\cdot ({\bf \omega \times r}_{\alpha }) 
\]
\begin{equation}
={\bf v}^{2}+2{\bf v(\omega \times r}_{\alpha })+({\bf \omega \times r}%
_{\alpha })^{2}.  \label{eq3a}
\end{equation}

\noindent
Then the total energy is 
\begin{eqnarray}
T& =\sum_{\alpha }T_{\alpha }=\sum_{\alpha }\frac{1}{2}m_{\alpha }{\bf v}
^{2}+\sum_{\alpha }m_{\alpha }\left[ {\bf v\cdot }\left( {\bf \omega \times r
}_{\alpha }\right) \right] + \nonumber \\
& +\frac{1}{2}\sum_{\alpha }m_{\alpha }({\bf \omega \times r}_{\alpha })^{2}~;
\nonumber \\
T& =\frac{1}{2}M{\bf v}^{2}+{\bf v\cdot }\left[ {\bf \omega \times }%
\sum_{\alpha }m_{\alpha }{\bf r}_{\alpha }\right] +\frac{1}{2}\sum_{\alpha}
m_{\alpha }\left( {\bf \omega \times r}_{\alpha }\right) ^{2}. \nonumber
\end{eqnarray}

\noindent
If the  origin is fixed to the solid body, we can take it in the center of mass.
Thus, 
\[
{\bf R=}\frac{\sum_{\alpha }m_{\alpha }{\bf r}_{\alpha }}{M}=0, 
\]
and therefore we get 
\begin{equation}
T=\frac{1}{2}M{\bf v}^{2}+\frac{1}{2}\sum_{\alpha }m_{\alpha }\left( {\bf %
\omega \times r}_{\alpha }\right) ^{2}  \label{eq4a}
\end{equation}
\begin{equation}
T=T_{trans}+T_{rot}  \label{eq5a}
\end{equation}
where 
\begin{equation}
T_{trans}=\frac{1}{2}\sum_{\alpha }m_{\alpha }{\bf v}^{2}=\frac{1}{2}
M{\bf v}^{2}  \label{eq6a}
\end{equation}
\begin{equation}
T_{rot}=\frac{1}{2}\sum_{\alpha }m_{\alpha }\left( {\bf \omega \times r}%
_{\alpha }\right) ^{2}.  \label{eq7a}
\end{equation}

\noindent
In Eq.~(\ref{eq7a}) we use the vectorial identity
\begin{equation}
({\bf A}\times {\bf B})^{2}={\bf A}^{2}{\bf B}^{2}-({\bf A\cdot B})^{2}
\label{eq8a}
\end{equation}
to get the following form of the equation  
\[
T_{rot}=\frac{1}{2}\sum_{\alpha}m_{\alpha}\left[ {\bf \omega }^{2}{\bf r}%
^{2}-({\bf \omega \cdot r}_{\alpha })^{2}\right]~, 
\]
which in terms of the components of ${\bf \omega }$ and ${\bf r}$ 
\[
{\bf \omega }={\bf (}\omega _{1},\omega _{2},\omega _{3})\;\;\;{\rm and}\;\;\;%
{\bf r}_{\alpha }=(x_{\alpha 1},x_{\alpha 2},x_{\alpha 3}) 
\]
can be written as follows
\[
T_{rot}=\frac{1}{2}\sum_{\alpha }m_{\alpha }\left\{ \left( 
\mathop{\textstyle\sum}
_{i}\omega _{i}^{2}\right) \left( 
\mathop{\textstyle\sum}
_{k}x_{\alpha k}^{2}\right) -\left( 
\mathop{\textstyle\sum}
_{i}\omega _{i}x_{\alpha i}\right) \left( 
\mathop{\textstyle\sum}
_{j}\omega _{j}x_{\alpha j}\right) \right\} . 
\]

\noindent Now, we introduce  
\[
\omega _{i}=
\mathop{\textstyle\sum}
_{j}\delta _{ij}\omega _{j} 
\]
\begin{equation}
T_{rot}=\frac{1}{2}\sum_{\alpha }\sum_{ij}m_{\alpha }\left\{ \omega
_{i}\omega _{j}\delta _{ij}\left( 
\mathop{\textstyle\sum}
_{k}x_{\alpha k}^{2}\right) -\omega _{i}\omega _{j}x_{\alpha i}x_{\alpha
j}\right\}  \nonumber
\end{equation}
\begin{equation}
T_{rot}=\frac{1}{2}\sum_{ij}\omega _{i}\omega _{j}\sum_{\alpha }m_{\alpha }
\left[ \delta _{ij}
\mathop{\textstyle\sum}
_{k}x_{\alpha k}^{2}-x_{\alpha i}x_{\alpha j}\right] .  \label{eq9a}
\end{equation}

\noindent
We can write $T_{rot}$ as follows 
\begin{equation}
T_{rot}=\frac{1}{2}\sum_{ij}I_{ij}\omega _{i}\omega _{j}  \label{eq10a}
\end{equation}
where 
\begin{equation}
I_{ij}=\sum_{\alpha }m_{\alpha }\left[ \delta _{ij}
\mathop{\textstyle\sum}
_{k}x_{\alpha k}^{2}-x_{\alpha i}x_{\alpha j}\right] .  \label{eq11a}
\end{equation}

\noindent
The nine quantities $I_{ij}$ are the components of a new mathematical
entity, denoted by $\left\{ I_{ij}\right\}$ and called 
{\it tensor of inertia}. 
It can be written in a 
convenient way as a ($3\times 3$) matrix 
\[
\left\{ I_{ij}\right\} =\left( 
\begin{array}{ccc}
I_{11} & I_{12} & I_{13} \\ 
I_{21} & I_{22} & I_{23} \\ 
I_{31} & I_{32} & I_{33}
\end{array}
\right)= 
\]
\begin{equation}
=\left( 
\begin{array}{ccc}
\sum_{\alpha }m_{\alpha }(x_{\alpha 2}^{2}+x_{\alpha 3}^{2}) & -\sum_{\alpha
}m_{\alpha }x_{\alpha 1}x_{\alpha 2} & -\sum_{\alpha }m_{\alpha }x_{\alpha
1}x_{\alpha 3} \\ 
-\sum_{\alpha }m_{\alpha }x_{\alpha 2}x_{\alpha 1} & \sum_{\alpha }m_{\alpha
}(x_{\alpha 1}^{2}+x_{\alpha 3}^{2}) & -\sum_{\alpha }m_{\alpha }x_{\alpha
2}x_{\alpha 3} \\ 
-\sum_{\alpha }m_{\alpha }x_{\alpha 3}x_{\alpha 1} & -\sum_{\alpha
}m_{\alpha }x_{\alpha 3}x_{\alpha 2} & \sum_{\alpha }m_{\alpha }(x_{\alpha
1}^{2}+x_{\alpha 2}^{2})
\end{array}
\right) .  \label{eq12a}
\end{equation}

\noindent
We note that $I_{ij}=I_{ji}$, and therefore $\left\{ I_{ij}\right\}$ is
a {\it symmetric tensor}, implying that only six of the components
are independent.
The diagonal elements of $\left\{ I_{ij}\right\}$ are called
{\it moments of inertia} with respect to the axes of coordinates, whereas the 
negatives of the nondiagonal elements are called the
{\it products of inertia}. For a continuous distribution of mass, of
density $\rho ({\bf r)}$, $\left\{ I_{ij}\right\}$ is written
in the following way
\begin{equation}
I_{ij}=\int_{V}\rho ({\bf r)}\left[ \delta _{ij}
\mathop{\textstyle\sum}
_{k}x_{k}^{2}-x_{i}x_{j}\right] dV.  \label{eq13a}
\end{equation}

\bigskip

\noindent \underline{{\it EXAMPLE}}:

\noindent
Find the elements $I_{ij}$ of the tensor of
inertia $\left\{ I_{ij}\right\}$ for a cube of uniform density of side $b$,
mass $M$, with one corner placed at the origin.
$$
I_{11} =\int\limits_{V}\rho \left[ x_{1}^{2}+x_{2}^{2}+x_{3}^{2}-x_{1}x_{1}
\right] dx_{1}dx_{2}dx_{3}
 =\rho
\int\limits_{0}^{b}\int\limits_{0}^{b}\int
\limits_{0}^{b}(x_{2}^{2}+x_{3}^{2})dx_{1}dx_{2}dx_{3}~.
$$

\noindent
The result of the three-dimensional integral is $I_{11}=\frac{2}{3}
(\rho b^{3})^{2}=\frac{2}{3}Mb^{2}$.
$$
I_{12} =\int\limits_{V}\rho (-x_{1}x_{2})dV 
 =-\rho
\int\limits_{0}^{b}\int\limits_{0}^{b}\int
\limits_{0}^{b}(x_{1}x_{2})dx_{1}dx_{2}dx_{3} 
 =-\frac{1}{4}\rho b^{5}=-\frac{1}{4}Mb^{2}~.
$$

\noindent
We see that all the other integrals are equal, so that 
\[
I_{11}=I_{22}=I_{33}=\frac{2}{3}Mb^{2} 
\]
\[
\mathrel{\mathop{I_{ij}}\limits_{i\neq j}}
=-\frac{1}{4}Mb^{2}~, 
\]
leading to the following form of the matrix  
\[
\left\{ I_{ij}\right\} =\left( 
\begin{array}{ccc}
\frac{2}{3}Mb^{2} & -\frac{1}{4}Mb^{2} & -\frac{1}{4}Mb^{2} \\ 
-\frac{1}{4}Mb^{2} & \frac{2}{3}Mb^{2} & -\frac{1}{4}Mb^{2} \\ 
-\frac{1}{4}Mb^{2} & -\frac{1}{4}Mb^{2} & \frac{2}{3}Mb^{2}
\end{array}
\right) . 
\]

\bigskip

\noindent{\bf 3.4 Angular Momentum}

\noindent
The angular momentum of a RB made of $N$
particles of masses $m_{\alpha }$ is given by
\begin{equation}
{\bf L=}\sum_{\alpha }{\bf r}_{\alpha }\times {\bf p}_{\alpha }~,
\label{eq14a}
\end{equation}
where 
\begin{equation}
{\bf p}_{\alpha }=m_{\alpha }{\bf v}_{\alpha }=m_{\alpha }({\bf \omega }%
\times {\bf r}_{\alpha })~.  \label{eq15a}
\end{equation}
Substituting (\ref{eq15a}) in (\ref{eq14a}), we get 
\[
{\bf L=}\sum_{\alpha }m_{\alpha }{\bf r}_{\alpha }\times ({\bf \omega }%
\times {\bf r}_{\alpha })~. 
\]
Employing the vectorial identity
\[
{\bf A}\times ({\bf B}\times {\bf A)=(A}\cdot {\bf A}){\bf B}-({\bf A}\cdot 
{\bf B}){\bf A}={\bf A}^{2}{\bf B}-({\bf A}\cdot {\bf B}){\bf A}~, 
\]
leads to
\[
{\bf L=}\sum_{\alpha }m_{\alpha }({\bf r}_{\alpha }^{2}{\bf \omega -r}%
_{\alpha }({\bf \omega \cdot r}_{\alpha }). 
\]

\noindent
Considering the $i$th component of the vector ${\bf L}$
\[
L_{i}=\sum_{\alpha }m_{\alpha }\left( \omega _{i}%
\mathop{\textstyle\sum}%
_{k}x_{\alpha k}^{2}\right) {\bf -}x_{\alpha i}\left( 
\mathop{\textstyle\sum}
_{j}x_{\alpha j}\omega _{j}\right)~, 
\]
and introducing the equation 
\[
\omega _{i}=
\mathop{\textstyle\sum}
_{j}\omega _{j}\delta _{ij}~, 
\]
we get
\begin{eqnarray}
L_{i}& =\sum_{\alpha }m_{\alpha }\left( 
\mathop{\textstyle\sum}
_{j}\delta _{ij}\omega _{j}
\mathop{\textstyle\sum}
_{k}x_{\alpha k}^{2}\right) {\bf -}\left( 
\mathop{\textstyle\sum}
_{j}x_{\alpha j}x_{\alpha j}\omega _{j}\right) \\
& =\sum_{\alpha }m_{\alpha }
\mathop{\textstyle\sum}
_{j}\omega _{j}\delta _{ij}\left( 
\mathop{\textstyle\sum}
_{k}x_{\alpha k}^{2}{\bf -}x_{\alpha i}x_{\alpha j}\right) \\
& =\sum_{j}\omega _{j}
\mathop{\textstyle\sum}
_{\alpha }m_{\alpha }\left( \delta _{ij}
\mathop{\textstyle\sum}
_{k}x_{\alpha k}^{2}{\bf -}x_{\alpha i}x_{\alpha j}\right)~.
\end{eqnarray}
Comparing with the equation (\ref{eq11a}) leads to 
\begin{equation}
L_{i}=\sum_{j}I_{ij}\omega _{j}~.  \label{eq16a}
\end{equation}
This equation can also be written in the form 
\begin{equation}
{\bf L=}\left\{ I_{ij}\right\} {\bf \omega }~,  \label{eq17a}
\end{equation}
or 
\begin{equation}
\left( 
\begin{array}{c}
L_{1} \\ 
L_{2} \\ 
L_{3}
\end{array}
\right) =\left( 
\begin{array}{ccc}
I_{11} & I_{12} & I_{13} \\ 
I_{21} & I_{22} & I_{23} \\ 
I_{31} & I_{32} & I_{33}
\end{array}
\right) \left( 
\begin{array}{c}
\omega _{1} \\ 
\omega _{2} \\ 
\omega _{3}
\end{array}
\right) .  \label{eq18a}
\end{equation}

\noindent
The rotational kinetic energy, $T_{rot}$, can be related to the 
angular momentum as follows: first, multiply the equation (
\ref{eq16a}) by $\frac{1}{2}\omega _{i}$
\begin{equation}
\omega _{i}\frac{1}{2}L_{i}=\frac{1}{2}\omega _{i}\sum_{j}I_{ij}\omega _{j}~,
\label{eq19a}
\end{equation}
and next summing over all the $i$ indices, gives
\[
\sum_{i}\frac{1}{2}L_{i}\omega _{i}=\frac{1}{2}\sum_{ij}I_{ij}\omega
_{i}\omega _{j}~. 
\]
Compararing this equation with (\ref{eq10a}), we see that the second term
is $T_{rot}$. Therefore
\begin{equation}
T_{rot}=\sum_{I}\frac{1}{2}L_{i}\omega _{i}=\frac{1}{2}{\bf L\cdot \omega}~ .
\label{eq20a}
\end{equation}

\noindent
Now, we substitute (\ref{eq17a}) in the equation (\ref{eq20a}), getting
the relationship between $T_{rot}$ and the tensor of inertia 
\begin{equation}
T_{rot}=\frac{1}{2}{\bf \omega \cdot }\left\{ I_{ij}\right\} {\bf \cdot
\omega .}  \label{eq21a}
\end{equation}

\bigskip

\noindent{\bf 3.5 Principal axes of inertia}

\noindent
Taking the tensor of inertia $\left\{ I_{ij}\right\} $ diagonal,
that is $I_{ij}=I_{i}\delta _{ij}$, the rotational kinetic energy
and the angular momentum are expressed as follows 
\[
T_{rot}=\frac{1}{2}\sum_{ij}I_{ij}\omega _{i}\omega _{j} 
\]
\[
=\frac{1}{2}\sum_{ij}\delta _{ij}I_{i}\omega _{i}\omega _{j} 
\]
\begin{equation}
T_{rot}=\frac{1}{2}\sum_{i}I_{i}\omega _{i}^{2}  \label{eq22a}
\end{equation}
and
\[
L_{i}=\sum_{j}I_{ij}\omega _{j} 
\]
\[
=\sum_{j}\delta _{ij}I_{i}\omega _{j}=I_{i}\omega _{i} 
\]
\begin{equation}
{\bf L}={\bf I\omega .}  \label{eq23a}
\end{equation}

\noindent
To seek a diagonal form of $\left\{ I_{ij}\right\} $ is equivalent to finding
a new system of three axes for which the kinetic energy and the angular momentum
take the form given by (\ref{eq22a})
and (\ref{eq23a}). In this case the axes are called
{\it principal axes of inertia}. That means that given an inertial reference
system within the body, we can pass from it to the principal axes by a 
particular orthogonal transformation, which is called {\it transformation to the
principal axes}.

\noindent
Making equal the components of (\ref{eq17a}) and (\ref{eq23a}), we have 
\begin{eqnarray}
L_{1}& =I\omega _{1}=I_{11}\omega _{1}+I_{12}\omega _{2}+I_{13}\omega _{3} \\
L_{2}& =I\omega _{2}=I_{21}\omega _{1}+I_{22}\omega _{2}+I_{23}\omega _{3} \\
L_{3}& =I\omega _{3}=I_{31}\omega _{1}+I_{32}\omega _{2}+I_{33}\omega _{3}
~.
\end{eqnarray}

\noindent
This is a system of equations that can be rewritten as 
\begin{eqnarray}
(I_{11}-I)\omega _{1}+I_{12}\omega _{2}+I_{13}\omega _{3}& =0  \label{eq24a}
\\
I_{21}\omega _{1}+(I_{22}-I)\omega _{2}+I_{23}\omega _{3}& =0  \nonumber \\
I_{31}\omega _{1}+I_{32}\omega _{2}+(I_{33}-I)\omega _{3}& =0~.  \nonumber
\end{eqnarray}

\noindent
To get nontrivial solutions, the determinant of the system should be zero
\begin{equation}
\left| 
\begin{array}{ccc}
(I_{11}-I)\omega _{1} & I_{12}\omega _{2} & I_{13}\omega _{3} \\ 
I_{21}\omega _{1} & (I_{22}-I)\omega _{2} & I_{23}\omega _{3} \\ 
I_{31}\omega _{1} & I_{32}\omega _{2} & (I_{33}-I)\omega _{3}
\end{array}
\right| =0~.  \label{eq25a}
\end{equation}

\bigskip

\noindent
This determinant leads to a polynomial of third order in $I$, known as
{\it the characteristic polynomial}. The equation (\ref{eq25a}) is called the
{\it secular equation} or {\it characteristic equation}. In practice,
 the principal moments of inertia, being the eigenvalues of ${\bf I}$,
 are obtained as solutions of the secular equation.

\bigskip

\noindent \underline{{\it EXAMPLE}}:

\noindent
Determine the principal axes of inertia for the cube of the previous 
example.

\bigskip

\noindent
Substituting the values obtained in the previous example
in the equation
(\ref{eq25a}) we get: 
\[
\left| \left( 
\begin{array}{ccc}
(\frac{2}{3}\beta -I) & -\frac{1}{4}\beta & -\frac{1}{4}\beta \\ 
-\frac{1}{4}\beta & (\frac{2}{3}\beta -I) & -\frac{1}{4}\beta \\ 
-\frac{1}{4}\beta & -\frac{1}{4}\beta & (\frac{2}{3}\beta -I)
\end{array}
\right) \right| =0 ~,
\]
where $\beta =Mb^{2}$. Thus, the characteristic equation will be
\[
\left( \frac{11}{12}\beta -I\right) \left( \frac{11}{12}\beta -I\right)
\left( \frac{1}{6}\beta -I\right) =0~. 
\]
The solutions, i.e., the principal moments of inertia are: 
\[
I_{1}=\frac{1}{6}\beta ,\;\;\;\;I_{2}=I_{3}=\frac{11}{12}\beta~, 
\]

\bigskip
\noindent
whose corresponding eigenvalues are given by 
\[
I=\frac{1}{6}\beta \leftrightarrow \frac{1}{\sqrt[2]{3}}\left( 
\begin{array}{c}
1 \\ 
1 \\ 
1
\end{array}
\right) ,\;\;\;\;I_{2},I_{3}=\frac{11}{12}\beta \leftrightarrow \frac{1}{%
\sqrt[2]{2}}\left\{ \left( 
\begin{array}{c}
-1 \\ 
1 \\ 
0
\end{array}
\right) ,\left( 
\begin{array}{c}
-1 \\ 
0 \\ 
1
\end{array}
\right) \right\}~. 
\]
The matrix that diagonalizes $\left\{ I_{ij}\right\}$ is: 
\[
\lambda =\sqrt[2]{\frac{1}{3}}\left( 
\begin{array}{ccc}
1 & -\sqrt[2]{\frac{3}{2}} & -\sqrt[2]{\frac{3}{2}} \\ 
1 & \sqrt[2]{\frac{3}{2}} & 0 \\ 
1 & 0 & \sqrt[2]{\frac{3}{2}}
\end{array}
\right)~. 
\]
The diagonalized $\left\{ I_{ij}\right\}$ will be 
\[
\left\{ I_{ij}\right\} _{diag}=\left( \lambda
\right) ^{*}\left\{ I_{ij}\right\} \lambda =\left(
\begin{array}{ccc}
\frac{1}{6}\beta  & 0 & 0 \\ 
0 & \frac{11}{12}\beta  & 0 \\ 
0 & 0 & \frac{11}{12}\beta 
\end{array}
\right) .
\]

\bigskip

\noindent{\bf 3.6 The theorem of parallel axes}

\noindent
We suppose that the system $x_{1},x_{2},x_{3}$ has the origin in the center of
mass of the RB. A second system $X_{1},X_{2},X_{3}$,
has the origin in another position w.r.t. the first system. The only imposed 
condition on them is to be parallel. We define the vectors ${\bf r}%
=(x_{1},x_{2},x_{3})$, ${\bf R}=(X_{1},X_{2},X_{3})$ y ${\bf a}%
=(a_{1},a_{2},a_{3})$ in such a way that ${\bf R}={\bf r}+{\bf a}$, or in 
component form
\begin{equation}
X_{i}=x_{i}+a_{i}.  \label{eq26a}
\end{equation}

\noindent
Let $J_{ij}$ be the components of the tensor of inertia w.r.t. the system $
X_{1}X_{2}X_{3}$, 
\begin{equation}
J_{ij}=\sum_{\alpha }m_{\alpha }\left[ \delta _{ij}
\mathop{\textstyle\sum}
_{k}X_{\alpha k}^{2}-X_{\alpha i}X_{\alpha j}\right]~.  \label{eq27a}
\end{equation}
We substitute (\ref{eq26a}) in (\ref{eq27a}), 
\[
J_{ij}=\sum_{\alpha }m_{\alpha }\left[ \delta _{ij}
\mathop{\textstyle\sum}
_{k}(x_{\alpha k}+a_{k})^{2}-(x_{\alpha i}+a_{i})(x_{\alpha j}+a_{j})\right] 
\]
\begin{eqnarray}
& =\left[ 
\mathop{\textstyle\sum}
_{\alpha }m_{\alpha }\left( \delta _{ij}
\mathop{\textstyle\sum}
_{k}(x_{\alpha k})^{2}-x_{\alpha i}x_{\alpha j}\right) \right] +
\mathop{\textstyle\sum}
_{\alpha }m_{\alpha }\left( \delta _{ij}
\mathop{\textstyle\sum}
_{k}a_{k}^{2}-a_{i}a_{j}\right)  \label{eq28a} \\
& +\left[ 
\mathop{\textstyle\sum}
_{k}2a_{k}\delta _{ij}\left( 
\mathop{\textstyle\sum}
_{\alpha }m_{\alpha }x_{\alpha k}\right) -a_{j}\left( 
\mathop{\textstyle\sum}
_{\alpha }m_{\alpha }x_{\alpha j}\right) -a_{i}\left( 
\mathop{\textstyle\sum}
_{\alpha }m_{\alpha }x_{\alpha i}\right) \right]~.  \nonumber
\end{eqnarray}
Since the center of mass coordinate is defined as 
\[
\bar{x}=\frac{
\mathop{\textstyle\sum}_{\alpha }m_{\alpha }x_{\alpha }}{M}
\]
we take into account that we have already set the origin in the center of 
mass, i.e., 
\[
(\bar{x}_{1},\bar{x}_{2},\bar{x}_{3})=(0,0,0)~. 
\]
Now, if we also compare the first term in (\ref{eq28a}) with the equation
(\ref{eq11a}), we have 
\begin{equation}
J_{ij}=I_{ij}+M(a^{2}\delta _{ij}-a_{i}a_{j})  \label{eq29a}
\end{equation}
and therefore the elements of the tensor of inertia $I_{ij}$ for the center of 
mass system will be given by: 
\begin{equation}
I_{ij}=J_{ij}-M(\delta _{ij}a^{2}-a_{i}a_{j})~.  \label{eq30a}
\end{equation}
This is known as the {\it theorem of the parallel axes}.

\bigskip

\noindent \underline{{\it EXAMPLE}}:

\noindent
Find $I_{ij}$ for the previous cube w.r.t. a reference system parallel to the 
system in the first example and with the origin in the center of mass.

\noindent
We already know from the previous example that: 
\[
\left\{ J_{ij}\right\} =\left( 
\begin{array}{ccc}
\frac{2}{3}\beta & -\frac{1}{4}\beta & -\frac{1}{4}\beta \\ 
-\frac{1}{4}\beta & \frac{2}{3}\beta & -\frac{1}{4}\beta \\ 
-\frac{1}{4}\beta & -\frac{1}{4}\beta & \frac{2}{3}\beta
\end{array}
\right)~. 
\]
Now, since the vector ${\bf a}=(\frac{b}{2},\frac{b}{2},\frac{b}{2})$ and 
${\bf a}%
^{2}=\frac{3}{4}b^{2}$, we can use the equation (\ref{eq30a}) and the fact that
$\beta =Mb^{2}$ to get, 
\begin{eqnarray}
I_{11}& =J_{11}-M(a^{2}-a_{1}^{2})=\frac{1}{6}Mb^{2} \\
I_{22}& =J_{22}-M(a^{2}-a_{2}^{2})=\frac{1}{6}Mb^{2} \\
I_{33}& =J_{33}-M(a^{2}-a_{3}^{2})=\frac{1}{6}Mb^{2} \\
I_{12}& =J_{12}-M(-a_{1}a_{2})=0 \\
I_{12}& =I_{13}=I_{23}=0~.
\end{eqnarray}
Therefore 
\[
\left\{ I\right\} =\left( 
\begin{array}{ccc}
\frac{1}{6}Mb^{2} & 0 & 0 \\ 
0 & \frac{1}{6}Mb^{2} & 0 \\ 
0 & 0 & \frac{1}{6}Mb^{2}
\end{array}
\right) . 
\]

\bigskip

{\bf \noindent } \underline{{\it EXAMPLE}}:

\noindent
We consider the case for which the vector $%
{\bf a}=(0,\frac{b}{2},\frac{b}{2})$ and $a^{2}=\frac{b^{2}}{2}$. Then, the new
tensor of inertia will be: 
\begin{eqnarray}
I_{11}& =J_{11}-M(a^{2}-a_{1}^{2})=\left( \frac{2}{3}Mb^{2}\right) -M\left( 
\frac{b^{2}}{2}-0\right) =\frac{1}{6}Mb^{2} \\
I_{22}& =J_{22}-M(a^{2}-a_{2}^{2})=\left( \frac{2}{3}Mb^{2}\right) -M\left( 
\frac{b^{2}}{2}-\frac{b^{2}}{4}\right) =\frac{5}{12}Mb^{2} \\
I_{33}& =J_{33}-M(a^{2}-a_{3}^{2})=\left( \frac{2}{3}Mb^{2}\right) -M\left( 
\frac{b^{2}}{2}-\frac{b^{2}}{4}\right) =\frac{5}{12}Mb^{2} \\
I_{12}& =J_{12}-M(-a_{1}a_{2})=\left( -\frac{1}{4}Mb^{2}\right) -M(0)
=-\frac{1}{4}Mb^{2} \\
I_{13}& =J_{13}-M(-a_{1}a_{3})=\left( -\frac{1}{4}Mb^{2}\right) -M(0)
=-\frac{1}{4}Mb^{2} \\
I_{23}& =J_{23}-M(-a_{2}a_{3})=\left( -\frac{1}{4}Mb^{2}\right)
-M(\frac{1}{4}Mb^{2})=0~.
\end{eqnarray}
It follows that $\{I_{ij}\}$ is equal to: 
\[
\{I_{ij}\}=\left( 
\begin{array}{ccc}
\frac{1}{6}Mb^{2} & -\frac{1}{4}Mb^{2} & -\frac{1}{4}Mb^{2} \\ 
-\frac{1}{4}Mb^{2} & \frac{5}{12}Mb^{2} & 0 \\ 
-\frac{1}{4}Mb^{2} & 0 & \frac{5}{12}Mb^{2}
\end{array}
\right)~. 
\]

\bigskip

\noindent {\bf 3.7 The dynamics of the rigid body}

\noindent
The rate of change in time of the angular momentum ${\bf L}$ 
is given by: 
\begin{equation}
\left( \frac{d{\bf L}}{dt}\right) _{inertial}={\bf N}^{(e)}.  \label{eq31a}
\end{equation}

\noindent
For the description w.r.t. the system fixed to the body we have to use the 
operator identity
\begin{equation}
{d \overwithdelims() dt}
_{inertial}=
{d \overwithdelims() dt}
_{body}+{\bf \omega }\times~.  \label{eq32a}
\end{equation}
Applying this operator to the equation (\ref{eq31a})
\begin{equation}
{d{\bf L} \overwithdelims() dt}
_{inertial}=
{d{\bf L} \overwithdelims() dt}
_{body}+{\bf \omega }\times {\bf L.}  \label{eq33a}
\end{equation}
Then, instead of (\ref{eq31a}) we shall have 
\begin{equation}
{d{\bf L} \overwithdelims() dt}
_{body}+{\bf \omega }\times {\bf L}={\bf N.}  \label{eq34a}
\end{equation}

\noindent
Now we project the equation (\ref{eq34a}) onto the principal axes of inertia,
that we call $(x_{1},x_{2},x_{3})$; then $T_{rot}$ and $%
{\bf L}$ take by far simpler forms, e.g., 
\begin{equation}
L_{i}=I_{i}\omega _{i}~.  \label{eq35a}
\end{equation}
The $i$th component of (\ref{eq34a}) is
\begin{equation}
\frac{dL_{i}}{dt}+\epsilon _{ijk}\omega _{j}L_{k}=N_{i}~.  \label{eq36a}
\end{equation}
Now projecting onto the principal axes of inertia and using the equation
(\ref{eq35a}), one can put (\ref{eq36a}) in the form: 
\begin{equation}
I_{i}\frac{d\omega _{i}}{dt}+\epsilon _{ijk}\omega _{j}\omega _{k}I_{k}=N_{i}
\label{eq37a}
\end{equation}
since the principal elements of inertia are independent of time.
Thus, we obtain the following system of equations known as Euler's equations
\begin{eqnarray}
I_{1}\dot{\omega}_{1}+\omega _{2}\omega _{3}(I_{2}-I_{3})& =N_{1}
\label{eq38a} \\
I_{2}\dot{\omega}_{2}+\omega _{3}\omega _{1}(I_{3}-I_{1})& =N_{2}  \nonumber
\\
I_{3}\dot{\omega}_{3}+\omega _{1}\omega _{2}(I_{1}-I_{2})& =N_{3}~.  
\nonumber
\end{eqnarray}

\bigskip

\noindent \underline{{\it EXAMPLE}}:

\noindent
For the rolling and sliding of a billiard ball,
prove that after a horizontal kick the ball slips
a distance 
\[
x_{_{1}}=\frac{12u_{0}^{2}}{49\mu g}~,
\]
where $u_0$ is the initial velocity. Then,
it starts rolling without gliding at the time  
\[
t_{1}=\frac{2u_{0}}{7\mu g}. 
\]
{\it SOLUTION}: When the impulsive force stops, the initial conditions are 
\begin{eqnarray*}
x_{0} &=&0,\;\;\;\;\;\;\;\;\;\;\dot{x}_{0}=u_{0} \\
\phi &=&0,\;\;\;\;\;\;\;\;\;\;\dot{\phi}=0~.
\end{eqnarray*}
The friction force is given by
\[
{\bf F}_{f}=-\mu g{\bf \hat{e}}_{1}~,
\]
and the equation of motion reads 
\begin{equation}
\ddot{x}=-\mu gM.  \label{eq39a}
\end{equation}

\noindent
The equation for ${\bf L}$ is 
\begin{equation}
\frac{dL_{3}}{dt}=I_{3}\ddot{\phi}=N_{3}  \label{eq40a}
\end{equation}
where $I_{3}$ is 
\[
I_{3}=\int \rho ({\bf r})\left[ x_{1}^{2}-x_{2}^{2}\right]
dx_{1}dx_{2}dx_{3}=\frac{2}{5}Ma^{2} 
\]
and 
\[
N_{3}=F_{f}a=\mu Mga~.
\]
Substituting $I_3$ and $N_3$ in (\ref{eq40a}), one gets
\begin{equation}
a\ddot{\phi}=\frac{5}{2}\mu g.  \label{eq41a}
\end{equation}

\noindent
Integrating once both (\ref{eq39a}) and (\ref{eq41a}), we get 
\begin{equation}
\dot{x}=-\mu gt+C_{1}  \label{eq42a}
\end{equation}
\begin{equation}
a\dot{\phi}=\frac{5}{2}\mu gt+C_{2}~.  \label{eq43a}
\end{equation}
To these equations we apply the initial conditions to put them in the form
\begin{equation}
\dot{x}(t)=-\mu gt+u_{0}  \label{eq44a}
\end{equation}
\begin{equation}
a\dot{\phi}(t)=\frac{5}{2}\mu gt.  \label{eq45a}
\end{equation}

\noindent
The condition of pure rolling (no friction) is
\begin{equation}
\dot{x}(t)=a\dot{\phi}(t).  \label{eq46a}
\end{equation}

\noindent
From (\ref{eq45a}) and (\ref{eq46a}) evaluated at $t_{1}$ we get
\[
\frac{5}{2}\mu gt_{1}=-\mu gt_{1}+u_{0} 
\]
\begin{equation}
\Rightarrow \;t_{1}=\frac{2u_{0}}{7\mu g}.  \label{eq47a}
\end{equation}

\noindent
Now we integrate (\ref{eq44a}) once again and applying the initial
conditions we get 
\begin{equation}
x(t)=-\mu g\frac{t^{2}}{2}+u_{0}t~.  \label{eq48a}
\end{equation}
By evaluating (\ref{eq48a}) and (\ref{eq44a}) at time $t_{1}$ we 
are led to
\[
x=\frac{12u^{2}}{49\mu g} 
\]
\[
\dot{x}=\frac{5}{7}u_{0}~.
\]

\bigskip

{\bf \noindent 3.8 Symmetrical top free of torques}

\noindent
A symmetric top is any solid of revolution. If the moments of inertia are 
$$
I_{1}=I_{2}=I_{3}\qquad \qquad {\rm spherical} \quad {\rm top}
$$
$$
I_{1}=I_{2}\neq I_{3}\qquad \qquad {\rm symmetric} \quad {\rm top}
$$
$$
I_{1}\neq I_{2}\neq I_{3}\qquad \qquad {\rm asymmetric} \quad
{\rm top.}
$$

\noindent
Let us consider the symmetric top $I_{1}=I_{2}\neq I_{3}$. In this case
the axis $X_{3}$ is the axis of symmetry. The Euler equations projected onto the
principal axes of inertia read 
\begin{equation}
I_{1}\dot{\omega}_{1}+\omega _{2}\omega _{3}(I_{2}-I_{3})=N_{1}
\label{eq49a}
\end{equation}
\begin{equation}
I_{2}\dot{\omega}_{2}+\omega _{3}\omega _{1}(I_{3}-I_{1})=N_{2}
\label{eq50a}
\end{equation}
\begin{equation}
I_{3}\dot{\omega}_{3}+\omega _{1}\omega _{2}(I_{1}-I_{2})=N_{3}.
\label{eq51a}
\end{equation}

\noindent
Since the system we consider here is free of torques 
\begin{equation}
N_{1}=N_{2}=N_{3}=0~,  \label{eq52a}
\end{equation}
we use $I_{1}=I_{2}$ in (\ref{eq52a}) to get 
\begin{equation}
I_{1}\dot{\omega}_{1}+\omega _{2}\omega _{3}(I_{2}-I_{3})=0  \label{eq53a}
\end{equation}
\begin{equation}
I_{2}\dot{\omega}_{2}+\omega _{3}\omega _{1}(I_{3}-I_{1})=0  \label{eq54a}
\end{equation}
\begin{equation}
I_{3}\dot{\omega}_{3}=0.  \label{eq55a}
\end{equation}

\noindent
The equation (\ref{eq55a}) implies that
\[
\omega _{3}={\rm const}. 
\]

\noindent
The equations (\ref{eq53a}) and (\ref{eq54a}) are rewritten as follows: 
\begin{equation}
\dot{\omega}_{1}=-\Omega \omega _{2}\;\;\;\;\;\;\;{\rm where}\;\;
\Omega =\omega _{3}\left( \frac{I_{3}-I_{1}}{I_{1}}\right)  \label{eq56a}
\end{equation}
\begin{equation}
\dot{\omega}_{2}=-\Omega \omega _{1}~.  \label{eq57a}
\end{equation}
Multiplying (\ref{eq57a}) by $i$ and summing it to (\ref{eq56a}), we have
\begin{eqnarray*}
(\dot{\omega}_{1}+i\dot{\omega}_{2}) &=&-\Omega (\omega _{2}-i\omega _{1}) \\
(\dot{\omega}_{1}+i\dot{\omega}_{2}) &=&i\Omega (\omega _{1}+i\omega _{2}).
\end{eqnarray*}

\noindent
If we write $\eta (t)=\dot{\omega}_{1}(t)+i\dot{\omega}_{2}(t)$, then 
\[
\dot{\eta}(t)-i\Omega \eta (t)=0~. 
\]
The solution is 
\[
\eta (t)=A\exp (i\Omega t)~. 
\]
This implies 
\[
(\omega _{1}+i\omega _{2})=A\cos (\Omega t)+i\sin (\Omega t)~. 
\]
Thus, 
\begin{equation}
\omega _{1}=A\cos (\Omega t)  \label{eq58a}
\end{equation}
\begin{equation}
\omega _{2}=A\sin (\Omega t).  \label{eq59a}
\end{equation}

\noindent
The modulus of the vector $\omega$ does not change in time
\[
\omega =\left| \left| {\bf \omega }\right| \right| =\sqrt[2]{\omega
_{1}+\omega _{2}+\omega _{3}}=\sqrt[2]{A^{2}+\omega _{3}^{2}}=const~. 
\]
This vector performs a precessional motion of precession frequency $\Omega$
given by 
\[
\Omega =\omega _{3}\left( \frac{I_{3}-I_{1}}{I_{1}}\right)~.
\]
Moreover, we notice that $\Omega$ is constant.

\noindent
If we denote by $\lambda $ the angle between ${\bf \omega }$ and $X_{3}$ the 
equations (\ref{eq58a}) and (\ref{eq59a}) take the form 
\[
\omega _{1}=\omega \sin \lambda \cos (\Omega t)
\]
\[
\omega _{2}=\omega \sin \lambda \sin (\Omega t)
\]
\[
\omega _{1}=\omega \cos \lambda~,
\]
where $A=\omega \sin \lambda$.

\noindent
For a flattened body of revolution $I_{1}=I_{2}=I_{12}$
and $I_{3}>I_{1}$. For example, for the case of the earth
\[
\Omega _{\bigoplus}
=\omega _{3}\left( \frac{I_{3}-I_{12}}{I_{12}}\right) \simeq
\frac{\omega _{3}}{305}.
\]

\noindent
The observations point to a mean value of fourteen months $\simeq$ 450
days. (This is due to the fact that the earth is not strictly
a RB; there is also an internal liquid structure).

\bigskip

\noindent {\bf 3.9 Euler angles}

\noindent
As we already know, a rotation can be described by a 
rotation matrix $\lambda$ by means of the equation
\begin{equation}
{\bf x}=\lambda {\bf x{\acute{}}}~.  \label{eq60a}
\end{equation}
${\bf x}$ represents the set of axes of the system 
rotated w.r.t. the system whose axes are represented by 
${\bf x{\acute{}}}$. The rotation
$\lambda$ can be accomplished through a set of ``partial" rotations
$\lambda =\lambda _{1}\lambda _{2}...\lambda _{n}$. There are many 
possibilities to choose these $\lambda$\'{}s. One of them is the set of
angles $\phi ,\theta $ and $\varphi$
called {\it Euler angles}. The partial rotations are in this case
the following:

\begin{itemize}
\item A rotation around the $X{\acute{}}
_{3}$ axis of angle $\varphi$ (in the positive trigonometric sense). 
The corresponding matrix is:
\[
\lambda _{\varphi }=\left( 
\begin{array}{ccc}
\cos \varphi & \sin \varphi & 0 \\ 
-\sin \varphi & \cos \varphi & 0 \\ 
0 & 0 & 1
\end{array}
\right)~. 
\]

\item  A rotation of angle $\theta$ around the $X
{\acute{}} 
{\acute{}}_{1}$ axis (positive sense).The associated  matrix is:
\[
\lambda _{\theta }=\left( 
\begin{array}{ccc}
1 & 0 & 0 \\ 
0 & \cos \theta & \sin \theta \\ 
0 & -\sin \theta & \cos \theta
\end{array}
\right) . 
\]

\item  A rotation of angle $\phi $ around the
$X{\acute{}}{\acute{}}{\acute{}}_{3}$ axis (positive sense); the 
assoiated matrix is:
\[
\lambda _{\phi }=\left( 
\begin{array}{ccc}
\cos \phi & \sin \phi & 0 \\ 
-\sin \phi & \cos \phi & 0 \\ 
0 & 0 & 1
\end{array}
\right) . 
\]
\end{itemize}

\noindent
The full transformation of the system of axes $\left\{ X{\acute{}}
_{1},X{\acute{}}_{2},X{\acute{}}_{3}\right\}$ to the system of axes
$\left\{ X_{1},X_{2},X_{3}\right\}$
is given by (\ref{eq60a}), where 
\[
\lambda =\lambda _{_{\phi }}\lambda _{_{\theta }}\lambda _{_{\varphi }}~. 
\]
Doing the product of matrices, we get 
\[
\lambda _{11}=\cos \varphi \cos \phi -\cos \theta \sin \phi \sin \varphi 
\]
\[
\lambda _{21}=-\sin \varphi \cos \phi -\cos \theta \sin \phi \cos \varphi 
\]
\[
\lambda _{31}=\sin \theta \sin \phi 
\]
\[
\lambda _{12}=\cos \varphi \sin \phi +\cos \theta \cos \phi \sin \varphi 
\]
\[
\lambda _{22}=-\sin \varphi \sin \phi +\cos \theta \cos \phi \sin \varphi 
\]
\[
\lambda _{32}=-\sin \varphi \cos \phi 
\]
\[
\lambda _{13}=\sin \varphi \cos \phi 
\]
\[
\lambda _{23}=\cos \varphi \sin \theta 
\]
\[
\lambda _{33}=\cos \theta 
\]
where 
\[
\lambda =\left( 
\begin{array}{ccc}
\lambda _{11} & \lambda _{12} & \lambda _{13} \\ 
\lambda _{21} & \lambda _{22} & \lambda _{23} \\ 
\lambda _{31} & \lambda _{32} & \lambda _{33}
\end{array}
\right) . 
\]

\noindent
Now, we take into account that:

\begin{itemize}
\item  ${\bf \dot{\phi}}$ is along the $X{\acute{}}_{3}$ (fixed) axis.

\item  ${\bf \dot{\theta}}$ is along the so-called {\it line of nodes}.

\item  ${\bf \dot{\varphi}}$ is along the $X_{3}$ axis
(of the body).
\end{itemize}

\bigskip

\noindent
This allows to write the three components of each of the three vectors in the
system $\left\{ X_{1},X_{2},X_{3}\right\}$ as follows:
\[
\begin{array}{ccc}
\dot{\phi}_{1}=\dot{\phi}\sin \theta \sin \varphi , & \dot{\theta}_{1}=\dot{%
\theta}\cos \varphi & \dot{\varphi}_{1}=0 \\ 
\dot{\phi}_{2}=\dot{\phi}\sin \theta \cos \varphi , & \dot{\theta}_{2}=-\dot{%
\theta}\sin \varphi & \dot{\varphi}_{2}=0 \\ 
\dot{\phi}_{3}=\dot{\phi}\cos \theta , & \dot{\theta}_{3}=0 & \dot{\varphi}%
_{3}=\dot{\varphi}~.
\end{array}
\]
Then, 
\begin{eqnarray*}
{\bf \omega } &=&{\bf \dot{\phi}}+{\bf \dot{\theta}}+{\bf \dot{\varphi}} \\
&=&\left[ \left( \dot{\phi}_{1}+\dot{\theta}_{1}+\dot{\varphi}_{1}\right)
,\left( \dot{\phi}_{2}+\dot{\theta}_{2}+\dot{\varphi}_{2}\right) ,\left( 
\dot{\phi}_{3}+\dot{\theta}_{3}+\dot{\varphi}_{3}\right) \right] .
\end{eqnarray*}

\noindent
Thus, we are led to the following components of ${\bf \omega}$: 
\begin{eqnarray*}
\omega _{1} &=&\dot{\phi}\sin \theta \sin \varphi +\dot{\theta}\cos \varphi
\\
\omega _{2} &=&\dot{\phi}\sin \theta \cos \varphi -\dot{\theta}\sin \varphi
\\
\omega _{3} &=&\dot{\phi}\cos \theta +\dot{\varphi}.
\end{eqnarray*}

\bigskip

\noindent {\bf 3.10 Symmetrical top with a fixed point}

\noindent
As a more complicated example of the methods used to describe the dynamics
of the rigid body, we shall consider the motion of a symmetric body in a unifom
gravitational field when a point of the axis of symmetry is fixed in the space.

\noindent
The axis of symmetry is of course one of the principal axes, and we shall take
it as the $z$ axis of the body-fixed reference system.
Since there is a fixed point, the configuration of the top will be determined
by the three  Euler angles: $\theta$ measuring the deviation of
$z$ from the vertical, $\phi$, giving the azimuth of the top w.r.t. the 
vertical, and $\varphi$, which is the rotation angle of the top 
w.r.t. its proper $z$. The distance from the center of gravity to the fixed point
will be denoted by $l$. To get a solution to the motion of the top
we shall use the method of Lagrange instead of the Euler equations.

\noindent
The kinetic energy is: 
\[
T=\frac{1}{2}I_{1}(\omega _{1}^{2}+\omega _{2}^{2})+\frac{1}{2}I_{3}\omega
_{3}^{2}~, 
\]
or, in terms of the Euler angles:
\[
T=\frac{1}{2}I_{1}(\dot{\phi}^{2}\sin ^{2}\theta +\dot{\theta}^{2})
+\frac{1}{2}I_{3}(\dot{\phi}\cos \theta +\dot{\varphi})^{2}~.
\]
According to an elementary theorem, in a constant gravitational field the
potential
energy of a body is the same with that of a material point of equal mass 
concentrated in its center of mass.
A formal proof is as follows. The potential energy of the body
is the sum of the potential energies of all its particles: 
\begin{equation}
V=-m_{i}{\bf r}_{i}\cdot {\bf g}~,  \label{eq61a}
\end{equation}
where ${\bf g}$ is the constant acceleration of gravity.
According to the definition of the center of mass, this is equivalent to 
\begin{equation}
V=-M{\bf R}_{i}\cdot {\bf g,}  \label{eq62a}
\end{equation}
thus proving the theorem. The potential energy is a function of the Euler angles:
\begin{equation}
V=Mgl\cos \theta ,  \label{eq63a}
\end{equation}
and the Lagrangian will be
\begin{equation}
L=\frac{1}{2}I_{1}(\dot{\phi}^{2}\sin ^{2}\theta +\dot{\theta}^{2})+\frac{1}{%
2}I_{3}(\dot{\phi}\cos \theta +\dot{\varphi})^{2}-Mgl\cos \theta .
\label{eq64a}
\end{equation}

\noindent
We note that $\phi$ and $\varphi$ are cyclic coordinates, and therefore  
$p_{\phi }$ and $p_{\varphi }$ are constants of motion. 
\begin{equation}
p_{\varphi }=\frac{\partial L}{\partial \dot{\varphi}}=I_{3}(\dot{\varphi}+%
\dot{\phi}\cos \theta )=const  \label{eq65a}
\end{equation}
and 
\begin{equation}
p_{\phi }=\frac{\partial L}{\partial \dot{\phi}}=I_{1}\dot{\phi}\sin
^{2}\theta +I_{3}(\dot{\phi}\cos ^{2}\theta +\dot{\varphi}\cos \theta )=const.
\label{eq66a}
\end{equation}

\noindent
From the equation (\ref{eq65a}) we get $\dot{\varphi}$%
\begin{equation}
\dot{\varphi}=\frac{p_{\varphi }-I_{3}\dot{\phi}\cos \theta }{I_{3}}~,
\label{eq67a}
\end{equation}
that we substitute in (\ref{eq66a}) 
\begin{eqnarray*}
p_{\phi } &=&\frac{\partial L}{\partial \dot{\phi}}=I_{1}\dot{\phi}\sin
^{2}\theta +I_{3}(\dot{\phi}\cos ^{2}\theta +\frac{p_{\varphi }-I_{3}\dot{%
\phi}\cos \theta }{I_{3}}\cos \theta )=const. \\
p_{\phi } &=&I_{1}\dot{\phi}\sin ^{2}\theta +p_{\varphi }\cos \theta~,
\end{eqnarray*}
where from we get 
\begin{equation}
\dot{\phi}=\frac{p_{\phi }-p_{\varphi }\cos \theta }{I_{1}\sin ^{2}
\theta}~.
\label{eq68a}
\end{equation}
Substituting it in (\ref{eq67a}) one gets
\begin{equation}
\dot{\varphi}=\frac{p_{\varphi }}{I_{3}}-\frac{p_{\phi }-p_{\varphi }\cos
\theta }{I_{1}\sin ^{2}\theta }\cos \theta .  \label{eq69a}
\end{equation}

\noindent
Now, since the system is conservative, another integral of motion is the 
energy 
\[
E=T+V=\frac{1}{2}I_{1}(\dot{\phi}^{2}\sin ^{2}\theta +\dot{\theta}^{2})+%
\frac{1}{2}I_{3}(\dot{\phi}\cos \theta +\dot{\varphi})^{2}+Mgl\cos \theta . 
\]

\noindent
The quantity $I_{3}\omega _{3}=p_{\varphi }$ is an integral of motion.
Multiplying this constant by $p_{\varphi }\omega _{3}$ we get
\begin{eqnarray*}
I_{3}p_{\varphi }\omega _{3}^{2} &=&p_{\varphi }^{2}\omega _{3} \\
I_{3}^{2}\omega _{3}^{3} &=&p_{\varphi }^{2}\omega _{3} \\
\frac{1}{2}I_{3}\omega _{3}^{2} &=&\frac{1}{2}\frac{p_{\varphi }^{2}}{I_{3}}~.
\end{eqnarray*}
The quantity $\frac{1}{2}I_{3}\omega _{3}^{2}$ is a constant. Therefore,
we can define the quantity
\begin{eqnarray*}
E{\acute{}}
&=&E-\frac{1}{2}I_{3}\omega _{3}^{2}=const. \\
&=&\frac{1}{2}I_{1}\dot{\theta}^{2}+\frac{1}{2}I_{1}\dot{\phi}^{2}\sin
^{2}\theta +Mgl\cos \theta ~,
\end{eqnarray*}
wherefrom we can identify 
\[
V(\theta )=\frac{1}{2}\dot{\phi}^{2}\sin ^{2}\theta +Mgl\cos \theta 
\]
\begin{equation}
V(\theta )=\frac{1}{2}I_{1}\left( \frac{p_{\phi }-p_{\varphi }\cos \theta }{%
I_{1}\sin ^{2}\theta }\right) ^{2}\sin ^{2}\theta +Mgl\cos \theta .
\label{eq70a}
\end{equation}

\noindent
Thus, $E{\acute{}}$ is:
\[
E{\acute{}}
=\frac{1}{2}I_{1}\dot{\theta}^{2}+V(\theta)~.
\]
From this equation we get  
$\dot{\theta}\equiv \frac{d\theta }{dt}=\left[ \frac{2}{I_{1}}\left( E
{\acute{}}-V(\theta )\right) \right] ^{1/2}$, which leads to 
\begin{equation}
t(\theta )=
\displaystyle\int 
\frac{d\theta }{\sqrt[2]{\left( \frac{2}{I_{1}}\right) \left( E
{\acute{}}
-V(\theta )\right) }}~.  \label{eq71a}
\end{equation}

\noindent
Performing the integral in (\ref{eq71a}) one gets
$t=f(\theta)$, and therefore, in principle, one can get $\theta (t)$.
Then, $\theta (t)$ is replaced by $\dot{\phi}$
and $\dot{\varphi}$ (in eqs. (\ref{eq68a}) and (\ref{eq69a})) and integrating
them we can obtain the complete solution of the problem.


\bigskip
\bigskip

\begin{center} {\bf References} \end{center}

\bigskip

\begin{itemize}
\item  H. Goldstein, {\it Classical Mechanics}, (Addison-Wesley, 1992).

\item  L. D. Landau \& E. M. Lifshitz, {\it Mechanics,} (Pergammon, 1976).

\item  J. B. Marion \& S.T. Thornton, 
  {\it Classical Dynamics of Particles and Systems}, (Harcourt Brace, 1995).

\item W. Wrigley \& W.M. Hollister, {\it The Gyroscope: Theory and
application}, Science 149, 713 (Aug. 13, 1965).
\end{itemize}


%% file: alben.tex
\pagestyle{plain}

\centerline{{\Large 4. SMALL OSCILLATIONS}}

\bigskip
\bigskip

\noindent
{\bf Forward}:
A familiar type of motion in mechanical and many other systems are the small
oscillations (vibrations). 
They can be met as atomic and molecular vibrations, electric 
circuits, acoustics, and so on. In general, any motion in the neighborhood of 
stable equilibria is vibrational.

\bigskip
\bigskip

{\bf CONTENTS}:

\bigskip

4.1 THE SIMPLE HARMONIC OSCILLATOR

\bigskip

4.2 FORCED HARMONIC OSCILLATOR

\bigskip

4.3 DAMPED HARMONIC OSCILLATORS

\bigskip

4.4 NORMAL MODES

\bigskip

4.5 PARAMETRIC RESONANCE

\newpage

\section*{\protect\smallskip 4.1 THE SIMPLE HARMONIC OSCILLATOR}

A system is at stable equilibrium when its potential energy
$U(q)$ is at minimum; when the system is slightly displaced from the 
equilibrium position, a force $-dU/dq$ occurs which acts to restore the 
equilibrium. Let $q_0$ be the value of the generalized coordinate corresponding
to the equilibrium position. Expanding $U(q) - U(q_{0})$ in a Taylor series
of $q - q_0$ for small deviations from the equilibrium
\setcounter{equation} {0}\\
\[
U(q)-U(q_0)\cong \frac 12k(q-q_0)^2~,
\]

\noindent
donde:
\begin{eqnarray*}
\frac{\partial U}{\partial q} &=&0 \\
U(q) &=&0~,
\end{eqnarray*}

\noindent
This means that there are no external forces acting on the system
and the zero has been chosen at the equilibrium position; moreover, higher-order
terms have been neglected. The coefficient $k$ represents the value
of the second derivative of $U(q)$ for q=q$_0$.
For simplicity reasons we denote

\[
x=q-q_0
\]

\noindent
for which the potential energy can be written as:

\begin{equation}
U(x)=\frac 12kx^2~.  \label{4.1.1}
\end{equation}

\noindent
For simplicity reasons we denote
\[
x=q-q_0
\]

\noindent
for which the potential energy can be written as:

\begin{equation}
U(x)=\frac 12kx^2~.  \label{4.1.1}
\end{equation}

\noindent
The kinetic energy of a system is

\begin{equation}
T=\frac 12m\stackrel{\cdot }{x}^2~,  \label{4.1.2}
\end{equation}

\noindent
and using (\ref{4.1.1}) and (\ref{4.1.2}) we get the 
Lagrangian of a system performing linear oscillations 
(so-called linear oscillator):

\begin{equation}
L=\frac 12m\stackrel{\cdot }{x}^2-\frac 12kx^2~.  \label{4.1.3}
\end{equation}

\noindent
The equation of motion corresponding to this $L$ is:

\[
m\stackrel{\cdot \cdot }{x}+kx=0~,
\]

\noindent
or

\begin{equation}
\stackrel{\cdot \cdot }{x}+w^2x=0~,  \label{4.1.4}
\end{equation}

\noindent
where $w^2=\sqrt{k/m}$. This differential equation has two independent
solutions: $\cos wt$ and ${\rm sin} wt$, from which one can form the general
solution:

\begin{equation}
x=c_1\cos wt+c_2\sin wt~,  \label{4.1.5}
\end{equation}

\noindent
or, we can also write the solution in the form:

\begin{equation}
x=a\cos (wt+\alpha )~.  \label{4.1.6}
\end{equation}

\noindent
Since $\cos(wt+\alpha )=\cos wt\cos \alpha -{\rm sin} wt
{\rm sin} \alpha$, by
comparing with (\ref{4.1.5}), one can see that the arbitrary constants $a$
and $\alpha$ are related to $c_1$ and $c_2$ as follows:
$$a=\sqrt{(c_1^2+c_2^2)},\;\;\;\; {\rm y}\;\;\;\; {\rm tan}\alpha =-c_1/c_2~.$$

\noindent
Thus, any system in the neighborhood of the stable equilibrium position
performs harmonic oscillatory motion. The $a$ coefficient in
(\ref{4.1.6}) is the amplitude of the oscillations, whereas the argument of the 
cosine function is the phase of the harmonic oscillation; $\alpha$ is the initial
value of the phase, which depends on the chosen origin of time.
The quantity $w$ is the angular frequency of the oscillations, which does not 
depend on the initial conditions of the system, being a proper characteristic of
the harmonic oscillations.

\noindent
Quite often the solution is expressed as the real part of a complex quantity
\[
x={\rm Re}\left[ A\exp(iwt)\right]
\]

\noindent
where $A$ is the complex amplitude, whose  modulus gives the
ordinary amplitude:

\[
A=a\exp(i\alpha )~.
\]

\noindent
The energy of a system in small oscillatory motion is:

\[
E=\frac 12m\stackrel{\cdot }{x}^2+\frac 12kx^2~,
\]

\noindent
or by substituting (\ref{4.1.6})

\[
E=\frac 12mw^2a^2~.
\]

\noindent
Now, we consider the case of $n$ degrees of freedom. In this case, taking the sum
of exterior forces as zero, the generalized force will be given by

\begin{equation}
Q_i=-\frac{\partial U}{\partial q_i}=0~.  \label{4.1.7}
\end{equation}

\noindent
Repeating the procedure for the case of a single degree of freedom,
we expand the potential energy in Taylor series
taking the minimum of the potential energy at $q_i=q_{i0}$.
Introducing small oscillation coordinates

\[
x_i=q_i-q_{i0}~,
\]

\noindent
we can write the series as follows

\begin{equation}
U(q_1,q_2,...,q_n)=U(q_{10},q_{20},...,q_{n0})+\sum \left( \frac{\partial U}
{\partial q_i}\right) _0x_i+\frac 1{2!}\sum \left( \frac{\partial ^2U}{
\partial q_i\partial q_j}\right) _0x_ix_j+....  \label{4.1.8}
\end{equation}

\noindent
Under the same considerations as given for (\ref{4.1.1}), we obtain:

\begin{equation}
U(q_1,q_2,...,q_n)=U=\frac 12\sum_{i,j}k_{ij}x_ix_j~.  \label{4.1.9}
\end{equation}

\noindent
From (\ref{4.1.8}) one notes that $k_{ij}=k_{ji}$, i.e., they are symmetric 
w.r.t. their subindices.

Let us look now to the kinetic energy, which, in general, is of the form

\[
\frac 12a_{ij}(q)\stackrel{\cdot }{x}_i\stackrel{\cdot }{x}_j~,
\]

\noindent
where the $a_{ij}$ are functions of the coordinates only. Denoting them by
$a_{ij}=m_{ij}$ the kinetic energy will be

\begin{equation}
T=\frac 12\sum_{i,j}m_{ij}\stackrel{\cdot }{x}_i\stackrel{\cdot }{x}_j~.
\label{4.1.10}
\end{equation}

\noindent
We can pass now to the Lagrangian for the system of $n$ degrees of freedom

\begin{equation}
L=T-U=\frac 12\sum_{i,j}(m_{ij}\stackrel{\cdot }{x}_i\stackrel{\cdot }{x}
_j-k_{ij}x_ix_j)~.  \label{4.1.11}
\end{equation}

\noindent
This Lagrangian leads to the following set of simultaneous
differential equations of motion

\begin{equation}
\frac d{dt}\frac{\partial L}{\partial \stackrel{\cdot }{x}_i}-\frac{\partial
L}{\partial x_i}=0  \label{4.1.12}
\end{equation}

\noindent
or

\begin{equation}
\sum (m_{ij}\stackrel{\cdot \cdot }{x}_j+k_{ij}x_j)=0~.  \label{4.1.13}
\end{equation}

\noindent
This is a linear system of homogeneous equations, which can be considered as the 
n components of the matricial equation

\begin{equation}
(M)(\stackrel{\cdot \cdot }{X})+(K)(X)=0~,  \label{4.1.14}
\end{equation}

\noindent
where the matrices are defined by:

\begin{equation}
(M)=\left(
\begin{array}{llll}
m_{11} & m_{12} & ... & m_{1n} \\
m_{21} & m_{22} & ... & m_{2n} \\
\vdots &  &  & \vdots \\
m_{n1} & m_{n2} & ... & m_{nn}
\end{array}
\right)  \label{4.1.15}
\end{equation}
\begin{equation}
(K)=\left(
\begin{array}{llll}
k_{11} & k_{12} & ... & k_{1n} \\
k_{21} & k_{22} & ... & k_{2n} \\
\vdots &  &  & \vdots \\
k_{n1} & k_{n2} & ... & k_{nn}
\end{array}
\right)  \label{4.1.16}
\end{equation}

\begin{equation}
(\stackrel{\cdot \cdot }{X})=\frac{d^2}{dt^2}\left(
\begin{array}{l}
x_1 \\
x_2 \\
\vdots \\
x_n
\end{array}
\right)  \label{4.1.17}
\end{equation}

\begin{equation}
(X)=\left(
\begin{array}{l}
x_1 \\
x_2 \\
\vdots \\
x_n
\end{array}
\right)~.  \label{4.1.18}
\end{equation}

\noindent
Similarly to the one dof system, we look for $n$
unknown functions $x_j(t)$ of the form

\begin{equation}
x_j=A_j\exp(iwt)~,  \label{4.1.19}
\end{equation}

\noindent
where A$_j$\ are constants to be determined. Substituting (\ref{4.1.19}) in 
(\ref{4.1.13}) and dividing by $\exp(iwt)$, one gets a linear system of 
algebraic and homogeneous equations, which should be fulfilled by
A$_j$.

\begin{equation}
\sum_j(-w^2m_{ik}+k_{ik})A_k=0~.  \label{4.1.20}
\end{equation}

\noindent
This system has nponzero solutions if the determinant of its coefficients is 
zero.
\begin{equation}
\left| k_{ij}-w^2m_{ij}\right| ^2=0~.  \label{4.1.21}
\end{equation}

\noindent
This is the characteristic equation of order $n$ w.r.t.
$w^2$. In general, it has $n$ different real and positive solutions $w_{\alpha}$
($\alpha =1,2,...,n$). The $w_{\alpha}$ are called proper frequencies of the 
system. Multiplying by A$_i^{*}$ and summing over $i$ one gets

\[
\sum_j(-w^2m_{ij}+k_{ij})A_i^{*}A_j=0~,
\]

\noindent
where from 

\[
w^2=\sum k_{ij}A_i^{*}A_i/\sum m_{ij}A_i^{*}A_i~.
\]

\noindent
Since the coefficients $k_{ij}$ and $m_{ij}$ are real and symmetric,
the quadratic forms of the numerator and denominator are real, and being essentially
positive one concludes that $w^2$ are equally positive.

\underline{EXAMPLE}

\noindent
As an example we model the equations of motion of a double pendulum. 
The potential energy of this system with two degrees of freedom is

\[
U=m_1gl_1(1-\cos\theta _1)+m_2gl_1(1-\cos\theta _1)+m_2gl_2(1-\cos\theta _2)~.
\]

\noindent
Applying (\ref{4.1.8}), one gets

\[
U=\frac 12(m_1+m_2)gl_1\theta _1^2+\frac 12m_2gl_2\theta _2^2~.
\]

\noindent
Comparing with (\ref{4.1.9}), we identify
\begin{eqnarray*}
k_{11} &=&(m_1+m_2)l_1^2 \\
k_{12} &=&k_{21}=0 \\
k_{22} &=&m_2gl_2~.
\end{eqnarray*}

\noindent
For the kinetic energy one gets 

\[
T=\frac 12(m_1+m_2)l_1^2\stackrel{.}{\theta }_1^2+\frac 12m_2l_2^2\stackrel{.%
}{\theta }_2^2+m_2l_1l_2\stackrel{.}{\theta }_1\stackrel{.}{\theta }_2~.
\]

\noindent
Identifying terms from the comparison with (\ref{4.1.10}) we find
\begin{eqnarray*}
m_{11} & = & (m_1+m_2)l_1^2 \\
m_{12} & = & m_{21}=m_2l_1l_2 \\
m_{22} & = & m_2l_2^2~.
\end{eqnarray*}

\noindent
Substituting the energies in (\ref{4.1.11}) one obtains the 
Lagrangian for the double pendulum oscillator and as the final result the 
equations of motion for this case:

\[
\left(
\begin{array}{ll}
m_{11} & m_{12} \\
m_{21} & m_{22}
\end{array}
\right) \left(
\begin{array}{l}
\stackrel{..}{\theta }_1 \\
\stackrel{..}{\theta }_2
\end{array}
\right) +\left(
\begin{array}{ll}
k_{11} & 0 \\
0 & k_{22}
\end{array}
\right) \left(
\begin{array}{l}
\theta _1 \\
\theta _2
\end{array}
\right) =0~.
\]

\section*{4.2 FORCED HARMONIC OSCILLATOR}

\noindent
If an external weak force acts on an oscillator system the oscillations of the 
system are known as forced oscillations.

\noindent
Besides its proper potential energy the system gets a supplementary potential
energy $U_{e}(x,t)$ due to the external field. Expanding the latter in a Taylor
series of small amplitudes $x$ we get:

\[
U_e(x,t)\cong U_e(0,t)+x\left[ \frac{\partial U_e}{\partial x}\right] _{x=0}~.
\]

\noindent
The second term is the external force acting on the system at its equilibrium 
position, that we denote by $F(t)$. Then, the Lagrangian reads

\begin{equation}
L=\frac 12m\stackrel{\cdot }{x}^2-\frac 12kx^2+xF(t)~.  \label{4.2.1}
\end{equation}

\noindent
The corresponding equation of motion is

\[
m\stackrel{\cdot \cdot }{x}+kx=F(t)~,
\]

\noindent
or

\begin{equation}
\stackrel{\cdot \cdot }{x}+w^2x=F(t)/m~,  \label{4.2.2}
\end{equation}

\noindent
where $w$ is the frequency of the proper oscillations. The general solution of this 
equation is the sum of the solution of the homogeneous equation and a particular
solution of the nonhomogeneous equation
\[
x=x_h+x_p~.
\]

\noindent
We shall study the case in which the external force is periodic in time of 
frequency $\gamma$
forma

\[
F(t)=f\cos(\gamma t+\beta )~.
\]

\noindent
The particular solution of (\ref{4.2.2}) is sought in the form
$x_1=b\cos(\gamma t+\beta )$ and by substituting it one finds that
the relationship 
$b=f/m(w^2-\gamma ^2)$ should be fulfilled. Adding up both solutions, one gets 
the general solution

\begin{equation}
x=a\cos(wt+\alpha )+\left[ f/m(w^2-\gamma ^2)\right] \cos(\gamma t+\beta )~.
\label{4.2.3}
\end{equation}

\noindent
This result is a sum of two oscillations: one due to the proper frequency
and another at the frequency of the external force.

\noindent
The equation (\ref{4.2.2}) can in general be integrated for an arbitrary
external force. Writing it in the form

\[
\frac d{dt}(\stackrel{\cdot }{x}+iwx)-iw(\stackrel{\cdot }{x}+iwx)=\frac
1mF(t)~,
\]

\noindent
and making $\xi =\stackrel{\cdot }{x}+iwx$, we have

\[
\frac d{dt}\xi -iw\xi =F(t)/m~.
\]

\noindent
The solution to this equation is $\xi =A(t)\exp(iwt)$; for
$A(t)$ one gets

\[
\stackrel{\cdot }{A}=F(t)\exp(-iwt)/m~.
\]

\noindent
Integrating it leads to the solution

\begin{equation}
\xi =\exp(iwt)\int_{0}^{t}\frac{1}{m}F(t)\exp(-iwt)dt+\xi _o~.  \label{4.2.4}
\end{equation}

\noindent
This is the general solution we look for; the function x(t) is given by the
imaginary part of the general solution divided by $w$.

\underline{EXAMPLE}

\noindent
We give here an example of employing the previous equation.

\noindent
Determine the final amplitude of oscillations of a system acted by an 
extenal force $F_0=const.$ during a limited time $T$. For this time interval
we have
$$\xi =  \frac{F_0}{m}\exp(iwt)\int_{0}^{T}\exp(-iwt)dt~,$$
$$\xi =  \frac{F_0}{iwm}[1-\exp(-iwt)]\exp(iwt)~. \nonumber$$
Using $\left| \xi
\right| ^2=a^2w^2$ we obtain

\[
a=\frac{2F_0}{mw^2}\sin (\frac 12wT)~.
\]

\section*{4.3 DAMPED HARMONIC OSCILLATOR}

\noindent
Until now we have studied oscillatory motions in free space (the vacuum), or
when the effects of the medium through which the oscillator moves are 
negligeable. However, when a system moves through a medium its motion is 
retarded by the reaction of the latter.
There is a dissipation of the energy in heat or other forms of energy.
We are interested in a simple description of the dissipation phenomena.

\noindent
The reaction of the medium can be imagined in terms of friction forces.
When they are small we can expand them in powers of the velocity.
The zero-order term is zero because there is no friction force acting on a body
at rest. Thus, the lowest order nonzero term is proportional to the velocity, and
moreover we shall neglect all higher-order terms

\[
f_r=-\alpha \stackrel{\cdot }{x}~,
\]

\noindent
where $x$ is the generalized coordinate and $\alpha$ is a positive 
coefficient; the minus sign shows the oposite direction to that of the moving
system. Adding this force to the equation of motion we get 

\[
m\stackrel{..}{x}=-kx-\alpha \stackrel{\cdot }{x}~,
\]

\noindent
or

\begin{equation}
\stackrel{..}{x}=-kx/m-\alpha \stackrel{\cdot }{x}/m~.  \label{4.3.1}
\end{equation}

\noindent
Writing $k/m=w_o^2$ and $\alpha /m=2\lambda $; where $w_o$ is the frequency of
free oscillations of the system and $\lambda$ is the damping coefficient.
Therefore

\[
\stackrel{..}{x}+2\lambda \stackrel{\cdot }{x}+w_o^2x=0~.
\]

\noindent
The solution of this equation is sought of the type $x=\exp(rt)$, which we
substitute back in the equation to get the characteristic equation for $r$. 
Thus

\[
r^2+2\lambda +w_o^2=0~,
\]

\noindent
where from

\[
r_{1,2}=-\lambda \pm \sqrt{\left( \lambda ^2-w_o^2\right) }~.
\]

\noindent
We are thus led to the following general solution of the equation of motion

\[
x=c_1\exp(r_1t)+c_2\exp(r_2t)~. \]

\noindent
Among the roots $r$ we shall look at the following particular cases:

(i) $\lambda <w_o$. One gets complex conjugate solutions. The solution is

\[
x={\rm Re}
\left\{ Aexp\left[ -\lambda t+i\sqrt{(w_o^2-\lambda ^2)}\right] \right\}~,
\]

\noindent
where $A$ is an arbitrary complex constant. The solution can be written of the
form

\begin{equation}
x=a\exp(-\lambda t)\cos(wt+\alpha ),
\;\;{\rm where}\;\;\;\;w=\sqrt{\left( w_o^2-\lambda
^2\right)}~,  \label{4.3.2}
\end{equation}

\noindent
where $a$ and $\alpha$ are real constants. Thus, one can say that a damped 
oscillation is a harmonic oscillation with an exponentially decreasing 
amplitude.
The rate of decreasing of the amplitude is determined by the exponent $\lambda$. 
Moreover, the frequency $w$ is smaller than that of free oscillations.

(ii) $\lambda > w_o$. Then, both $r$ are real and negative.
The general form of the solution is:

\[
x=c_1\exp\left\{ -\left[ \lambda -\sqrt{\left( \lambda ^2-w_o^2\right) }%
\right] t\right\} +c_2\exp\left\{ -\left[ \lambda +\sqrt{\left( \lambda
^2-w_o^2\right) }\right] t\right\}~.
\]

\noindent
If the friction is large, the motion is just a monotone decaying amplitude
asymptotically ($t
\rightarrow \infty$) tending to the equilibrium position (without oscillations).
This type of motion is called aperiodic.

(iii)
$\lambda =w_o$. Then $r=-\lambda$, whose general solution is

\[
x=(c_1+c_2t)\exp(-\lambda t)~.
\]

\noindent
If we generalize to systems of n degrees of freedom, the generalized friction 
forces corresponding to the coordinates $x_i$ are linear functions of the 
velocities

\begin{equation}
f_{r,i}=\sum_{j}\alpha _{ij}\stackrel{.}{x}_i~.  \label{4.3.3}
\end{equation}

\noindent
Using $\alpha _{ik}=\alpha _{ki}$, one can also write 

\[
f_{r,i}=-\frac{\partial F}{\partial \stackrel{.}{x}_i}~,
\]

\noindent
where $F=\frac{1}{2}\sum_{i,j}\alpha _{ij}\stackrel{.}{x}_i
\stackrel{.}{x}_j$ is called the dissipative function. The differential equation
is obtained by adding up all these forces to (\ref{4.1.13})

\begin{equation}
\sum (m_{ij}\stackrel{\cdot
\cdot }{x}_j+k_{ij}x_j)
=-\sum_{j}\alpha _{ij}\stackrel{.}{x}_i~.  \label{4.3.4}
\end{equation}

\noindent
Employing

\[
x_k=A_k\exp(rt)
\]

\noindent
in (\ref{4.3.4}) and deviding by $\exp(rt)$, one can obtain the following system
of linear algebraic equations for the constants $A_j$

\[
\sum_{j}(m_{ij}r^2+\alpha _{ij}r+k_{ij})A_j=0~.
\]

\noindent
Making equal to zero the determinant of this system, one gets the 
corresponding characteristic equation

\begin{equation}
\left| m_{ij}r^2+\alpha _{ij}r+k_{ij}\right| =0~.  \label{4.3.5}
\end{equation}

\noindent
This is an equation for $r$ of degree $2n$.
\section*{4.4 NORMAL MODES}

\noindent
Before  defining the normal modes, we rewrite (\ref
{4.1.14}) as follows

\[
M\left| \stackrel{..}{X}\right\rangle +K\left| X\right\rangle =0~,
\]

\noindent
where $\left| X\right\rangle $ is the $n$-dimensional vector whose matrix
representation is (\ref{4.1.18}); $M$ and $K$ are two operators having the 
matrix representation given by (\ref{4.1.15}) and (\ref
{4.1.16}), respectively. We have thus an operatorial equation.
Since M is a nonsingular and symmetric operator, 
the inverse operator M$^{-1}$ and the operators M$
^{1/2}$ and M$^{-1/2}$ are well defined. In this case, we can express
the operatorial equation in the form

\[
\frac{d^2}{dt^2}M^{1/2}\left| X\right\rangle =-M^{-1/2}KM^{-1/2}M^{1/2}\left|
X\right\rangle~,
\]

\noindent
or more compactly

\begin{equation}
\frac{d^2}{dt^2}\left| \stackrel{\_}{X}\right\rangle =-\lambda \left|
\stackrel{\_}{X}\right\rangle~,  \label{4.4.1}
\end{equation}

\noindent
where

\[
\left| \stackrel{\_}{X}\right\rangle =M^{1/2}\left| X\right\rangle
\]

\noindent
and 
\[
\lambda =M^{-1/2}KM^{-1/2}~.
\]

\noindent
Since M$^{-1/2}$ and K are symmetric operators, then $\lambda$ is also symmetric.
If we use orthogonal eigenvectors as a vectorial base
(for example, the three-dimensional Euclidean space), the matrix representation
of the operator can be diagonal, e.g.,

\[
\lambda _{ij}=\lambda _i\delta _{ij}~.
\]

\noindent
Let us consider the following eigenvalue problem

\begin{equation}
\lambda \left| \rho _i\right\rangle =\lambda _i\left| \rho _i\right\rangle~,
\label{4.4.2}
\end{equation}

\noindent
where $\left| \rho _i\right\rangle$ is an orthogonal set of eigenvectors. Or

\[
M^{-1/2}KM^{-1/2}\left| \rho _i\right\rangle =\lambda _i\left| \rho
_i\right\rangle~.
\]

\noindent
The eigenvalues are obtained by multiplying both sides by $\left\langle
\rho _i\right| $, leading to

\[
\lambda _i=\frac{\left\langle \rho _i\right| M^{-1/2}KM^{-1/2}\left| \rho
_i\right\rangle }{\langle \rho _i\left| \rho _i\right\rangle }~.
\]

\noindent
Since the potential and kinetic energies are considered positive quantities,
one should take

\[
\left\langle \rho _i\right| M^{-1/2}KM^{-1/2}\left| \rho _i\right\rangle
\rangle 0
\]

\noindent
and therefore

\[
\lambda _i> 0~.
\]

\noindent
This leads to the set

\[
\lambda _i=w_i^2~.
\]

\noindent
If we express the vector $\left| \stackrel{\_}{X}\right\rangle $\ in terms of 
these eigenvectors of $\lambda$,

\[
\left| \stackrel{\_}{X}\right\rangle =\sum_{i}y_i\left|
\stackrel{\_}{X}\right\rangle~,
\]

\noindent
where

\begin{equation}
y_i=\langle \rho _i\left| \stackrel{\_}{X}\right\rangle~.
\label{4.4.3}
\end{equation}

\noindent
Inserting this result in the equation of motion (\ref{4.4.1}),
we obtain

\[
\frac{d^2}{dt^2}\sum_{i}y_i\left| \rho _i\right\rangle
=-\lambda \left| \stackrel{\_}{X}\right\rangle =-\sum_{i}
\lambda _iy_i\left| \rho _i\right\rangle~.
\]

\noindent
The scalar product of this equation with the constant eigenvector
$\left\langle \rho _j\right|$ leads to the equation of motion for the 
generalized coordinate $y_j$

\[
\frac{d^2}{dt^2}y_j=-w_j^2y_j~.
\]

\noindent
The solution of this equation reads

\begin{equation}
y_j=A_j\cos(w_jt+\phi _j)~.  \label{4.4.4}
\end{equation}

\noindent
Use of these new generalized harmonic coordinates lead to a set of independent 
equations of motion. The relationship between 
$y_j$ and $\stackrel{\_}{x}_i$ is given by (\ref{4.4.3})

\[
y_j=\rho _{j1}\stackrel{\_}{x}_1+\rho _{j2}\stackrel{\_}{x}_2+...+\rho _{jn}%
\stackrel{\_}{x}_n~.
\]

\noindent
The components $\rho _{jl}$ ($l=1,2,..,n$) are determined by solving the 
eigenvalue problem given by (\ref{4.4.2}). The new coordinates are called
normal coordinates and the $w_j$
are known as the normal frequencies. The equivalent matrix form (\ref{4.4.4}) is

\begin{equation}
\left(
\begin{array}{l}
\stackrel{\_}{x}_1^{(j)} \\
\stackrel{\_}{x}_2^{(j)} \\
\vdots  \\
\stackrel{\_}{x}_n^{(j)}
\end{array}
\right) =A_j\cos(w_jt+\phi _j)\left(
\begin{array}{l}
\rho _{j1} \\
\rho _{j2} \\
\vdots  \\
\rho _{jn}
\end{array}
\right)~.   \label{4.4.5}
\end{equation}

\noindent
These are the normal vibrational modes of the system.
One reason for introducing the coordinates $y_j$ is found from the 
expression for the kinetic energy, which is seen to be invariant under the 
rotation to the new axes. 

\[
T=\frac{1}{2}\sum_{j=1}^{n}M_{j}\stackrel{.}{y}_j^2~.
\]

\underline{EXAMPLE}

\noindent
Apply the matricial procedure as already shown, given the following 
equations of motion 

\[
\frac{d^2}{dt^{2}}\left(
\begin{array}{c}
\stackrel{\_}{x}_{1} \\
\stackrel{\_}{x}_{2} \\
\stackrel{\_}{x}_{3}
\end{array}
\right) =-\left(
\begin{array}{ccc}
5 & 0 & 1 \\
0 & 2 & 0 \\
1 & 0 & 5
\end{array}
\right) \left(
\begin{array}{c}
\stackrel{\_}{x}_{1} \\
\stackrel{\_}{x}_{2} \\
\stackrel{\_}{x}_{3}
\end{array}
\right)~.
\]

\noindent
Comparing with (\ref{4.4.1}), we identify the 
operator $\lambda$. To find the eigenvectores,
we use (\ref{4.4.2}) getting

\[
\left(
\begin{array}{ccc}
5 & 0 & 1 \\
0 & 2 & 0 \\
1 & 0 & 5
\end{array}
\right) \left(
\begin{array}{c}
\rho _{1} \\
\rho _{2} \\
\rho _{3}
\end{array}
\right) =\lambda _{i}\left(
\begin{array}{c}
\rho _{1} \\
\rho _{2} \\
\rho _{3}
\end{array}
\right)~.
\]

\noindent
The characteristic equation for $\lambda _{i}$ is

\[
\det (\lambda -\lambda _{i}I)=0~,
\]

\noindent
and by substituting the values

\[
\left|
\begin{array}{ccc}
5-\lambda  & 0 & 1 \\
0 & 2-\lambda  & 0 \\
1 & 0 & 5-\lambda
\end{array}
\right| =0~.
\]
\newline

\noindent
Solving the equation one gets $\lambda _{i}=2,4,6$.
For $\lambda=4$

\[
\left(
\begin{array}{ccc}
5 & 0 & 1 \\
0 & 2 & 0 \\
1 & 0 & 5
\end{array}
\right) \left(
\begin{array}{c}
\rho _{1} \\
\rho _{2} \\
\rho _{3}
\end{array}
\right) =4\left(
\begin{array}{c}
\rho _{1} \\
\rho _{2} \\
\rho _{3}
\end{array}
\right)
\]

\noindent
we have the following set of equations

\begin{eqnarray*}
(5-4)\rho _{1}+\rho _{3} &=&0 \\
2\rho _{2}-4\rho _{2} &=&0 \\
\rho _{1}+(5-4)\rho _{3} &=&0~.
\end{eqnarray*}

\noindent
Taking into account the normalization condition, one is led to the following
values

\begin{eqnarray*}
\rho _{1} &=&-\rho _{3}=\frac{1}{\sqrt{2}} \\
\rho _{2} &=&0~.
\end{eqnarray*}

\noindent
Therefore

\[
\left| \rho _{\lambda =4}\right\rangle =\frac{1}{\sqrt{2}}\left(
\begin{array}{c}
1 \\
0\\
-1
\end{array}
\right)~.
\]

\noindent
By the same means one gets

\begin{eqnarray*}
\left| \rho _{\lambda =6}\right\rangle  &=&\frac{1}{\sqrt{2}}\left(
\begin{array}{c}
1 \\
0 \\
1
\end{array}
\right)  \\
\left| \rho _{\lambda =2}\right\rangle  &=&\left(
\begin{array}{c}
0 \\
1 \\
0
\end{array}
\right)~.
\end{eqnarray*}

\noindent
Thus, the new vectorial space is determined by 

\[
\left| \rho _{i}\right\rangle =\left(
\begin{array}{ccc}
\frac{1}{\sqrt{2}} & \frac{1}{\sqrt{2}} & 0 \\
0 & 0 & 1 \\
-\frac{1}{\sqrt{2}} & \frac{1}{\sqrt{2}} & 0
\end{array}
\right)~,
\]

\noindent
where from 

\[
\left\langle \rho _{i}\right| =\left(
\begin{array}{ccc}
\frac{1}{\sqrt{2}} & 0 & -\frac{1}{\sqrt{2}} \\
\frac{1}{\sqrt{2}} & 0 & \frac{1}{\sqrt{2}} \\
0 & 1 & 0
\end{array}
\right)~.
\]

\noindent
Thus, the normal coordinates are given by (\ref{4.4.3})

\[
\left(
\begin{array}{c}
y_{1} \\
y_{2} \\
y_{3}
\end{array}
\right) =\left(
\begin{array}{ccc}
\frac{1}{\sqrt{2}} & 0 & -\frac{1}{\sqrt{2}} \\
\frac{1}{\sqrt{2}} & 0 & \frac{1}{\sqrt{2}} \\
0 & 1 & 0
\end{array}
\right) \left(
\begin{array}{c}
\stackrel{\_}{x}_{1} \\
\stackrel{\_}{x}_{2} \\
\stackrel{\_}{x}_{3}
\end{array}
\right)~.
\]
\section*{4.5 PARAMETRIC RESONANCE}

\noindent
The important phenomenon of parametric resonance shows up for systems initially
at rest in un unstable equilibrium position, say $x=0$; 
thus, the slightest deviation from this position produces a displacement 
growing rapidly (exponentially) in time. This is different from the 
ordinary resonances, where the displacement grows only linearly in time.

\noindent
The parameters of a linear system are the coefficients $m$ and $k$ of the 
Lagrangian (\ref{4.1.3}); if they are functions of time, the equation of motion
is:

\begin{equation}
\frac d{dt}(m\stackrel{\cdot }{x})+kx=0~.  \label{5.7.1}
\end{equation}

\noindent
If we take a constant mass, the previous equation can be written in the form

\begin{equation}
\frac{d^2x}{dt^2}+w^2(t)x=0~.  \label{4.6.2}
\end{equation}

\noindent
The function $w(t)$ is given by the problem at hand. Assuming it a periodic 
function of frequency
$\gamma$ (of period $T=2\pi/\gamma $), i.e.,

\[
w(t+T)=w(t)~,
\]

\noindent
any equation of the type (\ref{4.6.2}) is invariant w.r.t.
the transformation $t\rightarrow t + T$. Thus, if $x(t)$ is one of its
solutions, $x(t+T)$ is also a 
solution. Let $x_1(t)$ and $x_2(t)$ be two independent 
solutions of {4.6.2}). They should change
to itselves in a linear combination when $t\rightarrow t + T$. 
Thus, one gets

\begin{eqnarray}
x_1(t+T) &=&\mu _1x(t)  \label{4.6.3} \\
x_2(t+T) &=&\mu _2x(t)~,  \nonumber
\end{eqnarray}

\noindent
or, in general

\begin{eqnarray*}
x_1(t) &=&\mu _1^{t/T}F(t) \\
x_2(t) &=&\mu _2^{t/T}G(t)~,
\end{eqnarray*}

\noindent
where $F(t)$ and $G(t)$ are periodical functions in time of period $T$.
The relationship between these constants can be obtained by manipulating 
the following equations

\begin{eqnarray*}
\stackrel{..}{x}_1+w^2(t)x_1 &=&0 \\
\stackrel{..}{x}_2+w^2(t)x_2 &=&0~.
\end{eqnarray*}

\noindent
Multiplying by $x_2$ and $x_1$, respectively, and substracting term by term, 
we get

\[
\stackrel{..}{x_1}x_2-\stackrel{..}{x}_2x_1=\frac d{dt}(\stackrel{.}{x_1}x_2-%
\stackrel{.}{x}_2x_1)=0~,
\]

\noindent
or

\[
\stackrel{.}{x_1}x_2-\stackrel{.}{x}_2x_1=const.~.
\]

\noindent
Substituting $t$ by $t+T$ in the previous equation, the right hand side
is multiplied by $\mu _1\mu _2$ (see eqs. (\ref{4.6.3})); thus, it is obvious
that the following condition holds

\begin{equation}
\mu _1\mu _2=1~,  \label{4.6.4}
\end{equation}

\noindent
where one should take into account (\ref{4.6.2})
and the fact that the coefficients are real.
If $x(t)$ is one integral of this equation, then $x^{*}(t)$
is also a solution. Therefore $\mu _1$, $\mu _2$ should coincide
with $\mu _1^{*}$, $\mu _2^{*}$. This leads to either
$\mu _1$=$\mu _2^{*}$ or $\mu _1$ and $\mu _2$ both real. 
In the first case, based on (\ref{4.6.4}) one gets $\mu _1=1/$ $\mu _1^{*}$,
that is equivalent to $\left| \mu _1\right| ^2=\left| \mu _2\right| ^2=1$. 
In the second case, the two solutions are of the form
\begin{eqnarray*}
x_1(t) &=&\mu ^{t/T}F(t) \\
x_2(t) &=&\mu ^{-t/T}G(t)~.
\end{eqnarray*}

\noindent
One of these functions grows exponentially in time, which is the characteristic
feature of the parametric resonance.

\bigskip
\bigskip

\begin{center} REFERENCES AND FURTHER READING \end{center}

\bigskip

* H. Goldstein, {\it Classical mechanics}, Second ed. (Addison-Wesley, 1981).

\bigskip

* L. D. Landau \& E. M. Lifshitz, {\it Mechanics}, (Pergammon, 1976).

\bigskip

* W. Hauser, {\it Introduction to the principles of mechanics}, (Wesley, 1965).

\bigskip

* E.I. Butikov, {\it Parametric Resonance}, Computing
in Science \&

Engineering, May/June 1999, pp. 76-83 (http://computer.org).


%% file: c11en.tex




\centerline{\Large 5. CANONICAL TRANSFORMATIONS}

\bigskip
\bigskip

\noindent
{\bf Forward}:
The main idea of canonical transformations
is to find all those coordinate systems in the phase space  
for which the form of the Hamilton eqs is invariant for whatever Hamiltonian.
In applications one chooses the coordinate system that allows a simple solution
of the problem at hand.\\

\bigskip

{\bf CONTENTS:}

\bigskip

5.1 Definitions, Hamiltonians and Kamiltonians

\bigskip

5.2 Necessary and sufficient conditions for canonicity 

\bigskip

5.3 Example of application of a canonical transformation

\newpage

\noindent
\section*{5.1 Definitions, Hamiltonians and Kamiltonians}
For the time-independent and time-dependent cases, respectively, one defines
a canonical transformation as follows 

\noindent
{\bf Definition 1:} A time-independent transformation 
$Q=Q(q,p)$, and $P=P(q,p)$ is called canonical 
if and only if there is a function $F(q,p)$ such that
\setcounter{equation} {0}\\
$$dF(q,p)=\sum_{i}p_{i}dq_{i}-\sum_{i}P_{i}(q,p)dQ_{i}(q,p)~.$$
{\bf Definition 2:} A time-dependent transformation $Q=Q(q,p,t)$, and 
$P=P(q,p,t)$ is called canonical if and only if there is a function  
$F(q,p,t)$ such that for an arbitrary fixed time $t=t_{0}$ 
$$dF(p,q,t_{0})= \sum_{i}p_{i}dq_{i}-\sum_{i}P_{i}(q,p,t_{0})
dQ_{i}(p,q,t_{0})~,$$
where
$$dF(p,q,t_{0})= \sum_{i}\frac{\partial F(p,q,t_{0})}{\partial q_{i}}dq_{i}
+ \sum_{i}\frac{\partial 
F(p,q,t_{0})}{\partial p_{i}}dp_{i}$$
and
$$dQ(p,q,t_{0})= \sum_{i}\frac{\partial Q(p,q,t_{0})}{\partial q_{i}}dq_{i}
+ \sum_{i}\frac{\partial
Q(p,q,t_{0})}{\partial p_{i}}dp_{i}$$

\noindent
{\bf Example:}
Prove that the following transformation is canonical
\begin{eqnarray*}
P &=& \frac{1}{2}(p^2+q^2)\\
Q &=& Tan^{-1}\left( \frac{q}{p}\right)~.
\end{eqnarray*}
{\bf Solution:}
According to the first definition we have to check that $pdq-PdQ$ is an exact
differential. Substituting $P$ and $Q$ in the definition we get
$$
pdq-PdQ = pdq - \frac{1}{2}(p^2+q^2)\frac{pdq-qdq}{p^2+q^2}=
d\left(\frac{pq}{2}\right)~.
$$
We can see that indeed the given transformation is canonical.
We know that a dynamical system is usually characterized by a
Hamiltonian $H=H(q,p,t)$, where $q=q(q_{1},q_{2},...,q_{n})$, and  
$p=p(p_{1},p_{2},...,p_{n})$. Therefore, the dynamics of the system fulfills a 
set of $2n$ first-order differential eqs (Hamilton's eqs.)
\begin{eqnarray}
\dot{q_{i}}=\frac{\partial H}{\partial p_{i}}\\
-\dot{p_{i}}=\frac{\partial H}{\partial q_{i}}~.
\end{eqnarray}
Let us denote the coordinate transformations in the phase space by 
\begin{eqnarray}
Q_{j}=Q_{j}(q,p,t)\\
P_{j}=P_{j}(q,p,t)~.
\end{eqnarray}
According to the aforementioned principle for the set of canonical 
transformations denoted by $(3)$ and $(4)$, analogously to $(1)$ and $(2)$, 
there is a function $K=K(Q,P,t)$ such that we can write 
\begin{eqnarray}
\dot{Q_{i}}=\frac{\partial K}{\partial P_{i}}\\
-\dot{P_{i}}=\frac{\partial K}{\partial Q_{i}}~.
\end{eqnarray}
The relationship between the Hamiltonian $H$ and the Kamiltonian 
$K$\footnote{Here we follow the terminology of 
Goldstein by referring to $K=K(Q,P,t)$, which is different from the  
Hamiltonian $H=H(p,q,t)$ by an additive time derivative, as the  
\it{Kamiltonian}.} can be obtained arguing as 
follows\footnote{An alternative derivation has been given by
G. S. S. Ludford and D. W. Yannitell, Am. J. Phys. 36,
231 (1968).}.\\
$\;$\\
According to Hamilton's principle, the real trajectory of a classical system
can be obtained from the variation of the action integral 
\begin{equation}
\delta \int (\sum_{i}p_{i}dq_{i} - Hdt) = 0~.
\end{equation}
If the transformation is canonical, the 
Kamiltonian $K$ should fulfill a relationship similar to (7). In other words,
for the new set of variables $Q$ and $P$ we still have
\begin{equation}
\delta \int (\sum_{i}P_{i}dQ_{i} - Kdt) = 0~.
\end{equation}
Moreover, according to the Legendre transformation, 
$\sum_{i}p_{i}dq_{i} - Hdt= L(q,\dot{q},t)dt$, (7) - like (8)- is equivalent to 
\begin{equation}
\delta \int_{t_{1}}^{t_{2}} L(q,\dot{q},t)dt = 0~.
\end{equation}
In addition, (9) does not change if $L$ is replaced by ${\cal L}= L +
\frac{dF(q,t)}{dt}$ because in this case
\begin{eqnarray}
\delta \int_{t_{1}}^{t_{2}} {\cal L}dt = \delta \int_{t_{1}}^{t_{2}} (L +
\frac{dF(q,t)}{dt})dt~,
\end{eqnarray}
or similarly
\begin{eqnarray}
\delta \int_{t_{1}}^{t_{2}} {\cal L}dt = \delta \int_{t_{1}}^{t_{2}}
L(q,\dot{q},t)dt + 
\delta F(q_{(2)},t_{2}) - \delta F(q_{(1)},t_{1})~.
\end{eqnarray}
Thus, (10) and (11) differ only by constant terms whose variation is zero
in Hamilton's principle.\\
$\;$
It follows that the Hamiltonian $H$ and the
Kamiltonian $K$ are related by the equation 
\footnote{Some authors add to the right hand side of this equation
a constant multiplicative factor $A$ that does not change (9). Here we use
$A=1$, that is,
we decided to work with the so-called {\it reduced canonical transformations},
since this simpler case is sufficient to illustrate the structure of the 
canonical transformations.}
\begin{eqnarray}
p_{i}\dot{q_{i}} - H = P_{i}\dot{Q_{i}} - K + \frac{dF}{dt}~.
\end{eqnarray}
The function $F$ is the so-called {\it generating function}. It can be expressed
as a function of any arbitrary set of independent variables.
However, some very convenient results are obtained if $F$ is expressed as a 
function of the $n$ old variables and the $n$ new ones, plus the time. The 
results are especially convenient if the $n$ old variables are exactly the
$n$ $q_{i}$ 
- or the $n$ $p_{i}$-, and if the new variables are all of them the $n$ 
$Q_{i}$ - or the $n$ $P_{i}$.\\
$\;$\\
Using these coordinates,
the possible combinations of $n$ old variables and $n$ 
new variables -including $t$- in the generating function 
are\footnote{We shall use the convention of Goldstein
to denote each of the different combinations of the new and old
variables in the generating function.}
\begin{eqnarray}
F_{1} & = & F_{1}(Q,q,t) \\ 
F_{2} & = & F_{2}(P,q,t) \nonumber \\
F_{3} & = & F_{3}(Q,p,t) \nonumber \\
F_{4} & = & F_{4}(P,p,t)~. \nonumber
\end{eqnarray}
On the other hand, if we multiply $(12)$ by $dt$ we get:
\begin{equation}
p_{i}dq_{i} - Hdt = PdQ_{i} - Kdt + dF~.
\end{equation}
Making the change $F \rightarrow F_{1}$ above, and recalling
that $dQ_{i}$, $dq_{i}$, and $dt$ are independent variables, we get:
\begin{eqnarray}
P_{i} &=& -\frac{\partial F_{1}}{\partial Q_{i}} \nonumber \\
p_{i} &=& \frac{\partial F_{1}}{\partial q_{i}} \nonumber \\
K     &=& H+\frac{\partial F_{1}}{\partial t}~. \nonumber
\end{eqnarray}
Using now some algebraic manipulation it is possible to obtain analogous
expressions to the previous one for the rest of the generating functions.
The results are the following:
\begin{center}
\begin{tabular}{clll}
$F_{2}:$ & $Q_{i} = \;\; \frac{\partial F_{2}}{\partial P_{i}}$ & $p_{i}
= \;\; \frac{\partial 
F_{2}}{\partial q_{i}}$ 
        & $K     = H+\frac{\partial F_{2}}{\partial t}$ \\
&&&\\
$F_{3}:$ & $P_{i} = -\frac{\partial F_{3}}{\partial Q_{i}}$ & $q_{i}
= -\frac{\partial F_{3}}{\partial 
p_{i}}$  
        & $K     = H+\frac{\partial F_{3}}{\partial t}$ \\
&&&\\
$F_{4}:$ & $Q_{i} = \;\; \frac{\partial F_{4}}{\partial P_{i}}$  & $q_{i}
= -\frac{\partial F_{4}}{\partial 
p_{i}}$ 
        & $K     = H+\frac{\partial F_{4}}{\partial t}~.$ \\
\end{tabular}
\end{center}
In practice, one usually applies a useful theorem (see below) that allows,
together with the definitions we gave in the introduction
for canonical transformations, to solve any mechanical problem of interest
\footnote{For an example, see the final section of this chapter.}.

\noindent
{\bf Theorem 5.1}
We consider a system acted by a given external force. We also suppose that the
dynamical state of the system is determined by a set of variables $q,p= 
q_{1},q_{2},...,q_{n},p_{1},p_{2},...,p_{n}$ and that the 
Hamiltonian of the system is $H=H(q,p,t)$. The time evolution of the 
variables $q$ and $p$ is given by Hamilton's eqs.
\begin{eqnarray}
\dot{q_{i}} &=& \;\; \frac{\partial H(q,p,t)}{\partial p_{i}} \nonumber \\
\dot{p_{i}} &=& -\frac{\partial H(q,p,t)}{\partial q_{i}}~. \nonumber 
\end{eqnarray}
If we now perform a transformation to the new variables 
$$Q=Q(q,p,t) \qquad ;\qquad P=P(q,p,t)$$
and if the transformation is canonical, i.e., there exists a function
$F(q,p,t)$ such that for a fixed arbitrary time $t=t_{0}$ we have
$$dF(q,p,t_{0})= \sum_{i}y_{i}dx_{i}-\sum_{i}Y_{i}dX_{i}~,$$
where $x_{i},y_{i}=q_{i},p_{i}$ or $p_{i}, -q_{i}$ y $X_{i},Y_{i}=Q_{i},P_{i}$,
or $P_{i}, -Q_{i}$, 
then the equations of motion in terms of the variables $Q$ and $P$ are
\begin{eqnarray}
\dot{Q_{i}} &=& \;\; \frac{\partial K(Q,P,t)}{\partial P_{i}} \nonumber \\
\dot{P_{i}} &=& -\frac{\partial K(Q,P,t)}{\partial Q_{i}}~, \nonumber
\end{eqnarray}
where
$$K \equiv H + \frac{\partial F(q,p,t)}{\partial t} + \sum_{i}Y_{i}
\frac{\partial X_{i}(q,p,t)}{\partial t}~.$$
Moreover, if the determinant of the matrix $[\frac{\partial X_{i}}
{\partial y_j}]$ is different of zero, then the latter equation takes the form
$$K \equiv H + \frac{\partial F(x,X,t)}{\partial t}~.$$
\section*{5.2 Necessary and sufficient conditions
for a transformation to be canonical}
We have already mentioned that by a canonical transformation we mean a 
transformation, which, independently of the form of the 
Hamiltonian, keeps unchanged the form of Hamilton's equations.
However, one should be very careful with this issue because some transformations
fulfill this requirement only for 
{\it a particular Hamiltonian} \footnote{See, for example, J. Hurley 
Am. J. Phys. {\bf 40}, 533 (1972).}. 
Some authors call such transformations {\it 
canonical transformations w.r.t. H}
\footnote{See, for example, R. A. Matzner and
L. C. Shepley, {\it Classical Mechanics}
(Prentice Hall, 1991).}.\\ $\;$\\
To illustrate this point we use the following example, given in the paper of
J. Hurley:
Let us consider a {\it  particular physical system} whose Hamiltonian is 
\begin{eqnarray}
H = \frac{p^2}{2m}
\end{eqnarray}
and the following transformations 
\begin{eqnarray} 
\begin{array}{lll}
P & = & p^{2}  \\
Q & = & q~. 
\end{array}
\end{eqnarray}
It is easy to show that the Kamiltonian $K$ given by 
\begin{eqnarray}
K=\frac{2P^{3/2}}{3m} \nonumber
\end{eqnarray}
leads to
\begin{eqnarray}
\dot{P}=2p\dot{p}=0=-\frac{\partial K}{\partial Q} \nonumber 
\end{eqnarray}
and 
\begin{equation}
\dot{Q}=\dot{q}=\frac{p}{m}=\frac{P^{1/2}}{m}=\frac{\partial K}{\partial P}~.
\nonumber
\end{equation}
On the other hand, if we choose the Hamiltonian
\begin{eqnarray}
H = \frac{p^2}{2m} + q^2~, \nonumber
\end{eqnarray}
then it is possible to find a Kamiltonian $K$ for which the usage of
the transformation equations $(16)$ maintains unchanged the form of 
Hamilton's equations. Thus,
the equations $(16)$ keeps unchanged the form of Hamilton's equations only for
a {\it particular Hamiltonian}.

\noindent
It can be shown that the necessary and sufficient conditions  
for the canonicity of transformations of the form $(3)$ and $(4)$,
that is, to keep unchanged the form of Hamilton's equations 
{\it whatever the Hamiltonian}, are the following
\begin{equation}
[Q_{i},P_{j}]= \alpha 
\end{equation}
\begin{equation}
[P_{i},P_{j}]  =  0\\
\end{equation}
\begin{equation}
[Q_{i},Q_{j}]  =  0~,
\end{equation}
where $\alpha$ is an arbitrary constant related to scale changes.
Finally, we would like to make a few important comments before closing this 
section. First, we should keep in mind that 
$Q$ and $P$ {\it are not variables defining the configuration of the system}, 
i.e., they are not in general a set of generalized coordinates 
\footnote{Except for the trivial case in which the canonical transformation
is $Q=q$ and $P=p$.}. To distinguish $Q$ and $P$ from the generalized
coordinates $q$ and $p$, one calls them {\it canonical variables}. 
In addition, the equations of motion -similar in form to the Hamiltonian ones-
for the generalized coordinates $q$ and $p$- that one gets for $Q$ and $P$
are called {\it canonical Hamilton equations}.
Second, although we did not check here, if the transformation
$Q=Q(q,p,t)$ and 
$P=P(q,p,t)$ is canonical, then its inverse $q=q(Q,P,t)$
and $p=p(Q,P,t)$ is also canonical \footnote{For a proof see, for example,
E. A. Desloge,
{\it Classical Mechanics, Volume 2} (John Wiley \& Sons, 1982).}.
\section*{5.3 Example of application of TC}
As we already mentioned in the introduction, the main idea in performing
a canonical transformation is to find a phase space coordinate system
for which the form of the Hamilton eqs is maintaind whatever the Hamiltonian 
and to choose the one that makes easy the solution of the problem. 
We illustrate this important fact with the following example.

\noindent
\underline{{\it EXAMPLE}}:

\noindent
The Hamiltonian of a physical system is given by $H=\omega^{2}p(q+t)^{2}$,
where 
$\omega$ is a constant. Determine $q$ as a function of time.

\noindent
{\bf Solution:}

\noindent
1. {\it Solving the Hamilton equations for the 
variables $q$ and $p$.} Applying $(1)$ and $(2)$ to the given Hamiltonian 
we get
\begin{eqnarray}
\omega^{2}(q+t)^{2}=\dot{q}, & \;\;\;\;\;\;\;\; 2\omega^{2}p(q+t)=-\dot{p}~.
\nonumber
\end{eqnarray}
This system is not easy to solve. However, we can get the solution by means
of an appropriate canonical transformation as we show in the following.

\noindent
2. {\it Using $Q=q+t$, $P=p$.}
According to the theorem given in section $(5.1)$, since
\begin{eqnarray}
\frac{\partial Q}{\partial  p} &=& 0  \nonumber \\
\frac{\partial P}{\partial (-q)} &=& 0~, \nonumber
\end{eqnarray}
then the Kamiltonian $K$ of the system is given by
\begin{equation}
K = H + \frac{\partial F(q,p,t)}{\partial t} + P\frac{\partial Q}{\partial t}
 - Q\frac{\partial P}{\partial t}~. 
\end{equation}
The form of the function $F(q,p,t)$ can be obtained from the canonical 
transformation given in section 5.1 (the case corresponds to a time-dependent
canonical transformation). Therefore, we substitute $Q=q+t$, $P=p$ in
$$dF(q,p,t)= pdq - PdQ~,$$
to get without difficulty
$$F(q,p,t)= c, \;\;\;\; \mbox{c= constant}~.$$
On the other hand,
\begin{eqnarray}
\frac{\partial P}{\partial t} &=& 0  \nonumber \\
\frac{\partial Q}{\partial t} &=& 1~.  \nonumber
\end{eqnarray}
Finally, substituting these results in $(21)$
(and also $Q=q+t$, $P=p$ in $H$)
we get
$$K= P(\omega^{2}Q^{2}+1)~.$$
Moreover, from $(5)$ we find
$$\dot{Q}= \omega^{2}Q^{2}+1~.$$
This differential equation is now easy to solve, leading to
$$q=\frac{1}{\omega}{\rm tan}(\omega t + \phi)-t~,$$
where $\phi$ is an arbitrary phase.


%% file: c22en.tex



\centerline{\Large 6. POISSON BRACKETS}

\bigskip

\noindent
{\bf Forward}:
The Poisson brackets are very useful analytical tools for the study of any
dynamical system. Here, we define them, give some of their properties, and 
finally present several applications. 

\bigskip
\bigskip

{\bf CONTENTS:}

\bigskip



1. Definition and properties

\bigskip

2. Poisson formulation of the equations of motion

\bigskip

3. The constants of motion in Poisson formulation

\newpage


\noindent
{\bf 1. Definition and properties of Poisson brackets}
\bigskip

\noindent
If $u$ and $v$ are any two quantities that depend on the 
dynamical state of a system, i.e., on $p$ and $q$) and possibly on time,
the Poisson bracket of $u$ and $v$ w.r.t. a set of canonical variables $q$ and
$p$ \footnote{As in the previous chapter, by
$q$ and $p$ we mean $q=q_{1}, q_{2},...,q_{n}$ y $p=p_{1},
p_{2},..., p_{n}.$} 
is defined as follows
\setcounter{equation}{0}\\
\begin{equation}
[u,v] \equiv \sum_{i} \left( \frac{\partial u(q,p,t)}{\partial q_{i}}
\frac{\partial v(q,p,t)}{\partial p_{i}}
- \frac{\partial u(q,p,t)}{\partial p_{i}}\frac{\partial v(q,p,t)}{\partial 
q_{i}} \right)~.
\end{equation}
The Poisson brackets have the following properties (for $u$, $v$,
and $w$ arbitrary functions of $q$, $p$, and $t$;
$a$ is an arbitrary constant, and $r$
is any of $q_{i}$, $p_{i}$ or $t$) \footnote{The proof of these properties can
be obtained by using the definition of the PBs in order to express each term of
these identities by means of partial derivatives of $u$, $v$, and $w$, and
noticing by inspection that the resulting equations do hold.}:
\begin{description}
\item[{\it 1.}] \hspace{2in} $[u,v] \equiv - [v,u]$
\item[{\it 2.}] \hspace{2in} $[u,u] \equiv 0$
\item[{\it 3.}] \hspace{2in} $[u,v+w] \equiv [u,v] + [u,w]$
\item[{\it 4.}] \hspace{2in} $[u,vw] \equiv v[u,w] + [u,v]w$
\item[{\it 5.}] \hspace{2in} $a[u,v] \equiv [au,v] \equiv [u,av]$
\item[{\it 6.}] \hspace{2in} $\frac{\partial [u,v]}{\partial r}
\equiv [\frac{\partial u}{\partial r},
v]+[u,\frac{\partial v}{\partial r}]$
\item[{\it 7.}] {\it The Jacobi identity},
\hspace{0.3in}  $[u,[v,w]]+[v,[w,u]]+[w,[u,v]] \equiv 0~.$
\end{description}
Another very important property of PBs is the content of the following theorem

\noindent
{\bf Theorem 6.1} If the transformation $Q=Q(q,p,t)$, $P=P(q,p,t)$
is a canonical transformation, the PB of $u$ and $v$ w.r.t. the variables 
$q$, $p$ is equal to the PB of $u$ and $v$ w.r.t. the set of  variables 
$Q$, $P$, i.e.,
$$
\sum_{i} \left( \frac{\partial u(q,p,t)}{\partial q_{i}}
\frac{\partial v(q,p,t)}{\partial p_{i}}
- \frac{\partial u(q,p,t)}{\partial p_{i}}\frac{\partial v(q,p,t)}{\partial q_{i}}
\right)=
$$
$$
\sum_{i} \left( \frac{\partial u(q,p,t)}{\partial Q_{i}}
\frac{\partial v(q,p,t)}{\partial P_{i}}
- \frac{\partial u(q,p,t)}{\partial P_{i}}
\frac{\partial v(q,p,t)}{\partial Q_{i}} \right)~.
$$

\bigskip
\noindent
{\bf 2. Poisson formulation of the equations of motion}
\bigskip

\noindent
In the following, we outline as theorem-like statements the most important 
results on the PB formulation of the eqs of motion of the dynamical systems 
\footnote{The proofs have been omitted as being well known.
See, for example, E. A. Desloge,
{\it Classical Mechanics}, Volume 2 (John Wiley \& Sons, 1982).}:

\noindent
{\bf Theorem 6.2} Consider a system whose dynamical state
is defined by the canonical variables $q$, $p$ and whose dynamical behaviour
is defined by the Hamiltonian $H=H(q,p,t)$. Let $F$ be an arbitrary quantity
depending on the dynamical state of the system, i.e., on $q$, $p$,
and possibly on $t$. The rate of change in time of $F$ is given by
$$\dot{F}=[F,H]+\frac{\partial F(q,p,t)}{\partial t}~,$$
where $[F,H]$ is the PB of $F$ and $H$.

\noindent
{\bf Theorem 6.3} {\bf (Poisson formulation of the eqs. of motion)}. 
Consider a system described in terms of the canonical variables
$q$, $p$, and whose Hamiltonian is $H=H(q,p,t)$. The motion of the system is 
governed in this case by the equations
\begin{eqnarray}
\dot{q_{i}} &=& [q_{i},H] \nonumber \\
\dot{p_{i}} &=& [p_{I},H]~. \nonumber
\end{eqnarray}

\noindent
{\bf 3. Constants of motion in Poisson's formulation}

\bigskip

\noindent
We shall use again a theorem-like sketch of the basic results on 
the constants of motion in Poisson's formulation. 
These results are the following.

\noindent
{\bf Theorem 6.4} If one dynamical quantity $F$ is not an explicit function of
time and if the PB of $F$ and $H$ is zero, $[F,H]=0$, then $F$ is a 
constant/integral of motion as one can see from the theorem 6.2.

\noindent
{\bf Corollary 6.4.} If the Hamiltonian is not an explicit function
of time, then it is a constant of motion.



%% file: c33en.tex


\centerline{\Large 7. HAMILTON-JACOBI EQUATIONS}

\bigskip
\bigskip

\noindent
{\bf Forward}:
It is known from the previous chapters that in principle it is possible
to reduce the complexity of many dynamical problems by choosing an appropriate
canonical transformation.
In particular, we can try to look for those canonical transformations for which
the Kamiltonian $K$ is zero, a situation leading to the
Hamilton-Jacobi equations.

\bigskip
\bigskip

{\bf CONTENTS:}

\bigskip

7.1 Introduction

\bigskip

7.2 Time-dependent Hamilton-Jacobi equations

\bigskip

7.3 Time-independent Hamilton-Jacobi equations

\bigskip

7.4 Generalization of the Hamilton-Jacobi equations

\bigskip

7.5 Example of application of the Hamilton-Jacobi equations

\newpage

\section*{7.1 Introduction}
In order to reach the goals of this chapter we need to make use of the 
following result allowing us to find the set
of canonical variables for which the Kamiltonian takes a particular form.

\noindent
{\bf Theorem 7.1} Consider a system whose dynamical state 
is defined by $p$, $q$ and whose behaviour under the action of a given force
is governed by the Hamiltonian $H=H(q,p,t)$. Let $K=K(Q,P,t)$ be a 
{\it known function} of the canonical variables $Q$, $P$, and time.  
Then, any function $F(q,Q,t)$ that satisfies the partial differential equation 
\setcounter{equation}{0}\\
\begin{eqnarray}
K\left[Q,-\frac{\partial F(q,Q,t)}{\partial Q},t\right]= H\left[q,\frac{\partial F(q,Q,t)}{\partial q},t\right]+
\frac{\partial F(q,Q,t)}{\partial t} \nonumber
\end{eqnarray}
and also the condition 
\begin{eqnarray}
\left| \frac{\partial ^{2} F(q,Q,t)}{\partial q_{j}\partial Q_{j}}\right|
\neq 0 \nonumber
\end{eqnarray}
is a generating function for a canonical transformation of 
$q$, $p$ to $Q$, $P$, and the corresponding Kamiltonian is $K=K(Q,P,t)$.

\noindent
In the following sections we shall use this theorem to find those canonical
transformations whose Kamiltonian is zero\footnote{More exactly,
$K\left[Q,-\frac{\partial F(q,Q,t)}{\partial Q},t\right]=0$.}, that leads us to
the Hamilton-Jacobi equations.
\section*{7.2 Time-dependent HJ equations.}
As a consequence of Theorem 7.1 and of requiring a zero Kamiltonian
we get the following theorem 

\noindent
{\bf Theorem 7.2} Consider a system of $f$ degrees of freedom
defined by the set of variables $q$, $p$ and of Hamiltonian 
$H=H(q,p,t)$. If we build the partial differential equation 
\begin{eqnarray}
H\left[ q,\frac{\partial S(q,t)}{\partial q}, t\right]
+ \frac{\partial S(q,t)}{\partial t}=0
\end{eqnarray}
and if we are able to find a solution of the form 
$$S=S(q,\alpha,t)~,$$
where $\alpha=\alpha_{1}, \alpha_{2},...,\alpha_{f}$ is a set of constants
and if in addition the solution satisfies the condition 
$$\left| \frac{\partial^{2}S(q,\alpha,t)}{\partial q_{i}\partial \alpha_{i}}
\right| \neq 0~,$$
then $q(t)$ can be obtained from the equations 
\begin{equation}
\frac{\partial S(q,\alpha,t)}{\partial \alpha_{i}}= \beta_{i}~,
\end{equation}
where $\beta= \beta_{1},\beta_{2},...,\beta_{f}$ is a set of constants.
The set of equations $(2)$ provide us with $f$ algebraic equations
in the $f$ unknown variables $q_{1}, q_{2},...,q_{f}$. The values of the 
constants $\alpha$ and 
$\beta$ are determined by the boundary conditions. Moreover, if $q(t)$ is given
it is possible to find $p(t)$ starting from
\begin{eqnarray}
p_{i} &=& \frac{\partial S(q,\alpha,t)}{\partial q_{i}}~.
\end{eqnarray}
The partial differential equation $(1)$ is called {\it the time-dependent
Hamilton-Jacobi equation}. The function $S(q,\alpha,t)$ is known as 
{\it Hamilton's principal function}.

\noindent
To achieve a better meaning of the theorem,
as well as of the constants $\alpha$ and $\beta$, we proceed with its proof.

\noindent
{\bf Proof of the Theorem 7.2.}
According to the Theorem 7.1, any function $F(q,Q,t)$ satisfying the partial
differential equation
$$H\left[q,\frac{\partial F(q,Q,t)}{\partial q},t\right]
+\frac{\partial F(q,Q,t)}{\partial t}=0$$
and also the condition
$$\left| \frac{\partial ^{2} F(q,Q,t)}{\partial q_{j}
\partial Q_{j}}\right| \neq 0 $$
should be a generating function of a set of canonical variables $Q$, $P$ 
for which the Kamiltonian $K$ is zero, i.e.,
$K(Q,P,t)=0$. The function
$$F(q,Q,t)=[S(q,\alpha,t)]_{\alpha=Q}\equiv S(q,Q,t)$$
belongs to this class. Then $S(q,Q,t)$ is the generating function for a 
canonical transformation leading to the new set of canonical variables $Q$, $P$,
for which the Kamiltonian 
$K$ is identically zero. The transformation equations associated to
$S(q,Q,t)$ are 
\begin{eqnarray}
p_{i} &=& \;\; \frac{\partial S(q,Q,t)}{\partial q_{i}} \\
P_{i} &=& -\frac{\partial S(q,Q,t)}{\partial Q_{i}}
\end{eqnarray}
and because $K(Q,P,t)\equiv 0$, the equations of motion are
$$\begin{array}{lllll}
\dot{Q}_{i} &=& \;\; \frac{\partial K(Q,P,t)}{\partial P_{i}} &=& 0 \\
&&&& \\
\dot{P}_{i} &=& -\frac{\partial K(Q,P,t)}{\partial Q_{i}} &=& 0
\end{array}~.$$
From these equations we infer that 
\begin{eqnarray}
Q_{i} &=& \;\; \alpha_{i} \\
P_{i} &=& -\beta~,
\end{eqnarray}
where $\alpha_{i}$ and $\beta_{i}$ are constants. The choice of the negative 
sign for $\beta$ in $(7)$ is only a convention.
If now we substitute the equations $(6)$ and $(7)$ 
in $(5)$ we get
$$ -\beta_{i} = -\left[ \frac{\partial S(q,Q,t)}{\partial Q_{i}}
\right]_{Q=\alpha}=  
   -\frac{\partial S(q,\alpha,t)}{\partial \alpha_{i}}~,$$
which reduces to $(2)$. If, in addition, we substitute $(6)$ in $(4)$
we get $(2)$. This complets the proof.
\section*{7.3 Time-independent HJ equations}
If the Hamiltonian does not depend explicitly on time, we can partially solve
the time-dependent Hamilton-Jacobi equation. This result can be spelled out 
as the following theorem

\noindent
{\bf Theorem 7.3} Consider a system of $f$ degrees of freedom
defined in terms of $q$, $p$, and whose behaviour under a given force
is governed by the time-independent Hamiltonian $H(q,p)$.

\noindent
If we build the partial differential equation  
\begin{equation}
H\left[q,\frac{\partial W(q)}{\partial q}\right]= E~,
\end{equation}
where $E$ is a constant whose value for a particular set of conditions
is equal to the value of the integral of motion $H(q,p)$ for the given boundary
conditions, and if we can find a solution to this equation of the form 
$$W=W(q,\alpha)~,$$
where $\alpha \equiv \alpha_{1},\alpha_{2},...,\alpha_{f}$ is a set of constants
that explicitly or implicitly include the constant $E$,
i.e., $E=E(\alpha)$, and if the solution satisfies the condition
$$\left| \frac{\partial^{2}W(q,\alpha)}{\partial q_{i} \partial
\alpha_{j}}\right| \neq 0~,$$
then the equations of motion are given by  
\begin{equation}
\frac{\partial S(q,\alpha,t)}{\partial \alpha_{i}}= \beta_{i}
\end{equation}
where
$$S(q,\alpha,t) \equiv W(q,\alpha)-E(\alpha)t$$
and $\beta = \beta_{1}, \beta_{2},...,\beta_{f}$ is a set of
constants. The set of equations $(9)$ provide $f$ algebraic equations in the $f$
unknown variables $q_{1}, q_{2},...,q_{f}$.
The values of the  constants $\alpha$ and
$\beta$ are determined by the boundary conditions. The partial differential 
equation $(8)$ is {\it the time-independent Hamilton-Jacobi equation}, and 
the function $W(q,\alpha)$ is known as 
{\it the characteristic Hamilton function}.
\section*{7.4 Generalization of the HJ equations}
The Hamilton-Jacobi equation can be generalized according to the 
following theorem allowing sometimes the simplification of 
some Hamilton-Jacobi problems.

\noindent
{\bf Theorem 7.4} Consider a system of $f$ degrees of freedom
whose dynamics is defined by $x$, $y$, where $x_{i},
y_{i}= q_{i}, p_{i}$
or $p_{i}, -q_{i}$, and whose behaviour under the action of a given force
is governed by the Hamiltonian $H(x,y,t)$. If we write the partial differential
equation $$ H\left[ x,\frac{\partial S(x,t)}{\partial x},t\right]+\frac{\partial
S(x,t)}{\partial t}=0$$
and if we can find a solution of this equation of the form
$$S=S(x,\alpha,t)$$
where $\alpha \equiv \alpha_{1},\alpha_{2},...,\alpha_{f}$ is a set of constants
and in addition the solution satisfies the condition
$$\left|\frac{\partial^{2}S(x,\alpha,t)}{\partial x_{j}\partial\alpha_{j}}
\right|\neq  0~,$$
then the laws of motion of the system can be obtained from the equations
\begin{eqnarray}
\frac{\partial S(x,\alpha,t)}{\partial x_{i}} &=&y_{i} \\
\frac{\partial S(x,\alpha,t)}{\partial \alpha_{i}} &=&\beta_{i}~,
\end{eqnarray}
where $\beta \equiv \beta_{1},\beta_{2},...,\beta_{f}$ is a set of constants.

\section*{7.5 Example of application of the HJ equations}
We shall solve the problem of the one-dimensional harmonic oscillator of mass
$m$, using the  Hamilton-Jacobi method.

\noindent
We know that the Hamiltonian of the system is 
\begin{equation}
H=\frac{p^2}{2m}+\frac{kx^2}{2}
\end{equation}
According to Theorem 7.2 the Hamilton-Jacobi equation for the system is
\begin{equation}
\frac{1}{2m}\left( \frac{\partial F}{\partial q}\right)+\frac{kq^2}{2}
+\frac{\partial F}{\partial t}=0
\end{equation}
We assume a solution of $(13)$ of the form $F=F_{1}(q)+F_{2}(t)$.
Therefore, $(13)$ converts to
\begin{equation}
\frac{1}{2m}\left( \frac{dF_1}{dq} \right) ^{2} + \frac{kq^2}{2}
= -\frac{dF_2}{dt}
\end{equation}
Making each side of the previous equation equal to $\alpha$, we find
\begin{eqnarray}
\frac{1}{2m}\left( \frac{dF_1}{dq} \right) ^{2} + \frac{kq^2}{2}&=&\alpha \\
\frac{dF_2}{dt}&=& -\alpha
\end{eqnarray}
For zero constants of integration, the solutions are
\begin{eqnarray}
F_{1}&=& \int \sqrt{2m(\alpha - \frac{kq^2}{2})}dq \\
F_{2}&=& -\alpha t
\end{eqnarray}
Thus, the generating function $F$ is
\begin{equation}
F= \int \sqrt{2m(\alpha - \frac{kq^2}{2})}dq  -\alpha t~.
\end{equation}
According to $(2)$, $q(t)$ is obtained starting from
\begin{eqnarray}
\beta &=& \frac{\partial}{\partial \alpha}\left\{ \int \sqrt{2m(\alpha
- \frac{kq^2}{2})}dq  -\alpha t \right\}\\
&=& \frac{\sqrt{2m}}{2}\int \frac{dq}{\sqrt{\alpha - \frac{kq^2}{2}}} - t
\end{eqnarray}
and effecting the integral we get
\begin{equation}
\sqrt{\frac{m}{k}}\sin ^{-1}(q\sqrt{k/2\alpha})= t+\beta~,
\end{equation}
from which $q$ is finally obtained in the form
\begin{equation}
q= \sqrt{\frac{2\alpha}{k}}\sin\sqrt{k/m}(t+\beta)~.
\end{equation}
In addition, we can give a physical interpretation to the constant
$\alpha$ according to the following argument.

\noindent
The factor $\sqrt{\frac{2\alpha}{k}}$ should correspond to the 
amplitude $A$ of the oscillator. On the other hand, the total energy $E$ 
of a one-dimensional harmonic oscillator of amplitude $A$ is given by 
$$
E = \frac{1}{2}kA^2
  = \frac{1}{2}k\left( \sqrt{\frac{2\alpha}{k}}\right) ^2
  = \alpha~.
$$

\noindent
In other words, $\alpha$ is physically the total energy $E$ of the 
one-dimensional harmonic oscillator.

\bigskip
\bigskip

\begin{center} FURTHER READING \end{center}

\bigskip

\noindent
C.C. Yan, {\it Simplified derivation of the HJ eq.}, Am. J. Phys. 52, 555
(1984)

\bigskip
\noindent
N. Anderson \& A.M. Arthurs, {\it Note on a HJ approach to the rocket pb.},
Eur. J. Phys. 18, 404 (1997)

\bigskip
\noindent
M.A. Peterson, {\it Analogy between thermodynamics and mechanics},
Am. J. Phys. 47, 488 (1979)

\bigskip
\noindent
Y. Hosotani \& R. Nakayama, {\it The HJ eqs for strings and p-branes},
hep-th/9903193 (1999)


%% file: c44en.tex


\centerline{\Large 8. ACTION-ANGLE VARIABLES}

\bigskip
\bigskip

\noindent {\bf Forward}:
The Hamilton-Jacobi equation provides a link to going from a set of 
canonical variables $q$, $p$ to a second set 
$Q$, $P$, where both are constants of motion.

\noindent
In this chapter, we briefly present another important procedure by which one
goes from an initial pair of canonical variables to a final one, where not both
variables are simultaneously constants of motion. 

\bigskip
\bigskip

{\bf CONTENS:}

\bigskip

8.1 Separable systems

\bigskip

8.2 Cyclic systems

\bigskip

8.3 Action-angle variables

\bigskip

8.4 Motion in action-angle variables

\bigskip

8.5 Importance of action-angle variables

\bigskip

8.6 Example: the harmonic oscillator

\newpage

\section*{8.1 Separable systems}
Separable systems are those ones for which the
Hamiltonian is not an explicit function of time, i.e.,
\setcounter{equation}{0}\\
$$H=H(q,p)~,$$
allowing, in addition, to find a solution of the time-independent
Hamilton-Jacobi of the form
$$W(q,\alpha)=\sum_{i}W_{i}(q_{i},\alpha)~.$$
\section*{8.2 Cyclic systems} We know that the dynamical state 
of a system is characterized by a set of generalized coordinates
$q\equiv q_{1}, q_{2},..., q_{f}$
and momenta $p\equiv p_{1}, p_{2},..., p_{f}$. A system in motion will describe
an orbit in the phase space $q$, $p$ 
At the same time, there is an orbit in each of the subspaces
$q_{i}, p_{i}$. In every plane
$q_{i}, p_{i}$, the orbit can be represented by an equation of the form
$p_{i}= p_{i}(q_{i})$ or a pair of equations  
$p_{i}= p_{i}(t)$, $q_{i}= q_{i}(t)$.  If for each value of $i$,
the orbit $p_{i}= p_{i}(q_{i})$ is a closed curve in the plane $q_{i}-p_{i}$, 
then we say that the system is cyclic. In the figure 8 we show the two 
possibilities for a system to be cyclic. In  
$8.1a$, the system is cyclic because $q_{i}$ oscillates between the limits 
defined by $q_{i}=a$ and $q_{i}=b$, whereas in figure $8.1b$, the system is
cyclic because $q_{i}$ moves from $q_{i}=a$ to $q_{i}=b$, and repeats the
same motion afterwards.
\vskip 1ex
\centerline{
\epsfxsize=190pt
\epsfbox{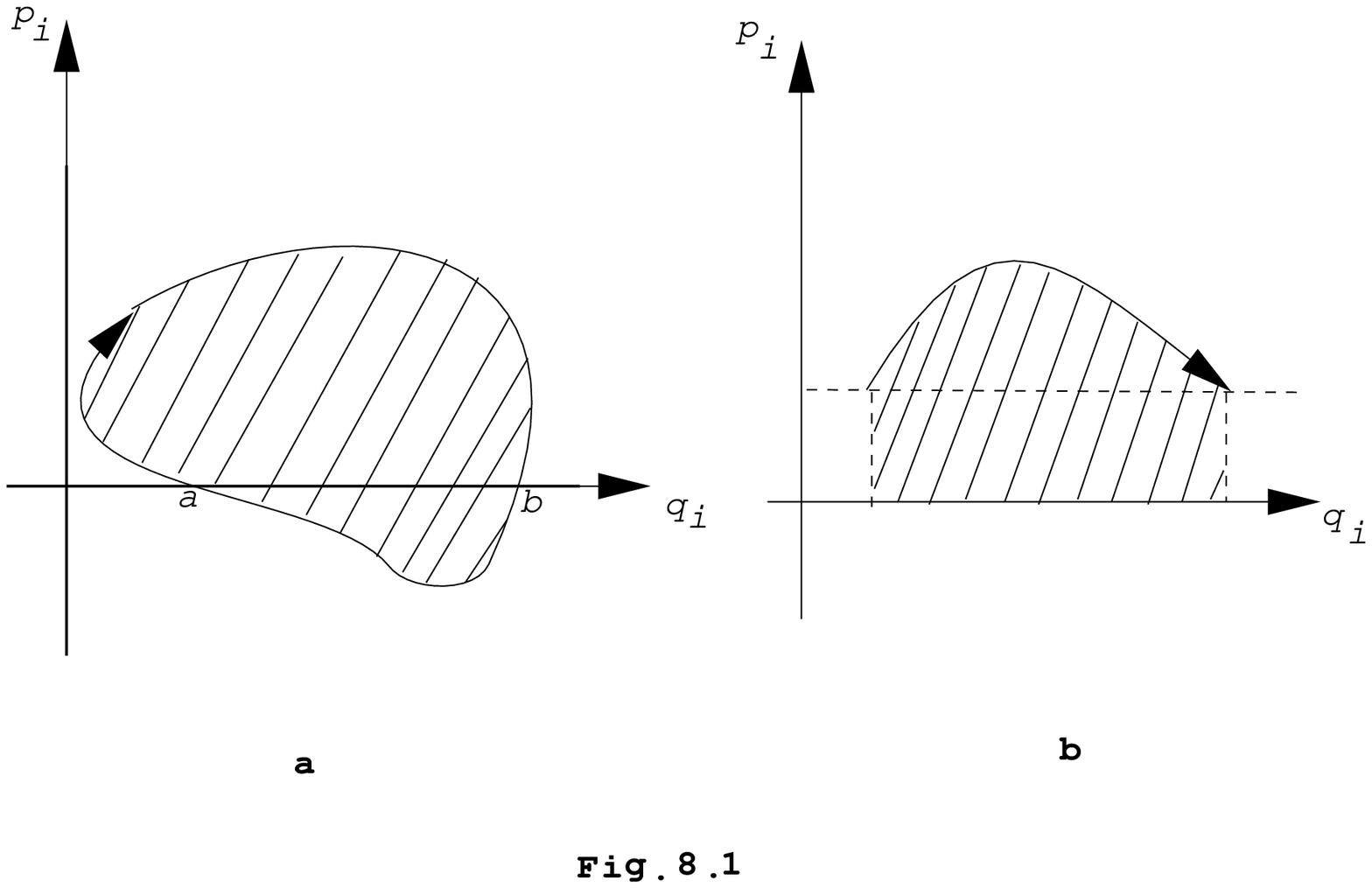}}
\vskip 2ex


\noindent
At this point, it is worthwhile to make two helpful remarks.

{\bf Remark 1}: The cyclic term has been introduced only for simplifying the 
notation in the next sections.
One should not interpret this term as if the system is cyclic in each subspace
$q_{i}, p{i}$. The system should come back to its initial state only in 
the global space $q, p$.

{\bf Remark 2}: If the cyclic system has only one degree of freedom, the time
required by the system to accomplish the cycle in $q-p$ is constant; 
therefore the motion in the space $q-p$ will be periodic in time. If the system
has more degrees of freedom, then, in general, the time required 
for a particular cycle in one of the subspaces $q_{i}, p_{i}$ will not be a
constant, but will depend on the motion of the other coordinates. 
As a result, the motion in the subspace $q_{i}, p_{i}$ will not be periodic 
in time. One should be careful with this point, since {\it not all the motions
in the subspaces $q_{i}, p_{i}$ are periodic.}
\section*{8.3 Action-angle variables}
We consider now a cyclic system of $f$ degrees of freedom, whose dynamical
state is characterized by the canonical set $q$, $p$. Let $H(q,p)$ be the 
Hamiltonian of the system and let
$$W(q,\alpha)\equiv \sum_{i} W_{i}(q_{i},\alpha)~,$$
(where $\alpha = \alpha_{1}, \alpha_{2},...,\alpha_{f}$ are constants)
be a solution of the time-independent Hamilton-Jacobi equation
$$H(q,\frac{\partial W}{\partial q}) = E~.$$
Let $J\equiv J_{1}, J_{2},..., J_{f}$ be the set of constants defined by the 
equations
\begin{equation}
J_{i}(\alpha) = \oint \frac{\partial W_{i}(q_{i},\alpha)}{\partial
q_{i}}dq_{i}~,
\end{equation}
where the integral is along a complete cycle in the variable $q_{i}$.
If we use the function
\begin{eqnarray*}
W(q,\alpha) &\equiv & W[q,\alpha (J)]\\
            &\equiv & \sum_{i}W_{i}[q_{i},\alpha (J)] \\
            &\equiv & \sum_{i}W_{i}(q_{i},J)
\end{eqnarray*}
as the generating function of a canonical transformation of $q$, $p$ to a new
set of coordinates $w$ and momenta $J$, i.e., if we define
the variables $w$ and $J$ by the transformation equations 
\begin{equation}
p_{i}  =  \frac{\partial W(q,\alpha)}{\partial q_{i}}  =  \frac{\partial 
W_{i}(q_{i},J)}{\partial q_{i}} 
\end{equation}
\begin{equation}
w_{i}   =  \frac{\partial W(q,J)}{\partial J_{i}}~,  
\end{equation}
then the new coordinates $w_{1}, w_{2},..., w_{f}$ are
called {\it angle variables}, and the new momenta $J_{1}, J_{2},..., J_{f}$ are
called {\it action variables}.

\noindent
From $(2)$ we get
\begin{equation}
p_{i}(q_{i},\alpha)= \frac{\partial W_{i}[q_{i},J(\alpha)]}{\partial q_{i}}
=\frac{\partial 
W_{i}(q_{i},\alpha)}{\partial q_{i}}~.
\end{equation}
Substituting $(4)$ in $(1)$ one gets
\begin{equation}
J_{i}(\alpha)=\oint p_{i}(q_{i},\alpha)dq_{i}~.
\end{equation}
The equation $p_{i}=p_{i}(q_{i},\alpha)$ gives the projected orbit
$p=p(q)$ on the subspace $p_{i}, q_{i}$. The integral in the right hand side
of the equation $(5)$ is thus the area bordered by the closed orbit, or beneath
the orbit, as shown in figure 8.1. Thus, the function $J_{i}(\alpha)$ has a
geometric interpretation as the area covered in the subspace  $q_{i}, p_{i}$ 
during a complete cycle in the subspace.
This area depends on the constants $\alpha$ or equivalently on the initial
conditions and can be arbitrary \footnote{Historically,
the first intents to pass from the classical mechanics to quantum mechanics
was related to the assumption that the value of $J_{i}$ could be only a multiple
of $h/2\pi$, where $h$ is Planck's constant.}.
\section*{8.4 Motion in terms of action-angle variables}
We give the following theorem-like statement for the motion of a system
in terms of action-angle variables.

\noindent
{\bf Theorem 8.4}

\noindent
Consider a separable cyclic system of $f$ degrees of freedom whose motion is
described by the variables $q, p \equiv q_{1}, q_{2},...,q_{f},
p_{1}, p_{2},..., p_{f}$, together with the Hamiltonian $H(q,p)$. If we 
transform the motion to the action-angle variables $J, w$, then the 
Hamiltonian $H$ is a function of $J$ alone, i.e.,
$$H=H(J)$$
and the equations of motion will be 
\begin{eqnarray*}
J_{i} &=& \gamma_{i} \\
w_{i} &=& \nu_{i}t+\phi_{i}~,
\end{eqnarray*}
where $\gamma_{i}$ y $\phi_{i}$ are constants determined by the initial 
conditions, whereas the $\nu_{i}$ are constants known as 
{\it the frequencies of the system} being defined as follows
\begin{eqnarray*}
\nu_{i} = \left[ \frac{\partial H(J)}{\partial J_{i}}\right]_{J=\gamma_{i}}~.
\end{eqnarray*}
\begin{center}
\section*{8.5 Importance of the action-angle variables}
\end{center}
The importance of the action-angle variables resides in providing a 
powerful technique for directly getting the frequencies of periodic motions
without solving for the equations of motion of the system. 

\noindent
This important conclusion can be derived through the following argument.
Consider the change of $w$ when $q$ describes a complete cycle
\begin{eqnarray*}
\Delta w= \oint \frac{\partial w}{\partial q}dq
\end{eqnarray*}
On the other hand, we know that
\begin{eqnarray*}
w=\frac{\partial W}{\partial J}~,
\end{eqnarray*}
and therefore
\begin{eqnarray*}
\Delta w &=& \oint \frac{\partial^{2}W}{\partial q\partial J}dq\\
         &=& \frac{d}{dJ}\oint \frac{\partial W}{\partial q}dq \\
         &=& \frac{d}{dJ}\oint pdq\\
         &=& 1~.
\end{eqnarray*}
This result shows that $w$ changes by a unity when $q$ varies within a complete
period.

\noindent
From the relationship
\begin{eqnarray*}
w = \nu t + \phi~,
\end{eqnarray*}
we infer that in a period $\tau$
\begin{eqnarray*}
\Delta w &=& 1 \\
         &=& \nu \tau~.
\end{eqnarray*}
This means that we can identify the constant $\nu$ with the inverse of the 
period
\begin{eqnarray*}
\nu = \frac{1}{\tau}~.
\end{eqnarray*}
\section*{8.6 Example: The simple harmonic oscillator}
Using the action-angle formalism prove that the frequency $\nu$ 
of the simple one-dimensional harmonic oscillator is given by 
$\nu = \sqrt{k/m}/2\pi$.

\noindent
Since $H$ is a constant of motion, the orbit in the space $q-p$ is given
\begin{eqnarray*}
\frac{p^2}{2m} + \frac{kq^2}{2} = E~,
\end{eqnarray*}
where $E$ is the energy. This is the equation of an ellipse of semiaxes
$\sqrt{2mE}$ and $\sqrt{2E/k}$. The area enclosed by the ellipse
is equal to the action $J$. Therefore,
\begin{displaymath}
J=\pi \sqrt{2mE}\sqrt{\frac{2E}{k}}= 2\pi \sqrt{\frac{m}{k}}E~.
\end{displaymath}
It follows that
\begin{displaymath}
H(J)=E=\frac{\sqrt{k/m}}{2\pi}J~.
\end{displaymath}
Thus, the frequency will be
\begin{displaymath}
\nu= \frac{\partial H(J)}{\partial J}= \frac{\sqrt{k/m}}{2\pi}~.
\end{displaymath}



%% file: peren.tex
\oddsidemargin  0in
\evensidemargin 0in
\marginparwidth 1.2in
\marginparsep   0.1in
\marginparpush  5pt
\topmargin  0in
\headheight   0pt
\headsep    0in
\topskip=10pt
\footheight  12pt
\footskip   0.6in
\textheight   8.5in
\textwidth  6.5in
\columnsep 10pt
\columnseprule 0pt
\renewcommand{\baselinestretch}{1.2}
\def\pa{\\$\;\;\;$\\}

\def\t{\'}
\def\ti{\'{\i}}
\def\n{\~n}



\newpage
$\;$\\
\vspace{0.3in}

\begin{center}
{\Large  9. CANONICAL PERTURBATION THEORY}
\end{center}

\vspace{0.4in}


\noindent
{\bf Forward}:
The great majority of problems that we have to solve in Physics are not 
exactly solvable.
Because of this and taking into account that we live in the epoch of computers
the last decades have seen a lot of progress in developing techniques leading
to approximate solutions.
The perturbation method is used for not exactly solvable Hamiltonian problems
when the Hamiltonian differs slightly from an exactly solvable one.
The difference between the two Hamiltonians is known as the perturbation
Hamiltonian.
All perturbation methods are based on 
the smallness of the latter with respect to both Hamiltonians.

$\;\;$\\

\bigskip
\bigskip

{\bf CONTENTS:}

\bigskip

9.1 Time-dependent perturbation theory (with two examples)

\bigskip

9.2 Time-independent perturbation theory (with an example)

\newpage
$\;\;$\\
{\Large \bf 9.1 Time-dependent perturbation theory}

$\;\;$\\ 
The most appropriate formulation of classical mechanics for the development of
perturbation methods is the Hamilton-Jacobi approach.
Thus, we take $H_0 (p,
q,t)$ as the Hamiltonian corresponding to the solvable (unperturbed) problem and
consider the Hamilton principal function $S(q, \alpha _0 , t)$, 
as the generating function of a canonical transformation of $(p,q)$ to the new 
canonical pair 
$(\alpha _0 , \beta _0)$ for which the new
Hamiltonian (or Kamiltonian) $K_0$ of the unperturbed system is zero
\setcounter{equation} {0}\\
\begin{eqnarray}
\frac{\partial S}{\partial t} + H_0(\frac{\partial S}{\partial q},q,t)
= K_0 = 0 \; .
\end{eqnarray}

$\;$\\
This is the Hamilton-Jacobi equation, where we used $p
= \partial S / \partial q$. The whole set of new canonical coordinates
$(\alpha _0 , \beta _0 )$ are constant in the unperturbed case because 
$K_0 =0$ and:

\begin{eqnarray}
\dot{\alpha} _0 &=& - \frac{\partial K_0}{\partial \beta _0 }, \nonumber \\ 
\dot{\beta} _0 &=& \;\; \frac{\partial K_0}{\partial \alpha _0 }. 
\end{eqnarray}  

$\;$\\
We now consider the Hamiltonian of the perturbed system written as follows:

\begin{eqnarray}
H(q,p,t) = H_0 (q,p,t) + \Delta H(q,p,t) ;  \;\;\;\;\;\;\;\;  (\Delta H \ll
H_0).
\end{eqnarray}

$\;$\\
Although $(\alpha _0 , \beta _0 )$ are still canonical coordinates (since the
transformation generated by $S$ is
independent of the particular form of the Hamiltonian),
they will not be constant and the  
Kamiltonian $K$ of the perturbed system will not be zero. In order not to
forget that in the perturbed system the transformed coordinates are not
constant, we denote them by $\alpha$ and $\beta$ instead of $\alpha _0$ and 
$\beta _0$, which are the corresponding constants in the unperturbed system.
Thus, for the perturbed system we have:

\begin{eqnarray}
K(\alpha, \beta, t) = H + \frac{\partial S}{\partial t} = (H_0 + \frac{\partial S}{\partial t}) + 
\Delta H = \Delta H(\alpha, \beta, t).
\end{eqnarray}

$\;$\\
The equations of motion for the transformed variables in the perturbed system
will be:

\begin{eqnarray}
\dot{\alpha_i} &=& - \frac{\partial \Delta H(\alpha,\beta,t)}{\partial \beta_i}
\nonumber \\ 
\dot{\beta_i} &=& \;\; \frac{\partial \Delta H(\alpha, \beta, t)}{\partial
\alpha_i},
\end{eqnarray}
 
$\;$\\
where $i = 1, 2, ... , n$ and $n$ is the number of degrees of freedom of the 
system. These are rigorous equations. If the system of  
$2n$ equations could be solved for  
$\alpha_i$ and $\beta_i$ as functions of time, the transformation equation
$(p,q) \rightarrow (\alpha,\beta)$ would give $p _i$ and $q _i$ as functions
of time and the problem would be solved.
However, the exact solution of the equations $(5)$ is not easier to get w.r.t.
the original equations.
From $(5)$ we see that when $\alpha$ and $\beta$ are not  
constant their variation in time is slow if we assume that 
$\Delta H$ changes infinitesimally  w.r.t. $\alpha$ y $\beta$. A first 
approximation for the temporal variations of
$(\alpha, \beta)$ can be obtained by substituting $\alpha$ y $\beta$
in the second terms of $(5)$ by their constant unperturbed values

\begin{eqnarray}
\dot{\alpha} _{i1} &=& - \left. \frac{\partial \Delta H(\alpha,
\beta, t)}{\partial \beta _i} \right 
\vert _0 \nonumber \\
\dot{\beta} _{i1} &=& \;\;\; \left. \frac{\partial
\Delta H(\alpha, \beta, t)}{\partial \alpha _i} \right 
\vert _0,
\end{eqnarray}

$\;$\\
where $\alpha_{i1}$ and $\beta_{i1}$ are the first-order solutions, i.e., in the
first power of the perturbation to 
 $\alpha_i$ and $\beta_i$, and the vertical bars with zero subindices
show that after the derivation one should substitute  
$\alpha$ and $\beta$ by their constant unperturbed values. 
Once this is done, the equations $(6)$ can be integrated
leading to $\alpha_i$ and $\beta_i$ as functions of time (in the first order).
Next, using the equations of transformation one can get 
$p$ and $q$ as functions of time in the same first-order approximation.
The second-order approximation can be obtained now by substituting
in the second terms of $(6)$ the first-order approximation
of $\alpha$ and $\beta$ w.r.t. time. In general,
the perturbation solution of order $N$ is obtained by integrating the following
equations

\begin{eqnarray}
\dot{\alpha} _{iN} &=& - \left. \frac{\partial \Delta
H(\alpha, \beta, t)}{\partial \beta _i} \right 
\vert _{N-1} \nonumber \\
\dot{\beta} _{iN} &=& \;\;\; \left. \frac{\partial \Delta
H(\alpha, \beta, t)}{\partial \alpha _i} \right 
\vert _{N-1}~.
\end{eqnarray}

$\;$\\
\underline{{\bf Example 1}}

$\;$\\
Let us consider the simple case of a free particle that next is subject to a
harmonic perturbation. Although this example is trivial can be used to
illustrate the aforementioned procedure.
The unperturbed Hamiltonian is

\begin{eqnarray}
H_0 = \frac{p^2}{2m}.
\end{eqnarray}

$\;$\\
Since $H_0 \neq H_0 (x)$, $x$ is a cyclic variable and $p=
\alpha _0$ is a constant of motion in the unperturbed system. 
Recalling that 
$p = \partial S / \partial x$, we substitute in $(1)$:

\begin{eqnarray}
\frac{1}{2m} \left( \frac{\partial S}{\partial x} \right)^2
+ \frac{\partial S}{\partial t} = 0.
\end{eqnarray}

$\;$\\
Since the system is conservative, it is convenient to consider the principal
function of the form 

\begin{eqnarray}
S = {\cal \bf S} (x) + F(t). 
\end{eqnarray}

$\;$\\
This type of separation of variables is quite useful when the 
Hamiltonian does not depend explicitly on time. Then, one writes
$F(t) = -Et$, where $E$ is the total energy of the system
\footnote{See, M.R. Spiegel,
{\it Theoretical Mechanics}, pp. 315, 316.}. Putting $(10)$ in $(9)$ 
we obtain

\begin{equation} \label{1}
\frac{1}{2m} \left( \frac{d {\cal \bf S}}{dx} \right)^2 = E,\qquad \qquad
{\cal \bf S} = \sqrt{2mE} x = \alpha _0 x.
\end{equation}

$\;$\\
Substituting $(11)$ in $(10)$, together with the fact that in this case
the Hamiltonian is equal to the energy, 
we can write the principal function of Hamilton as follows

\begin{eqnarray}
S = \alpha _0 x - \frac{\alpha _0 ^2 t}{2m}.
\end{eqnarray}

$\;$\\
If the transformed momentum is $\alpha_0$, the transformed coordinate (which is
also constant in the unperturbed system) is

\begin{eqnarray}
\beta _0 = \frac{\partial S}{\partial \alpha _0} = x - \frac{\alpha _0 t}{m}~.
\nonumber
\end{eqnarray}

$\;$\\

Therefore, the transformation generated by $S$ is given by the 
following eqs

\begin{eqnarray}
p &=& \alpha _0, \nonumber \\
x &=& \frac{\alpha _0 t}{m} + \beta _0.
\end{eqnarray}

$\;$\\
They represent the solution for the motion of free particle.
What we have given up to now is only the procedure to obtain the equations of
motion using the Hamilton-Jacobi formulation. Now, we introduce the perturbation

\begin{eqnarray}
\Delta H = \frac{kx^2}{2} = \frac{m \omega ^2 x^2}{2},
\end{eqnarray}

$\;$\\
or, in terms of the transformed coordinates, using $(13)$

\begin{eqnarray}
\Delta H = \frac{m \omega^2}{2} \left( \frac{\alpha t}{m} + \beta \right) ^2.
\end{eqnarray}

$\;$\\
Notice that we have renounced at the subindices $0$ for the transformed
coordinates since we already study the perturbed system.

$\;$\\
Substituting $(15)$ in $(5)$ we get

\begin{eqnarray}
\dot{\alpha} &=& - m \omega^2 \left( \frac{\alpha t}{m} + \beta \right),
\nonumber \\
\dot{\beta} &=& \;\; \omega^2 t \left( \frac{\alpha t}{m} + \beta \right).
\end{eqnarray}

$\;$\\
As one may expect, these equations have an exact solution of the harmonic type.
To be sure of that we perform the time derivative of the first equation
allowing us to conclude that $\alpha$ has a simple harmonic variation. The same
holds for $x$ as a consequence of the transformations $(13)$, which are invariant
in form in the perturbed system 
(up to the subindices of the transformed coordinates).  
However, we are interested to illustrate 
the perturbation method, so that we consider that $k$ (the elastic constant) 
is a small parameter. We seek approximate solutions in different perturbative
orders, without missing the fact that the transformed variables
$(\alpha, \beta)$ in the perturbed system are not constants of motion. 
In other words, even though $(\alpha, \beta)$ contain information on the 
unperturbed system, the effect of the perturbation is to make these parameters
varying in time.

$\;$\\
The first-order perturbation is obtained in general as given by $(6)$. 
Thus, we have to substitute  $\alpha$ and  
$\beta$ by their unperturbed values in the second terms of $(16)$.  
To simplify, we take $x(t=0)=0$ and therefore $\beta_0 = 0$, leading to

\begin{eqnarray}
\dot{\alpha}_1 &=& - \omega^2 \alpha_0 t, \nonumber \\ 
\dot{\beta}_1 &=& \alpha_0 \frac{\omega^2 t^2}{m}, 
\end{eqnarray}

$\;$\\
which integrated reads

\begin{eqnarray}
\alpha_1 &=& \alpha_0 - \frac{\omega^2 \alpha_0 t^2}{2}, \nonumber \\ 
\beta_1 &=& \frac{\alpha_0 \omega^2 t^3}{3m}.
\end{eqnarray}

$\;$\\
The first-order solutions for $x$ and $p$ are obtained by putting 
$\alpha_1$ and $\beta_1$ in the transformation eqs. $(13)$, where from

\begin{eqnarray}
x &=& \frac{\alpha_0}{m \omega} \left ( \omega t - \frac{\omega^3 t^3}{6}
\right), \nonumber \\
p &=& \alpha_0 \left( 1 - \frac{\omega^2 t^2}{2} \right).
\end{eqnarray}

$\;$\\
To obtain the approximate solution in the second perturbative order
we have to find 
$\dot{\alpha_2}$ and $\dot{\beta_2}$, as was indicated in $(7)$,
by substituting in the second terms of  
$(16)$, $\alpha$ and $\beta$ by $\alpha_1$ and $\beta_1$ 
as given in $(18)$. 
Integrating $\dot{\alpha_2}$ and $\dot{\beta_2}$ and using again the transformation
eqs. $(13)$, we get the second-order solutions
for $x$ and $p$:

\begin{eqnarray}
x &=& \frac{\alpha_0}{m \omega} \left( \omega t - \frac{\omega^3 t^3}{3!}
+ \frac{\omega^5 t^5}{5!} 
\right), \nonumber \\
p &=& \alpha_0 \left( 1 - \frac{\omega^2 t^2}{2!}
+ \frac{\omega^4 t^4}{4!} \right).
\end{eqnarray}

$\;$\\
In the limit in which the perturbation order $N$ tends to infinity,
we obtain the expected solutions compatible with the initial conditions

\begin{eqnarray}
x  \rightarrow  \frac{\alpha_0}{m \omega} \sin{\omega t}, & \;\;\;\; p
\rightarrow \alpha_0 \cos{\omega t}.
\end{eqnarray}

$\;$\\
The transformed variables $(\alpha, \beta)$ contain information about the
unperturbed orbit parameters. For example, if we consider as unperturbed system
that corresponding to the Kepler problem, a convenient coordinate pair
$(\alpha, \beta)$ could be the $(J, \delta)$ variables, which are the action
and the phase angle of the angle $w$, respectively (remember that $w = \nu
t + \delta$, where $\nu$ is the frequency). These variables are related to 
the set of orbital parameters such as semimajor axis, eccentricity, 
inclination, and so on. 

\noindent
The effect of the perturbation is to produce a time variation of all
these parameters. If the perturbation is small, the variation of the parameters 
during a period of the unperturbed motion will also be small. In such a case,
for small time intervals, the system moves along the so-called 
{\em osculating} orbit, having the same functional form as the orbit of the 
unperturbed system; the difference is that the parameters of the osculating 
curve vary in time.

$\;$\\
The osculating parameters can vary in two ways

\begin{itemize}

\item Periodic variation: if the parameter comes back to its initial value
after a time interval that in first approximation is usually the unperturbed
period. These periodic effects of the perturbation do not alter the mean values
of the parameters. Therefore, the trajectory is quite similar to the unperturbed
one. These effects can be eliminated by taking the average of the perturbations
in a period of the unperturbed motion.\\

\item Secular variation: At the end of every succesive orbital period there is 
a net increment of the value of the parameter. Therefore, at the end of many
periods, the orbital parameters can be very different of their unperturbed 
values. The instantaneous value of the variation of a parameter, for 
example the frequency, is seldom of interest, 
because its variation is very small in almost all cases in which the perturbation
formalism works.
(This variation is so small that it is practically impossible
to detect it in a single orbital period. This is why the secular variation is
measured after at least several periods.)\\

\end{itemize} 

\bigskip

\underline{{\bf Example 2}}

\bigskip

\noindent
From the theory of the Kepler two-body problem it is known that
if we add a potential $1/r^2$, the orbit of the motion of negative energy
is a rotating ellipse whose periapsis is precessing. 
In this example, we find the precession velocity for a more general perturbation

\begin{eqnarray}
V = - \frac{k}{r} - \frac{h}{r^n} \; ,
\end{eqnarray}

$\;$\\
where $n\geq 2$ is an entire number, and $h$ is such that the second term of 
the potential is a small perturbation of the first one.
The  perturbative Hamiltonian is

\begin{eqnarray}
\Delta H = - \frac{h}{r^n} \; .
\end{eqnarray}

$\;$\\
In the unperturbed problem, the angular position of the periapsis in the orbit 
plane is given by the constant $\omega = 2 \pi w_2$. In the perturbed case, we
have

\begin{eqnarray}
\dot{\omega} = 2 \pi \frac{\partial \Delta H}{\partial J_2}
= \frac{\partial \Delta H}{\partial l} \; , 
\end{eqnarray}

$\;$\\
where we have used $J_2 = 2 \pi l$. Moreover, $J_2$ and $w_2$ are two of the 
five integrals of motion that can be obtained when one uses the action-angle
variables to solve the Kepler problem.

$\;$\\
We need to know the mean of $\dot{\omega}$ in a period $\tau$ of the unperturbed 
orbit

\begin{eqnarray}
\langle \dot{\omega} \rangle \equiv \frac{1}{\tau} \int _0 ^{\tau}
\frac{\partial \Delta H}{\partial l} dt =
\frac{\partial }{\partial l} \left( \frac{1}{\tau} \int _0 ^{\tau}
\Delta H \; dt \right)
= \frac{\partial \langle \Delta H \rangle}{\partial l} \; .
\end{eqnarray}

$\;$\\
The temporal mean of the unperturbed Hamiltonian is

\begin{eqnarray}
\langle \Delta H \rangle = - h \langle \frac{1}{r^n} \rangle =
- \frac{h}{\tau} \int _0 ^{\tau} \frac{dt}{r^n} \;.
\end{eqnarray}

$\;$\\
On the other hand, since $l = mr^2 (d\theta / dt)$, we can get $dt$ 
and plunge it in $(27)$. This leads to

\begin{eqnarray}
\langle \Delta H \rangle &=& - \frac{mh}{l\tau} \int _0 ^{2 \pi}
\frac{d\theta}{r ^{n-2}} \nonumber \\
&=& - \frac{mh}{l\tau} \left( \frac{mk}{l^2} \right) ^{n-2} \int _0 ^{2 \pi}
[1 + e \cos{(\theta - \eta)} ] ^{n-2} d \theta \; .
\end{eqnarray}

$\;$\\
where $\eta$ is a constant phase, $e$ is the eccentricity, and where we expressed
$r$ as a function of $\theta$ using the general equation of the orbit with the 
origin in one of the focal points of the corresponding conic

\begin{eqnarray}
\frac{1}{r} = \frac{mk}{l^2} \Big[1 + e \cos{(\theta - \eta)} \Big]
\end{eqnarray}~.

$\;$\\
For $n=2$:

\begin{eqnarray}
\langle \Delta H \rangle &=& - \frac{2 \pi mh}{l \tau} \; , \nonumber \\
\langle \dot{\omega} \rangle &=& \frac{2 \pi mh}{l^2 \tau} \; .
\end{eqnarray}

$\;$\\
For $n=3$:

\begin{eqnarray}
\langle \Delta H \rangle &=& - \frac{2 \pi m^2 hk}{l^3 \tau} \; , \nonumber \\
\langle \dot{\omega} \rangle &=& \frac{6 \pi m^2 hk}{l^4 \tau} \; .
\end{eqnarray}
 
$\;$\\
The latter case, $n=3$, is of particular importance because the theory of
General Relativity predicts a correction of the Newtonian motion precisely of
$r^{-3}$ order. This prediction is related to the famous problem of the 
precession of the orbit of Mercury. Substituting the appropriate values of the 
period, mass, great semiaxis, that is included in $h$, and so on,
$(30)$ predicts a mean precession velocity of

$$ \langle \dot{\omega} \rangle = 42.98 \;\; {\rm arcsec./century} \;. $$

$\;$\\
The mean value is by far larger than the aforementioned one (by a factor larger
than one hundred). But before making any comparison one should eliminate from
the mean value the contributions due to the following factors:
{\it a)} the effect known as the precession of the equinoxes
(the motion of the reference point of longitudes w.r.t. the Milky Galaxy), 
{\it b)} the perturbations of the Mercury orbit due to the interaction with the 
other planets. Once these are eliminated, of which the first is the most 
significant, one may hopefully obtain the contribution of the relativistic 
effect. In 1973, this contribution has been estimated as $41.4 \pm 0.9$ 
${\rm arcsec./century}$. This is consistent with the prediction given by $(30)$.

$\;$\\
{\Large \bf 9.2 Time-independent perturbation theory}

$\;$\\
While in the time-dependent perturbation theory one seeks the time dependence of
the parameters of the unperturbed system initially considered as constant, the 
aim of the 
time-independent approach is to find the constant quantities of the perturbed
system. This theory can be applied only to conservative periodic systems  
(both in the perturbed and unperturbed state).
For example, it can be applied to planetary motion when one 
introduces any type of conservative perturbation to the Kepler problem, in 
which case it is known as the von Zeipel or Poincar\'e method.

$\;$\\
Here, we consider here the case of a periodic system of one degree of freedom 
and time-independent Hamiltonian of the form 

\begin{eqnarray}
H = H(p,q, \lambda), 
\end{eqnarray}

$\;$\\
where $\lambda$ is a small constant specifying the 
strength of the perturbation. We assume that

\begin{eqnarray}
H_0 (p,q) = H(p,q,0)
\end{eqnarray}

$\;$\\
corresponds to a system that has an exact (closed-form) unperturbed solution
in the action-angle variables $(J_0 , w_0 )$, i.e.,  

\begin{eqnarray}
H_0 (p,q) &=& K_0 (J_0) \nonumber \\
\nu_0 &=& \dot{w}_0 = \frac{\partial K_0}{\partial J_0} \; ;  \;\;\;\;\;
(w_0 = \nu_0 t + \delta_0 ). 
\end{eqnarray}

$\;$\\
Since the canonical transformation from $(p,q)$ to $(J_0 , w_0 )$ is
independent of the particular form of the  
Hamiltonian, the perturbed Hamiltonian 
$H(p,q,\lambda)$ can be written  
as $H(J_0, w_0, \lambda)$.  Due to the fact that the perturbed 
Hamiltonian depends on $w_0$, $J_0$, it is not constant any more. 
On the other hand, in principle, one can get new action-angle variables
$(J,w)$ that may be more appropriate for the perturbed system, such as

\begin{eqnarray}
H(p,q,\lambda) &=& E(J,\lambda) \nonumber \\
\nu &=& \dot{w} = \frac{\partial E}{\partial J} \\
\dot{J} &=& - \frac{\partial E}{\partial w} = 0 \; ;  \;\;\; (J= constant).
\nonumber 
\end{eqnarray}

$\;$\\
Since the transformation connecting $(p,q)$ to $(J_0, w_0)$ is known, 
we have to find the canonical transformation $S$ connecting $(J_0, w_0)$ to
$(J,w)$. If we assume that $\lambda$ is small the transformation we look for 
should not differ much from the identity transformation. 
Thus, we write the following expansion

\begin{eqnarray}
S = S(w_0, J, \lambda) = S_0 (w_0, J) + \lambda S_1 (w_0, J)
+ \lambda ^2 S_2 (w_0, J) + ...~.
\end{eqnarray}

$\;$\\
For $\lambda =0$ we ask $S$ to provide an identity transformation that leads to 

\begin{eqnarray}
S_0 = w_0 J~.
\end{eqnarray}

$\;$\\
The canonical transformations generated by $S$ read

\begin{eqnarray}
w &=& \frac{\partial S}{\partial J} = w_0 + \lambda \frac{\partial
S_1}{\partial J} (w_0,J) + \lambda ^2 \frac{\partial S_2}
{\partial J} (w_0,J) + ...~. \nonumber \\
J_0 &=& \frac{\partial S}{\partial w_0} = J + \lambda \frac{\partial
S_1}{\partial w_0} (w_0,J) + \lambda ^2 \frac{\partial S_2}
{\partial w_0} (w_0,J) + ...~. 
\end{eqnarray}

$\;$\\
Due to the fact that $w_0$ is an angle variable of the unperturbed system
we know that
$\Delta w_0 = 1$ over a cycle.  On the other hand, we know that the canonical
transformations have the property to conserve the phase space volume. Therefore,
we can write: 

\begin{eqnarray}
J = \oint pdq = \oint J_0 dw_0~.
\end{eqnarray}

$\;$\\
Integrating the second equation of $(37)$ along an orbit of the perturbed system,
we get

\begin{eqnarray}
\oint J_0 dw_0 = \oint J dw_0 + \sum_{n=1} \lambda^n \oint \frac{\partial S_n}
{\partial w_0} dw_0 ,
\end{eqnarray}

$\;$\\
and substituting $(39)$ in $(38)$ leads to

\begin{eqnarray}
J = J \Delta w_0 + \sum_{n=1} \lambda^n \oint \frac{\partial S_n}
{\partial w_0} dw_0.
\end{eqnarray}

$\;$\\
Since $\Delta w_0 = 1$, one gets 

\begin{eqnarray}
\sum_{n=1} \lambda ^n \oint \frac{\partial S_n}{\partial w_0} dw_0 = 0, 
\end{eqnarray}   

$\;$\\
or

\begin{eqnarray}
\oint \frac{\partial S_n}{\partial w_0} dw_0 = 0.
\end{eqnarray}

$\;$\\
Moreover, the Hamiltonian can be expanded in $\lambda$ as a function of 
$w_0$ and $J_0$:

\begin{eqnarray}
H(w_0, J_0, \lambda) = K_0(J_0) + \lambda K_1(w_0,J_0)
+ \lambda^2 K_2 (w_0,J_0) + ... \; ,
\end{eqnarray}

$\;$\\
where the $K_i$ are known because $H$ is a known function of
$w_0$ and $J_0$ for a given $\lambda$. On the other hand, one can write

\begin{eqnarray}
H(p,q, \lambda) &=& H(w_0, J_0, \lambda) \nonumber \\
&=& E(J,\lambda) \; ,
\end{eqnarray}

$\;$\\
which is the expression for the energy in the new action-angle coordinates
(where $J$ will be constant and $w$ will be a linear function of time).  

$\;$\\
$E$ can also be expanded in powers of $\lambda$:

\begin{eqnarray}
E(J, \lambda) = E_0 (J) + \lambda E_1 (J) + \lambda ^2 E_2 (J) + ... \;.
\end{eqnarray}

$\;$\\
Taking into account $(44)$ we can obtain equalities for the 
coefficients of the same powers of $\lambda$ in 
$(43)$ and $(45)$. However, these expressions 
for the energy depend on two different sets of variables.
To solve this issue we express $H_0$ in terms of $J$ by writing a Taylor 
expansion of $H (w_0,J_0,\lambda)$ w.r.t. $J_0$
in the infinitesimal neighborhood of $J$:

\begin{eqnarray}
H(w_0,J_0,\lambda) = H(w_0, J, \lambda) + (J_0 - J) \frac{\partial H}
{\partial J}
+ \frac{(J_0  
- J)^2}{2} \frac{\partial ^2 H}{\partial J^2} + ... \; ,
\end{eqnarray}

$\;$\\
The derivatives of this Taylor expansion, which in fact are the 
derivatives w.r.t. $J_0$ calculated for
$J_0 = J$, can also be written as 
derivatives w.r.t. $J$, once we substitute $J_0$ by $J$ in $H_0 (J_0)$.  
In the previous equation, all the terms containing $J_0$ should be rewritten
in terms of $J$ by using the transformation defined by $(37)$ connecting the 
coordinates 
$(J_0, w_0)$ and $(J,w)$. Thus, from the second equation in $(37)$
we obtain 
$(J_0 - J)$, which introduced in $(46)$ gives:

\begin{eqnarray}
H(w_0,J_0,\lambda) = H(w_0, J, \lambda) + \frac{\partial H}
{\partial J} \left( \lambda 
\frac{\partial S_1}{\partial w_0} + \lambda ^2
\frac{\partial S_2}{\partial w_0} + ... \right)
+ \frac{1}{2} \frac{\partial ^2 H}{\partial J^2} \lambda ^2
\left( \frac{\partial S_1}{\partial w_0} \right) ^2 
+ O (\lambda ^3)  \; .
\end{eqnarray}

$\;$\\
Next, we can make use of $(43)$ to write $H(w_0,J,\lambda) =
H(w_0, J_0, \lambda) 
\mid _{J_0 = J}$, that we put in $(47)$ in order to get

\begin{eqnarray}
H(w_0,J_0,\lambda) &=& K_0(J) + \lambda K_1(w_0,J) + \lambda ^2 K_2 (w_0,J)
+ ... \nonumber \\
&+&  \lambda \frac{\partial S_1}{\partial w_0}
\left( \frac{\partial K_0(J)}{\partial J}
+ \lambda \frac{\partial K_1(w_0,J)}{\partial J} + ... \right) \nonumber \\
&+&  \lambda ^2 \left[ \frac{\partial K_0(J)}{\partial J} \frac{\partial S_2}
{\partial w_0} + \frac{1}{2} \frac{\partial ^2 K_0}{\partial J^2} \left(
\frac{\partial S_1}{\partial w_0} \right) ^2 + ... \right] \nonumber \\
&\equiv& E(J,\lambda) \\
&=& E_0 (J) + \lambda E_1 (J) + \lambda ^2 E_2(J) + ... \; . \nonumber
\end{eqnarray}

$\;$\\
Now, we can solve the problem in terms of the coefficients $E_i(J)$ allowing us
to calculate the frequency of the perturbed motion in various 
perturbation orders. Since the expansion of the terms of $E_i$ does not imply
a dependence on $w_0$, then the occurrence of $w_0$ in $(48)$ is artificial.
The $K_i (w_0,J)$ of $(48)$ are known functions, whereas the $S_i (w_0,J)$ and
$E_i(J)$ are the unknown quantities.

$\;$\\
Making equal the corresponding powers of $\lambda$, we get

\begin{eqnarray}
E_0 (J) &=& K_0 (J) \nonumber \\
E_1 (J) &=& K_1 (w_0,J) + \frac{\partial S_1}{\partial w_0} \frac{\partial
K_0 (J)}{\partial J} \nonumber \\
E_2 (J) &=& K_2 (w_0,J) + \frac{\partial S_1}{\partial w_0} \frac{\partial
K_1 (w_0,J)}{\partial J} \nonumber \\
&+& \frac{1}{2} \left( \frac{\partial S_1}{\partial w_0} \right) ^2
\frac{\partial ^2 K_0(J)}
{\partial J^2} + \frac{\partial S_2}{\partial w_0} \frac{\partial K_0 (J)}
{\partial J} \;~. 
\end{eqnarray}

\vspace{0.5in}

$\;$\\
We can see that to obtain $E_1$ we need to know not only  $K_1$ 
but also $S_1$. Moreover, we should keep in mind that the $E_i$ are constant, 
being functions of $J$ only, which is a constant of motion.
We also have to notice that $\partial K_0 / \partial J$ does not depend
on $w_0$ (since $K_0 = K_0 (J)= K_0 (J_0) \mid _{J_0=J}$).  
Averaging over $w_0$ on both sides of the second equation in $(49)$,
one gets

\begin{eqnarray}
E_1 &=& \langle E_1 \rangle \nonumber \\
&=& \langle K_1 \rangle + \frac{\partial K_0}{\partial J} \langle
\frac{\partial S_1}{\partial 
w_0} \rangle \; . 
\end{eqnarray}

$\;$\\
But we have already seen that $\langle \partial S_i / \partial w_0 \rangle
= \oint (\partial S_i / 
\partial w_0 ) dw_0 = 0$. Therefore,

\begin{eqnarray}
E_1 = \langle E_1 \rangle = \langle K_1 \rangle~. 
\end{eqnarray}

$\;$\\
Introducing $(51)$ in the left hand side of the second equation of $(49)$
we get $(\partial S_1 / \partial w_0)$ as follows

\begin{eqnarray}
\frac{\partial S_1}{\partial w_0} = \frac{\langle K_1 \rangle - K_1}{\nu _0
(J)} \; ,
\end{eqnarray}

$\;$\\
where we used $\nu _0 = \partial K_0 / \partial J$.

$\;$\\
The solution for $S_1$ can now be found by a direct integration. In general,
once we assume that we already have $E_{n-1}$),
the procedure to obtain $E_n$ is the following 

\begin{itemize}

\item 
Perform the average on both sides of the $n$th equation of $(49)$.

\item
Introduce the obtained mean value of $E_n$, in the complete equation for $E_n$ 
given by $(49)$ (the one before averaging).

\item
The only remaining unknown $S_n$ can be obtained
by integrating the following relationship

$$\frac{\partial S_n}{\partial w_0} = \;\; {\rm known} \;\;
{\rm function} \;\; {\rm of} \;\;  w_0 \;\; {\rm and} \;\; J \; .$$

\item 
Substitute $S_n$ in the complete equation for $E_n$.

\end{itemize}

$\;$\\
Once all this has been done, the procedure can be repeated  for $n+1$.

$\;$\\
As one can see, getting the energy at a particular order
$n$ is possible if and  only if one has obtained the explicit form of 
$S_{n-1}$. On the other hand, $S_n$ can be obtained only when $E_n$ has been 
already found.

$\;$\\
The time-independent perturbation theory is very similar to the 
Rayleigh-Schroedinger perturbation scheme in wave mechanics, where one can get  
$E_n$ only if the wavefunction is known at the $n-1$ order. 
Moreover, the wavefunction of order $n$ can be found only if $E_n$ has been 
calculated. 





\input bilioen


%% file: bilioen.tex
$\;$\\
\vspace{1.0in}

\begin{center}
{\Large \bf Bibliography}
\end{center}

\vspace{0.5in}

\begin{itemize}

\item
H. Goldstein, {\it Classical Mechanics}, Second ed., Spanish version,  
Editorial Revert\'e S.A., 1992.\\

\item
R.A. Matzner and L.C. Shepley, {\it Classical Mechanics},  Prentice-Hall Inc.,
U.S.A., 1991.\\

\item
R. Murray Spiegel, {\it Theoretical Mechanics}, Spanish version,
McGraw-Hill S.A. de C.V.  M\'exico, 1976.\\

\item
L.D. Landau and E.M. Lifshitz, {\it Mechanics},  3rd. ed.  
Course of Theoretical Physics, Volume 1.  Pergammon Press, Ltd.  1976.\\

\end{itemize}

%% file: inv22en.tex


\centerline{\Large 10. ADIABATIC INVARIANTS}

\bigskip
\bigskip

\noindent
{\bf Forward}:
An adiabatic invariant is a function of the parameters and constants of 
motion of a system, which remains almost constant in the limit in which the
parameters change infinitesimally in time, even though in the end they may 
change by large amounts.

\bigskip
\bigskip

{\bf CONTENTS}:

\bigskip

10.1 BRIEF HISTORY

\bigskip

10.1 GENERALITIES

\newpage

{\bf 10.1 BRIEF HISTORY}\\

\noindent
The notion of adiabatic invariance goes back to the early years of quantum 
mechanics. Around 1910 people studying the emission and absorption of radiation
noticed that the atoms could live in quasi-stable states in which their energy
was alsmost constant.
At the 1911 Solvay Congress, the problem of adiabatic invariance became widely
known due to Einstein that draw the attention of the physics community
to the adiabatic invariant $E/ \nu$ of the one-dimensional pendulum of slowly 
varying length.
He suggested that similar invariants could occur for atomic systems setting 
their stability limits. Later, Ehrenfest was able to find such adiabatic 
invariants and their employment led to the first quantum approach of Bohr and 
Sommerfeld.


$\;$\\
\noindent
The method of adiabatic invariants resurged after several decades
in the area of magnetospheric physics. The Scandinavian scientists were 
especially 
interested in boreal aurora phenomena, i.e., the motion of electrons and 
ions in the terrestrial magnetosphere.
One of them, H. Alfven,
showed in his book {\it Cosmic Electrodynamics} that under appropriate
conditions a certain combination of dynamical parameters of the charged 
particles remains constant in the first order . 
Apparently, Alfven did not realized that he found an adiabatic invariant.
This was pointed out by L. Landau and E. Lifshitz who discussed in detail
this issue.

$\;$

{\bf 10.2 GENERALITIES}\\

\noindent
We first show in a simple way which is the adiabatic invariant in the case of
the one-dimensional harmonic oscillator. The employed method will be to prove
that in the limit of the infinitesimally slow variation of the parameters, the 
adiabatic invariant of the one-dimensional system goes to a quantity that is
exactly conserved in a corresponding two-dimensional system.\\

\bigskip

\noindent
Consider a particle at the end of a rope of negligible mass that rotates 
uniformly on a table. Let $a$ be the radius of the circle, $m$ the mass of 
the particle, and $\omega$ the angular frequency. If the origin is in the center
of the circle the projection of the motion on the $x$ axis corresponds to the 
motion of a simple harmonic oscillator. The motion continues to be harmonic
even when the rope is slowly made shorter by pulling it through a small hole
drilled at the origin. The problem is to find the invariant quantity in this
case. It is by far more difficult to get the answer for the circular motion 
in comparison with the corresponding projected oscillator. Since there are
only central forces even when the rope is slowly shortened the conserved 
quantity is the angular momentum $l$. Therefore, in the slow limit one can write
\setcounter{equation}{0}\\
\begin{eqnarray}
l=m\omega a^2~.
\end{eqnarray}
From this equation one can see that although $l$ is an {\it exact} invariant
$m\omega a^2$ is only a slow limit, i.e., adiabatic invariant. More exactly,
it is only in the slow limit that we can claim that the two quantities are 
equal and we are allowed to think of the right hand side as an invariant.
In the slow limit, the projected one-dimensional oscillator has a slowly 
varying amplitude, a 
frequency $\nu=\omega/2\pi$, and a total energy $E=\frac{1}{2}m\omega ^2a^2$.
Therefore, the adiabatic invariance of the right hand side of (1) implies
the adiabatic invariance of the quotient $E/\omega$ for the projected oscillator.

\bigskip
\noindent
A short analytic proof of this fact is the following. 
One writes the two-dimensional Hamiltonian 
\begin{eqnarray}
H=(2m)^{-1}(p_{x}^{2}+p_{y}^{2})+\frac{1}{2}m\omega ^2(x^2+y^2)~.
\end{eqnarray}
Hamilton's equations lead to
\begin{eqnarray}
\ddot{x}+\omega ^2x=o,\quad \quad \ddot{y}+\omega ^2y=0~.
\end{eqnarray}
Changing to polar coordinates we note that $l=m\dot{\theta}r^2$. The 
previous
Hamilton equations can be written as a single complex equation 
$\ddot{z}+\omega ^2 z=0$, where $z=x+iy$. We consider a solution of the form
$z=a\exp i(\omega t+b)$, where $a$, $b$, and $\omega$ are real functions that
are slowly varying in time. The real and imaginary parts are $x=a\cos 
(\omega t+b)$ and $y=a\sin(\omega t+b)$, respectively. Since $\omega$, $a$, and
$b$ vary slowly, $l$ can be written in a very good approximation as 
$l=m\omega a^2$. The energy of the projected oscillator is
$E=\frac{1}{2}m\dot{x}^2+\frac{1}{2}m\omega ^2x^2$. In the slow varying 
limit of the parameters, we can substitute $x$ in $E$ by the cosine function to
get $E=\frac{1}{2}m\omega ^2 a^2$. Thus, if $m\omega a^2$ is an adiabatic 
invariant then $E/\omega$ is an adiabatic invariant for the one-dimensional
oscillator. 

$\;$\\

\noindent
We now clarify the meaning of the term almost constant in the 
definition of the adiabatic invariant in the forward to this chapter.
$\;$\\
\noindent
Consider a system of one degree of freedom, which initially is consevative and 
periodic and depends on an initially constant parameter $a$.
The slow variation of this parameter due, for example, to
a low amplitude perturbation does not 
alter the periodic nature of 
the motion. By a slow variation we mean the one for which $a$ varies 
slowly during a period $\tau$ of the motion:
\begin{eqnarray}
\tau (da / dt) \ll a~.
\end{eqnarray}

$\;$\\
However, even when the variations of $a$ are small during a given period,
after a sufficently long time the motion may display large changes.

$\;$\\
When the parameter $a$ is constant, the system will be described by the 
action-angle variables $(w_0, J_0)$ and the Hamiltonian $H = H(J_0,a)$. 
Assume now that the generating function of the transformation
$(q,p)\rightarrow (w_0,J_0)$ is of the form $W^* (q, w_0,a)$.

$\;$\\
When $a$ is allowed to vary in time, $(w_0, J_0)$ will still be valid 
canonical variables, but $W^*$ turns into a function of time through $a$. 
Then, neither $J_0$ will be a constant nor $w_0$ a linear function of
time. The appropriate Hamiltonian will be

\begin{eqnarray}
K(w_0,J_0,a) &=& H(J_0,a) + \frac{\partial W^*}{\partial t} \nonumber \\
&=& H(J_0,a) + \dot a \frac{\partial W^*}{\partial a}~.
\end{eqnarray}

$\;$\\
The second term in (5) can be seen as a perturbation. Then, the 
temporal dependence of $J_0$ is given by

\begin{eqnarray}
\dot{J_0} = - \frac{\partial K}{\partial w_0} = - \dot a \frac{\partial}
{\partial w_0} \left( \frac{\partial W^*}{\partial a} \right)~.
\end{eqnarray}

$\;$\\
Proceeding similarly to the time-dependent perturbation theory, we 
seek the variation to first-order of the mean value of $\dot{J_0}$ during the 
period of the unperturbed motion. Since $a$ varies slowly, we can think of it as 
constant during this interval. Thus, we can write

\begin{eqnarray}
\langle \dot{J_0} \rangle = - \frac{1}{\tau} \int_{\tau} \dot{a}
\frac{\partial}{\partial w_0} \left( \frac{\partial W^*}{\partial a} \right) dt
= - \frac{\dot{a}}{\tau} \int_{\tau} \frac{\partial}{\partial w_0}
\left( \frac{\partial W^*}{\partial a} \right) dt + O(\dot{a}^2, \ddot{a}).
\end{eqnarray}

$\;$\\
One can prove that $W^*$ is a periodic function of $w_0$, and therefore,
can be written, together with its derivative w.r.t. $a$, as a Fourier series

\begin{eqnarray}
\frac{\partial W^*}{\partial a} = \sum_k A_k e^{2 \pi ikw_0}.
\end{eqnarray}

$\;$\\
Substituting $(8)$ in $(7)$ we get

\begin{eqnarray}
\langle \dot{J_0} \rangle = - \frac{\dot{a}}{\tau} \int_{\tau} \sum_k 2 \pi
ik A_k e^{2 \pi ikw_0} dt + O(\dot{a}^2, \ddot{a}).
\end{eqnarray}

$\;$\\
Since the integrant does not contain any constant term the integral is zero.
This leads to

\begin{eqnarray}
\langle \dot{J_0} \rangle = 0 + O(\dot{a}^2, \ddot{a}).
\end{eqnarray}

$\;$\\
Thus, $\langle \dot{J_0} \rangle$ will not display secular variations in the 
first order, i.e., in $\dot{a}$, which is one of the basic properties of the
adiabatic invariance. In this way, the term {\it almost constant} 
in the definition of the adiabatic invariant should be interpreted
as {\it constant in the first order}.

\input bilio2


%% file: bilio2.tex
\vspace{0.1in}

\begin{center}
{\Large Further reading}
\end{center}


\begin{itemize}





\item
L. Parker, {\it Adiabatic invariance in
simple harmonic motion}, Am. J. Phys. 39 (1971) pp. 24-27.\\

\item
A.E. Mayo, {\it Evidence for the adiabatic invariance of the black hole
horizon area}, Phys. Rev. D58 (1998) 104007 [gr-qc/9805047].

\end{itemize}

%% file: clasen.tex
\pagestyle{plain}

\begin{center}\section*{\Large 11. MECHANICS OF CONTINUOUS SYSTEMS}\end{center}

\bigskip

\noindent
{\bf Forward}:
All the formulations of mechanics up to now have dealt with systems having
a finite number of degrees of freedom or infinitely countable. However,
many physical systems are continuous; for example, an elastic solid in 
vibrational motion. Every point of such a solid participates in the 
oscillation(s) and the total motion can be described only by specifying 
the coordinates of all points.
It is not so difficult to modify the previous formulations in order to get
a formalism that works for continuous media.
The most direct method consists in approximating the continuum by a collection
of discrete subunits (`particles') and then study how the equations of motion
change when one goes to the continuous limit.

\bigskip
\bigskip

{\bf CONTENTS}:

\bigskip
11.1 Lagrangian formulation: from the discrete to the continuous case

\bigskip
11.2 Lagrangian formulation for the continuous systems

\bigskip
11.3 Hamiltonian formulation and Poisson brackets


\bigskip
11.4  Noether's theorem

\newpage

\section*{11.1 Lagrangian formulation: from discrete to the continuous case}

\noindent
As one of the most simple case for which one can go easily from a 
discrete system to a continuous counterpart we consider an infinitely long
elastic rod doing longitudinal vibrations, i.e., oscillations of its 
points along the rod axis. A system made of discrete particles
that may be considered as a discrete approximation of the rod is an infinite
chain of material points separated by the same distance $a$ and connected
through identical massless resorts of elastic constant $k$ (see the figure).

\vskip 1ex
\centerline{
\epsfxsize=220pt
\epsfbox{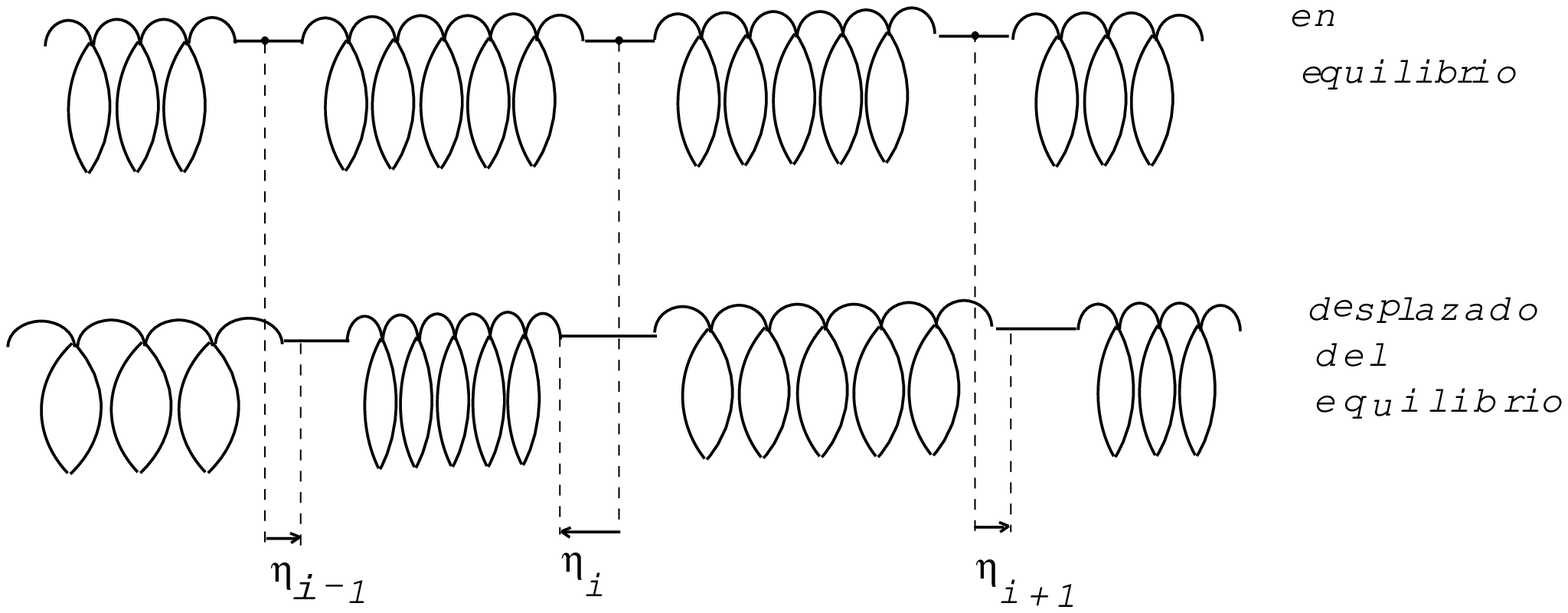}}
\vskip 2ex

\noindent
We suppose that the material points can move only along the chain.
Thus, the discrete system is an extension of the polyatomic lineal
molecule presented in chapter 6 of Goldstein's textbook.
Thus, the equations of motion of the chain can be obtained by applying
the common techniques used in the study of small oscillations. 
Denoting by $\eta _i$ the displacement of the
$i$th particle with respect to its equilibrium position,
the kinetic energy is
\setcounter{equation} {0}\\
\begin{equation}
T=\frac 12\sum_im\stackrel{.}{\eta }_i^2,  \label{1}
\end{equation}
where $m$ is the mass of each particle. The corresponding potential energy is
the sum of the potential energies of each of the resorts:
\begin{equation}
V=\frac 12\sum_ik\left( \eta _{i+1}-\eta _i\right) ^2.  \label{2}
\end{equation}
From the equations (\ref{1}) and (\ref{2}) we get the Lagrangian of the system
\begin{equation}
L=T-V=\frac 12\sum_i\left( m\stackrel{.}{\eta }_i^2-k\left( \eta _{i+1}-\eta
_i\right) ^2\right) ,  \label{3}
\end{equation}
which can also be written in the form 
\begin{equation}
L=\frac 12\sum_ia\left[ \frac ma\stackrel{.}{\eta }_i^2-ka\left( \frac{\eta
_{i+1}-\eta _i}a\right) ^2\right] =\sum_iaL_i,  \label{4}
\end{equation}
where $a$ is the equilibrium distance between the points. The Euler-Lagrange 
equations of motion for the $\eta _i$ coordinates read 
\begin{equation}
\frac ma\stackrel{\cdot \cdot }{\eta }_i-ka\left( \frac{\eta _{i+1}-\eta _i}{%
a^2}\right) +ka\left( \frac{\eta _i-\eta _{i-1}}{a^2}\right) =0.  \label{5}
\end{equation}
The particular form of $L$ in equation (\ref{4}) and the corresponding equations
of motion have been chosen as being the appropriate ones for passing to the 
continuous limit by means of $a\rightarrow 0$. it is clear that $m/a$ turns into
mass per lenght unit $\mu$ of the continuous system,
but the limiting value of $ka$ is not so obvious.
We recall that in the case of an elastic rod for which Hooke's law holds,
the enlargement of the rod per unit of length is proportional to the force 
(tension) according to
\[
F=Y\xi , 
\]
where $\xi$ is the enlargement per length unit and $Y$ is the Young modulus.
On the other hand, the relative enlargement of the length $a$ of a discrete 
system is given by $\xi =\left( \eta
_{i+1}-\eta _i\right) /a$. The corresponding force required to act on the resort 
will be 
\[
F=k\left( \eta _{i+1}-\eta _i\right) =ka\left( \frac{\eta _{i+1}-\eta _i}%
a\right) , 
\]
and therefore $\kappa a$ should correspond to the Young modulus  de Young of the 
continuous rod. When one passes from the discrete to the continuous case the 
integer index $i$ for the particular material point of the system
turns into the continuous position coordinate $x$; the discrete variable
$\eta _i$ is replaced by $\eta
\left( x\right)$. Moreover, the following quantity
\[
\frac{\eta _{i+1}-\eta _i}a=\frac{\eta \left( x+a\right) -\eta \left(
x\right) }a 
\]
entering $L_i$ goes obviously to the limit 
\[
\frac{d\eta }{dx}, 
\]
when $a\rightarrow 0$.
Finally, the sum over the number of discrete particles
turns into an integral over $x$, the length of the rod, and the Lagrangian 
(\ref{4}) takes the form
\begin{equation}
L=\frac 12\int \left( \mu \stackrel{\cdot }{\eta }^2-Y\left( \frac{d\eta }{dx%
}\right) ^2\right) dx.  \label{6}
\end{equation}
In the limit $a\rightarrow 0$, the last two terms
in the equation of motion (\ref{5}) are 
\[
{\rm Lim}_{a\rightarrow 0}
\;\frac{-Y}{a}\left\{ \left( \frac{d\eta }{dx}%
\right) _x-\left( \frac{d\eta }{dx}\right) _{x-a}\right\} , 
\]
and taking again the limit $a \rightarrow 0$ the last expression  defines
the second derivative of $\eta$. Thus, the equation of motion for the elastic
rod will be 
\begin{equation}
\mu \frac{d^2\eta }{dt^2}-Y\frac{d^2\eta }{dx^2}=0,  \label{7}
\end{equation}
which is the known one-dimensional wave equation of propagational velocity
\begin{equation}
\upsilon =\sqrt{\frac Y\mu }.  \label{8}
\end{equation}
The equation (\ref{8}) is the well-known formula for
the sound velocity (the propagational 
velocity of the elastic longitudinal waves).

\noindent
This simple example is sufficient to illustrate the main features of the 
transition from a discrete to a continuous system. The most important fact
that we have to understand is the role of the coordinate $x$. It is not a 
generalized coordinate; it is merely a continuous index substituting the 
discrete one $i$. To any value of $x$ corresponds a generalized 
coordinate $\eta\left( x\right)$. Since \thinspace $\eta$ also depends on 
the continuous variable $t$, maybe we have to write it in a more precise way
than $\eta
\left( x.t\right)$ by showing that $x$, equally to $t$, can be viewed as a 
parameter of the Lagrangian. If the system would have been 3-dimensional and 
not one-dimensional, the generalized coordinates would have to be distinguished
by three continuous indices $x,y,z$ and write in the form 
$\eta \left( x,y,z,t\right)$. We note that the quantities $x,y,z,t$ are
totally independent and show up only in $\eta$ as explicit variables.
The derivatives of $\eta$
with respect to any of them could thus be written as total derivatives without
any ambiguity.
The equation (\ref{6}) also shows the Lagrangian is an integral over the 
continuous index $x$; in the three-dimensional case, the Lagrangian
would write 
\begin{equation}
L=\int \int \int{\cal L} dx dy dz,  \label{9}
\end{equation}
where $\cal{L}$ is called the Lagrangian density. In the case of 
the longitudinal vibrations of the continuous rod, the Lagrangian density is
\begin{equation}
{\cal L} =\frac{1}{2}\left\{ \mu \left( \frac{d\eta }{dt}\right)
^2-Y\left( \frac{d\eta }{dx}\right) ^2\right\} ,  \label{10}
\end{equation}
and corresponds to the continuous limit of $L_i$ appearing in 
equation (\ref{4}). Thus, it is more the Lagrangian density than the 
Lagrangian that we used to describe the motion in this case.

\section*{11.2 Lagrangian formulation for continuous systems}

We note that in eq.
(\ref{9}) the $\cal{L}$ for the elastic rod
dependes on $\stackrel{\cdot }{\eta }=\partial \eta /\partial t$, the spatial
derivative of $\eta$, $\partial \eta /\partial x$; $x$ and $t$ play a role 
similar to its parameters. If besides the
interactions between first neighbours there would have been some local
forces,
$\cal{L}$ would have been a function of $\eta$. In general, for any
continuous system $\cal{L}$ can be an explicit function of $x$ and $t$.
Therefore, the Lagrangian density for any one-dimensional continuous system 
should be of the form
\begin{equation}
{\cal L} = {\cal L} \left( \eta ,\frac{d\eta }{dx},\frac{d\eta }{dt}%
,x,t\right) .  \label{11}
\end{equation}
The total Lagrangian, following eq.(\ref{9}), will be
\[
L=\int {\cal L}dx,
\]
and Hamilton's principle in the limit of the continuous system takes the form 
\begin{equation}
\delta I=\delta \int_1^2\int {\cal L}dxdt=0.  \label{12}
\end{equation}
From Hamilton's principle for the continuous system one would expect to get
the continuous limit of the equations of motion. For this (as in section 2-2
of Goldstein) we can use a varied path
for a convenient integration, by choosing $\eta$ from a family of functions of
$\eta$ depending on a parameter as follows 
\begin{equation}
\eta \left( x,t;\alpha \right) =\eta \left( x,t;0\right) +\alpha \zeta
\left( x,t\right) .  \label{13}
\end{equation}
where $\eta \left( x,t;0\right)$ is 
the correct function satisfying Hamilton's principle and $\zeta$ is
an arbitrary function of `good' behaviour that is zero in the extreme points
of $t$ and $x$. If we consider $I$ as a function of $\alpha$, 
then in order to be an extremal solution for the derivative of
$I$ with respect to $\alpha$ it should become zero in $\alpha =0$. By directly 
deriving $I$ we get 
\begin{equation}
\frac{dI}{da}=\int_{t_1}^{t_2}\int_{x_1}^{x_2}dxdt\left\{ \frac{\partial 
{\cal L}}{\partial \eta }\frac{\partial \eta }{\partial \alpha }+\frac{%
\partial {\cal L}}{\partial \frac{d\eta }{dt}}\frac \partial {\partial
\alpha }\left( \frac{d\eta }{dt}\right) +\left( \frac{\partial {\cal L}}{%
\partial \frac{d\eta }{dx}}\right) \frac \partial {\partial \alpha }\left( 
\frac{d\eta }{dx}\right) dt\right\} .  \label{14}
\end{equation}
Since the variation of $\eta$, $\alpha \zeta$, should be zero at the end
points, by integrating by parts on $x$ and $t$ we obtain the relationships 
\[
\int_{t_1}^{t_2}\frac{\partial {\cal L}}{\partial \frac{d\eta }{dt}}\frac
\partial {\partial \alpha }\left( \frac{d\eta }{dt}\right)
dt=-\int_{t_1}^{t_2}\frac d{dt}\left( \frac{\partial {\cal L}}{\partial 
\frac{d\eta }{dt}}\right) \frac{d\eta }{d\alpha }dt,
\]
and 
\[
\int_{x_1}^{x_2}\frac{\partial {\cal L}}{\partial \frac{d\eta }{dx}}\frac
\partial {\partial \alpha }\left( \frac{d\eta }{dx}\right)
dx=-\int_{x_1}^{x_2}\frac d{dx}\left( \frac{\partial {\cal L}}{\partial 
\frac{d\eta }{dx}}\right) \frac{d\eta }{d\alpha }dx.
\]
From this, Hamilton's principle could be written as follows
\begin{equation}
\int_{t_1}^{t_2}\int_{x_1}^{x_2}dxdt\left\{ \frac{\partial {\cal L}}
{\partial \eta }-\frac d{dt}\left( \frac{\partial {\cal L}}{\partial
\frac{
d\eta }{dt}}\right) -\frac d{dx}\left( \frac{\partial {\cal L}}{\partial
\frac{d\eta }{dx}}\right) \right\} \left( \frac{\partial \eta }{\partial
\alpha }\right) _0=0~.  \label{15}
\end{equation}
Since the varied path is arbitrary the expression in the curly brackets is zero: 
\begin{equation}
\frac d{dt}\left( \frac{\partial {\cal L}}{\partial \frac{d\eta }{dt}}%
\right) +\frac d{dx}\left( \frac{\partial {\cal L}}{\partial \frac{d\eta }
{dx}}\right) -\frac{\partial {\cal L}}{\partial \eta }=0.  \label{16}
\end{equation}
This equation is precisely the right equation of motion as given by 
Hamilton's principle.

\noindent
In the particular case of longitudinal vibrations along an elastic rod,
the form of the Lagrangian density given by  equation (\ref{10}) shows that 
\[
\frac{\partial {\cal L}}{\partial \frac{d\eta }{dt}}=\mu \frac{d\eta }{dt}~
,\qquad \frac{\partial {\cal L}}{\partial \frac{d\eta }{dx}}=-Y\frac{d\eta }{
dx}~,\qquad \frac{\partial {\cal L}}{\partial \eta }=0.
\]
Thus, as we would have liked, the Euler-Lagrange equation
(\ref{16}) can be reduced to the equation of motion (\ref{7}).

\noindent
The Lagrange formulation that we presented up to now is valid for continuous 
systems. It can be easily generalized to two, three, and more dimensions.
It is convenient to think of a four-dimensional space of coordinates
$x_o=t,x_1=x,x_2=y,x_3=z.$

\noindent
In addition, we introduce the following notation
\begin{equation}
\eta _{\rho ,\nu }\equiv \frac{d\eta _\rho }{dx_\nu };\qquad \eta
_{,j}\equiv \frac{d\eta }{dx_j};\qquad \eta _{i,\mu \nu }\equiv \frac{%
d^2\eta _i}{dx_\mu dx_\nu }.  \label{17}
\end{equation}
Employing this notation and the four $x$ coordinates the Lagrangian density
(\ref{11}) takes the form: 
\begin{equation}
\cal{L}={\cal L}\left( \eta _\rho ,\eta _{\rho ,\nu },x_{\nu} \right) .
\label{18}
\end{equation}
Thus, the total Lagrangian is an integral extended to all three-dimensional 
space: 
\begin{equation}
L=\int {\cal L}\left( dx_i\right) .  \label{19}
\end{equation}
In the case of Hamilton's principle the integral is extended to a region of the
four-dimensional space 
\begin{equation}
\delta I=\delta \int {\cal L}\left( dx_\mu \right) =0,  \label{20}
\end{equation}
where the variation of the $\eta _\rho$ nullify on the surface $S$ entailing
the integration region. The symbolic calculation needed to obtain the
corresponding Euler-Lagrange equations of motion is similar to the previous
symbolic exercise. Let a set of variational functions be
\[
\eta _\rho \left( x_\nu ;\alpha \right) =\eta _\rho \left( x_\nu \right)
+\alpha \zeta \left( x_\nu \right)~. 
\]
They depend on a single parameter and reduce to $\eta _\rho \left( x_\nu
\right)$ when the parameter $\alpha$ goes to zero. The variation of
$I$ is equivalent to put to zero the derivative of $I$ with respect to
$\alpha$, i.e.: 
\begin{equation}
\frac{dI}{d\alpha }=\int \left( \frac{\partial {\cal L}}{\partial \eta _\rho 
}\frac{\partial \eta _\rho }{\partial \alpha }+\frac{\partial {\cal L}}{%
\partial \eta _{\rho ,\nu }}\frac{\partial \eta _{\rho ,\nu }}{\partial
\alpha }\right) \left( dx_\mu \right) =0.  \label{21}
\end{equation}
Integrating by parts equation (\ref{21}), we get 
\[
\frac{dI}{d\alpha }=\int \left[ \frac{\partial  {\cal L}} {\partial \eta _\rho 
}-\frac d{dx_\nu }\left( \frac{\partial {\cal L}}{\partial \eta _{\rho ,\nu }%
}\right) \right] \frac{\partial \eta _\rho }{\partial \alpha }\left( dx_\mu
\right) +\int \left( dx_\mu \right) \frac d{dx_\nu }\left( \frac{\partial 
{\cal L}}{\partial \eta _{\rho ,\nu }}\frac{\partial \eta _{\rho ,\nu }}{%
\partial \alpha }\right) =0,
\]
and taking the limit $\alpha \rightarrow 0$ the previous expression turns into: 
\begin{equation}
\left( \frac{dI}{d\alpha }\right) _0=\int \left( dx_\mu \right) \left[ \frac{%
\partial {\cal L}}{\partial \eta _\rho }-\frac d{dx_\nu }\left( \frac{%
\partial {\cal L}}{\partial \eta _{\rho ,\nu }}\right) \right] \left( \frac{%
\partial \eta _\rho }{\partial \alpha }\right) _0=0.  \label{22}
\end{equation}
Since the variations of each $\eta _\rho$ is arbitrary and independent
equation (\ref{22}) is zero when each term in the brackets is zero separately: 
\begin{equation}
\frac d{dx_\nu }\left( \frac{\partial {\cal L}}{\partial \eta _{\rho ,\nu }}%
\right) -\frac{\partial {\cal L}}{\partial \eta _\rho }=0.  \label{23}
\end{equation}
The equations (\ref{23}) are a system of partial differential equations for 
the field quantities, with as many equations as $\rho$'s are.

\bigskip

\noindent
\underline{{\LARGE Example:}}
Given the Lagrangian density of an acoustical field 
\[
{\cal L}=\frac 12\left( \mu _0\stackrel{.}{\vec{\eta}}^2+2P_0\nabla
\cdot \vec{\eta} -\gamma P_0\left( \nabla \cdot \vec{\eta}\right)
^2\right) . 
\]
$\mu _0$ is the equilibrium mass density and $P_0$ is the equilibrium
pressure of the gas. The first term of ${\cal L}$ is the kinetic energy density,
while the rest of the terms represent the change in the potential energy of the 
gas per volume unit due to the work done on the gas
 o
por el curso de las contracciones y expansiones que son la marca de las
vibraciones ac\'{u}sticas, $\gamma$ es el cociente entre los calores
molares a presi\'{o}n y a volumen constante obtener las ecuaciones de
movimiento.

\noindent
\underline{{\Large Solution:}}

\noindent
In the four-dimensional notation, the form of the Lagrangian density is

\begin{equation}
{\cal L}=\frac 12\left( \mu _0\eta _{i,0}\eta _{i,0}+2P_0\eta _{i,i}
-\gamma P_0\eta _{i,i}\eta _{j,j}\right)~.   \label{24}
\end{equation}
From the equation (\ref{23}) the following equations of motion are obtained
\begin{equation}
\mu _0\eta _{j,00}-\gamma P_0\eta _{i,ij}=0,\qquad j=1,2,3.  \label{25}
\end{equation}
Coming back to the vectorial notation the equations (\ref{25}) can be written as
follows
\begin{equation}
\mu _0\frac{d^2}{\vec{\eta}}{dt^2}-\gamma P_0\nabla \nabla \cdot
{\vec{\eta}}=0.  \label{26}
\end{equation}
Using the fact that the vibrations are of small amplitude
the relative variation of the gas density is given by the relationship 
\[
\sigma =-\nabla \cdot \vec{\eta}~.
\]
Applying the divergence and using the previous equation
we obtain 
\[
\nabla ^2\sigma -\frac{\mu _0}{\gamma P_0}\frac{d^2\sigma}{dt^2}=0
\]
which is a three-dimensional wave equation, where
\[
\upsilon =\sqrt{\frac{\gamma P_0}{\mu _0}}
\]
is the sound velocity in gases.

\section*{11.3  Hamiltonian formulation and Poisson brackets}

\subsection*{11.3.1 Hamiltonian formulation}

Hamilton's formulation for continuous systems is similar to that for 
discrete systems. To show the procedure we go back to the chain of material
points we considered at the beginning of the chapter, where for each
$\eta _i$ one introduces a canonical momentum

\begin{equation}
p_i=\frac{\partial L}{\partial \stackrel{.}{\eta }_i}=a\frac{\partial L_i}{%
\partial \stackrel{.}{\eta }_i}.  \label{27}
\end{equation}
The Hamiltonian of the system will be
\begin{equation}
H\equiv p_i\stackrel{.}{\eta }_i-L=a\frac{\partial L_i}{\partial \stackrel{.%
}{\eta }_i}\stackrel{.}{\eta }_i-L,  \label{28}
\end{equation}
or 
\begin{equation}
H=a\left( \frac{\partial L_i}{\partial \stackrel{.}{\eta }_i}\stackrel{.}
{\eta }_i-L_i\right)~.   \label{29}
\end{equation}
Recalling that in the limit $a\rightarrow 0$, $L\rightarrow {\cal L}$ and the
sum in the  equation (\ref{29}) turns into an integral 
the Hamiltonian takes the form:
\begin{equation}
H=\int dx\left( \frac{\partial {\cal L}}{\partial \stackrel{.}{\eta }}
\dot{\eta}-{\cal L}\right)~.   \label{30}
\end{equation}
The individual canonical momenta $p_i$, given by equation
(\ref{27}), go to zero in the continuous limit, nevertheless we can define
a momentum density $\pi$ that remains finite: 
\begin{equation}
{\rm Lim}_{a\rightarrow 0}\frac{p_i}a\equiv \pi =\frac{
\partial {\cal L}}{\partial \stackrel{.}{\eta }}~.  \label{31}
\end{equation}
The equation (\ref{30}) has the form of a spatial integral of the
Hamiltonian density ${\cal H}$ defined as
\begin{equation}
{\cal H}=\pi \stackrel{.}{\eta }-{\cal L}~.  \label{32}
\end{equation}
Even when one can introduce a Hamiltonian formulation in this direct way for 
classical fields, we should keep in mind the the procedure has to give a 
special treatment to the time variable. This is different of the 
Lagrangian formulation where all the independent variables 
were considered on the same foot. This is why the Hamilton method
will be treated in a distinct manner.

\noindent
The obvious way to do a three-dimensional generalizeation of the field 
$\eta _\rho$ is the following.

\noindent
We define a canonical momentum 
\begin{equation}
\pi _{_\rho }\left( x_{_\mu }\right) =\frac{\partial {\cal L}}{\partial 
\stackrel{.}{\eta }_{_\rho }}.  \label{33}
\end{equation}
where $\eta _{_\rho }\left( x_i,t\right) ,\pi _{_\rho }\left(
x_i,t\right)$ together, define the phase space of infinite dimensions describing
the classical field and its time evolution.

\noindent
Similarly to a discrete system we can seek a conservation theorem for
$\pi$ that look like the corresponding canonical momentum
of the discrete systems. If a given field $\eta _\rho $ is a cyclic variable
(${\cal L}$ does not present an explicit dependence on $\eta _\rho $), 
the Lagrange field equation has the form of a conservation of a current: 
\[
\frac d{dx_{_\mu }}\frac{\partial {\cal L}}{\partial \eta _{_{\rho ,\mu }}}=0
\]
that is 
\begin{equation}
\frac{d\pi _\rho }{dt}-\frac d{dx_{_i}}\frac{\partial {\cal L}}{\partial
\eta _{_{\rho ,i}}}=0~.  \label{34}
\end{equation}
Thus, if $\eta _\rho$ is cyclic there is a conservative integral quantity 
\[
\Pi _\rho =\int dV\pi _{_\rho }\left( x_i,t\right) .
\]

\noindent
The generalization for the density (eq. (\ref{32})) in the case of the 
Hamiltonian density is
\begin{equation}
{\cal H}\left( \eta _{{\rho }},\eta _{_{\rho ,i}},\pi _{{ %
\rho }},x_{{ \mu }}\right) =\pi _{_\rho }\stackrel{.}{\eta }_{_\rho }-
\cal{L,}  \label{35}
\end{equation}
where it is assumed that the functional dependence of 
$\stackrel{.}{\eta }_\rho$ can be eliminated by inverting the eqs. 
(\ref{33}). From this definition, one gets
\begin{equation}
\frac{\partial \cal{H}}{\partial \pi _{{ \rho }}}=\stackrel{.}{%
\eta }_{_\rho }+\pi _{_\lambda }\frac{\partial \stackrel{.}{\eta }_{_\lambda
}}{\partial \pi _{_\rho }}-\frac{\partial {\cal L}}{\partial \stackrel{.}{%
\eta }_{_\lambda }}\frac{\partial \stackrel{.}{\eta }_{_\lambda }}{\partial
\pi _{_\rho }}=\stackrel{.}{\eta }_{_\rho }  \label{36}
\end{equation}
as a consequence of eq. (\ref{33}). In a similar way, we obtain
\begin{equation}
\frac{\partial \cal{H}}{\partial \eta _{{ \rho }}}=\pi _{_\lambda }
\frac{\partial \stackrel{.}{\eta }_{_\lambda }}{\partial \eta _{_\rho }}-
\frac{\partial {\cal L}}{\partial \stackrel{.}{\eta }_{_\lambda }}\frac{
\partial \stackrel{.}{\eta }_{_\lambda }}{\partial \eta _{_\rho }}-\frac{
\partial {\cal L}}{\partial \eta _{_\rho }}=-\frac{\partial {\cal L}}{
\partial \eta _{_\rho }}.  \label{37}
\end{equation}
Now, using Lagrange equations, eq. (\ref{37}) turns into 
\begin{equation}
\frac{\partial \cal{H}}{\partial \eta _{{ \rho }}}=-\frac
d{dx_{_\mu }}\left( \frac{\partial {\cal L}}{\partial \eta _{_{\rho ,\mu }}}
\right) =-\stackrel{.}{\pi }_{_\rho }-\frac d{dx_{_i}}\left( \frac{\partial 
{\cal L}}{\partial \eta _{_{\rho ,i}}}\right) .  \label{38}
\end{equation}
Due to the occurrence of ${\cal L}$, we still do not have a useful form.
However, by a similar procedure as used for getting the terms
$\frac{\partial {\cal H}}{\partial \pi _\rho }$ and $\frac{\partial
{\cal H}}{\partial \eta _\rho }$ for $\ \frac{\partial {\cal H}}{
\partial \eta _{\rho ,i}}$ we have
\begin{equation}
\frac{\partial \cal{H}}{\partial \eta _{{ \rho ,i}}}=\pi
_{_\lambda }\frac{\partial \stackrel{.}{\eta }_{_\lambda }}{\partial \eta
_{_{\rho ,i}}}-\frac{\partial {\cal L}}{\partial \stackrel{.}{\eta }%
_{_\lambda }}\frac{\partial \stackrel{.}{\eta }_{_\lambda }}{\partial \eta
_{_{\rho ,i}}}-\frac{\partial {\cal L}}{\partial \eta _{_{\rho ,i}}}=-\frac{%
\partial {\cal L}}{\partial \eta _{_{\rho ,i}}}~.  \label{39}
\end{equation}
Thus, by substituting (\ref{39}) in (\ref{38}) we get 
\begin{equation}
\frac{\partial \cal{H}}{\partial \eta _{{ \rho }}}-\frac
d{dx_i}\left( \frac{\partial \cal{H}}{\partial \eta _{{ \rho ,i}}}%
\right) =-\stackrel{.}{\pi }_{_\rho}~.  \label{40}
\end{equation}

\noindent
The equations (\ref{36}) y (\ref{40}) can be rewritten using a notation closer
to the Hamilton ones for a discrete system. This is possible by employing the
concept of functional derivative
\begin{equation}
\frac \delta {\delta \psi }=\frac \partial {\partial \psi }-\frac
d{dx_i}\frac \partial {\partial \psi _{,i}}.  \label{41}
\end{equation}
Since $\cal{H}$ is not a function of $\pi _{_{\rho ,i}}$ the equations (
\ref{36}) and (\ref{40}) can be written  
\begin{equation}
\stackrel{.}{\eta }_{_\rho }=\frac{\delta \cal{H}}{\delta \pi _{_\rho }}
,\qquad \stackrel{.}{\pi }_{_\rho }=-\frac{\delta \cal{H}}{\delta \eta _{
{ \rho }}}.  \label{42}
\end{equation}
Now, by employing the same notation, the Lagrange eqs. (\ref{23}) can be written
as follows 
\begin{equation}
\frac d{dt}\left( \frac{\partial {\cal L}}{\partial \stackrel{.}{\eta }
_{_\rho }}\right) -\frac{\delta {\cal L}}{\delta \eta _{{ \rho }}}=0.
\label{43}
\end{equation}

\noindent
It is fair to say that the almost unique advantage of the functional derivative
is the similarity of the formulas with the discrete case.
Moreover, one can see the parallel treatment of space and time variables.

\subsection*{11.3.2 Poisson brackets}

\noindent
We can get other properties of $\cal{H}$ by developing the total time derivative
of eq. (\ref{35}), remembering that
$\stackrel{.}{\eta }_{_\rho }$ es funci\'{o}n de $\eta
_{_\rho },\eta _{_{\rho ,j}},\pi _{_\rho }$ y $\pi _\mu $. Thus, we have that
\[
\frac{d{\cal H}}{dt}=\stackrel{.}{\pi }_{_\rho }\stackrel{.}{\eta }
_{_\rho }+\pi _{_\rho }\frac{d\stackrel{.}{\eta }_{_\rho }}{dt}-\frac{
\partial {\cal L}}{\partial \eta _{{ \rho }}}\stackrel{.}{\eta }
_{_\rho }-\frac{\partial {\cal L}}{\partial \stackrel{.}{\eta }_{_\rho }}
\frac{d\stackrel{.}{\eta }_{_\rho }}{dt}-\frac{\partial {\cal L}}{\partial
\eta _{_{\rho ,i}}}\frac{d\eta _{_{\rho ,i}}}{dt}-\frac{\partial {\cal L}}{
\partial t}. 
\]

\noindent
In this expression, the second and the forth terms nullifie each other because
of the definition (\ref{33}). The derivative simplifies to 
\begin{equation}
\frac{d{\cal H}}{dt}=\stackrel{.}{\pi }_{_\rho }\stackrel{.}{\eta }
_{_\rho }-\frac{\partial {\cal L}}{\partial \eta _{{ \rho }}}\stackrel{
.}{\eta }-\frac{\partial {\cal L}}{\partial \eta _{_{\rho ,i}}}\frac{d\eta
_{_{\rho ,i}}}{dt}-\frac{\partial {\cal L}}{\partial t}.  \label{44}
\end{equation}

\noindent
On the other hand, considering $\cal{H}$ as a function of $\eta _{_\rho
},\eta _{_{\rho ,j}},\pi _{_\rho }$ and $\pi _\mu ,$ the total time derivative 
is 
\begin{equation}
\frac{d{\cal H}}{dt}=\stackrel{.}{\pi }_{_\rho }\frac{\partial {\cal H}
}{\partial \pi _{{ \rho }}}+\frac{\partial {\cal H}}{\partial \eta
_{{ \rho }}}\stackrel{.}{\eta }_{_\rho }+
\frac{\partial {\cal H}}{
\partial \eta _{{ \rho ,i}}}\frac{d\eta _{_{\rho ,i}}}{dt}+\frac{
\partial {\cal H}}{\partial t}~,  \label{45}
\end{equation}
where we wrote the expression in such a manner to get an easy comparison with
the second terms of eq. (\ref{44}), and where using
the eqs. (\ref{36}), (\ref{37}) and (\ref{39}) we obtain 
\begin{equation}
\frac{\partial {\cal H}}{\partial t}=-\frac{\partial {\cal L}}{\partial t},
\label{46}
\end{equation}
which is an analog to the corresponding one for discrete systems.

\noindent
On the other hand, the equality of the total and partial time derivatives does
not hold. Using Hamilton's equations of motion (eq. (\ref{36}) yand(\ref{40}))
and interchanging the order of derivation, the eq.(\ref{45}) can be 
written as follows
\[
\frac{d{\cal H}}{dt}=\frac{\partial {\cal H}}{\partial \pi _{{
\rho }}}\frac d{dx_i}\left( \frac{\partial {\cal H}}{\partial \eta _{
{ \rho ,i}}}\right) +\frac{\partial {\cal H}}{\partial \eta _{
{ \rho ,i}}}\frac{d\stackrel{.}{\eta }_{_\rho }}{dx_i}+\frac{\partial 
{\cal H}}{\partial t}~.
\]
Now, employing eq. (\ref{46}) and combining the terms we finally have 
\begin{equation}
\frac{d{\cal H}}{dt}=\frac d{dx_i}\left( \stackrel{.}{\eta }_{_\rho }
\frac{\partial {\cal H}}{\partial \eta _{{ \rho ,i}}}\right) +\frac{
\partial {\cal H}}{\partial t},  \label{47}
\end{equation}
which is the closest we can approximate to the corresponding equations
for discrete systems.

\noindent
When ${\cal L}$ does not depend explicitly on time $t$,
it will not be in $\cal{H}$ as well. This implies the existence of a 
conservative current and consequently the conservation of an integral quantity,
which in this case is
\begin{equation}
H=\int {\cal H}dV~.  \label{48}
\end{equation}
Thus, if ${\cal H}$ is not an explicit function of time,
the conserved quantity is not ${\cal H}$, but the integral $H$.

\noindent
The Hamiltonian
is just an example of functions that are volume integrals of densities.
A general formalism can be provided for the time derivatives of such
integral quantities.
Consider a given density $\cal{U}$ and let it be a function of the 
coordinates of the phase space $\left( \eta _\rho ,\pi _\rho \right)$,
of its spatial gradients and possibly on $x_\mu$:
\begin{equation}
{\cal U}={\cal U}\left( \eta _{{ \rho }},\pi _{_\rho },\eta _{{ \rho
,i}},\pi _{_{\rho ,i}},x_\mu \right)~.   \label{49}
\end{equation}
The corresponding integral quantity is 
\begin{equation}
U\left( t\right) =\int {\cal U}dV  \label{50}
\end{equation}
where the volume integral is extended to all the space
limited by the contour surface on which  $\eta _\rho $ y $\pi _\rho $ take zero 
values.
Doing the time derivative of $U$ we have in general, 
\begin{equation}
\frac{dU}{dt}=\int \left\{ \frac{\partial \cal{U}}{\partial \eta _{%
{ \rho }}}\stackrel{.}{\eta }_{_\rho }+\frac{\partial \cal{U}}{%
\partial \eta _{{ \rho ,i}}}\stackrel{.}{\eta }_{{ \rho ,i}}+%
\frac{\partial \cal{U}}{\partial \pi _{{ \rho }}}\stackrel{.}{\pi }%
_{{ \rho }}+\frac{\partial {\cal U}}{\partial \pi _{{ \rho ,i}%
}}\stackrel{.}{\pi }_{{ \rho ,i}}+\frac{\partial {\cal U}}{\partial
t}\right\} dV~.  \label{51}
\end{equation}
Let us consider the term
\[
\int dV\frac{\partial {\cal U}}{\partial \eta _{{ \rho ,i}}}
\stackrel{.}{\eta }_{{ \rho ,i}}=\int dV\frac{\partial {\cal U}}{
\partial \eta _{{ \rho ,i}}}\frac{d\stackrel{.}{\eta }_{_\rho }}{dx_i}~.
\]
Integrating by parts and taking into account the nullity of
$\eta _\rho$ andd its derivatives on the contour surface, we have 
\[
\int dV\frac{\partial {\cal U}}{\partial \eta _{{ \rho ,i}}}%
\stackrel{.}{\eta }_{{ \rho ,i}}=-\int dV\stackrel{.}{\eta }_{_\rho
}\frac d{dx_i}\left( \frac{\partial {\cal U}}{\partial \eta _{{
\rho ,i}}}\right) .
\]

\noindent
For the term in $\stackrel{.}{\pi }_{\rho ,i}$ one uses a similar technique.
Substituting the obtained expressions and grouping appropriately the coefficients 
of $\stackrel{.}{\eta }$ and $\stackrel{.}{\pi }_\rho$, respectively,
and using the functional derivative notation 
equation (\ref{51}) is reduced to

\begin{equation}
\frac{dU}{dt}=\int dV\left\{ \frac{\delta \cal{U}}{\delta \eta _{{
\rho }}}\stackrel{.}{\eta }_{_\rho }+\frac{\delta \cal{U}}{\delta \pi _{
{ \rho }}}\stackrel{.}{\pi }_{{ \rho }}+\frac{\partial {\cal U
}}{\partial t}\right\}~.  \label{52}
\end{equation}
Finally, introducing the canonical equations of motion (\ref{42}), we have 
\begin{equation}
\frac{dU}{dt}=\int dV\left\{ \frac{\delta \cal{U}}{\delta \eta _{{
\rho }}}\frac{\delta \cal{H}}{\delta \pi _{{ \rho }}}-\frac{\delta
\cal{H}}{\delta \eta _{{ \rho }}}\frac{\delta \cal{U}}{\delta
\pi _{{ \rho }}}\right\} +\int dV\frac{\partial {\cal U}}{\partial t
}~.  \label{53}
\end{equation}

\noindent
The first integral in the rhs corresponds clearly to the Poisson brackets.
If $\cal{U}$ and $\cal{W}$ are two functions of density, these considerations 
allows us to take as the definition of the Poisson bracket for integral
quantities as
\begin{equation}
\left[ U,W\right] =\int dV\left\{ \frac{\delta \cal{U}}{\delta \eta _{%
{ \rho }}}\frac{\delta \cal{W}}{\delta \pi _{{ \rho }}}-%
\frac{\delta \cal{W}}{\delta \eta _{{ \rho }}}\frac{\delta
\cal{U}}{\delta \pi _{{ \rho }}}\right\} .  \label{54}
\end{equation}
We define the partial time derivative of $U$ through the following expression 
\begin{equation}
\frac{\partial U}{\partial t}=\int dV\frac{\partial {\cal U}}{\partial t}.
\label{55}
\end{equation}

\noindent
Thus, the eq. (\ref{53}) podr\'{a} could be written
\begin{equation}
\frac{dU}{dt}=\left[ U,H\right] +\frac{\partial U}{\partial t},  \label{56}
\end{equation}
which exactly corresponds, in this notation, to the equation for the 
discrete systems. 
Since by definition the Poisson bracket of $H$ with itself is zero,
the eq. (\ref{46}) turns into 
\begin{equation}
\frac{dH}{dt}=\frac{\partial H}{\partial t}~,  \label{57}
\end{equation}
which is the integral form of eq. (\ref{47}).
Thus, the Poisson bracket formalism ocurr as a consequence of the Hamiltonian
formulation.
However, one cannot perform a description in terms of Poisson brackets
for field theories by a step by step correspondence with the discrete case.

\noindent
Nevertheless there is one way to work out the classical fields 
which includes almost all the ingredients of Hamilton's formulation and
Poisson brackets of the discrete case.
The basic idea is to replace the continuous spatial variable or the 
continuous index by a countable discrete index.

\noindent
The requirement that $\eta$ be zero at the end points is a contour condition
that might be achieved physically only by fixing the rod between two rigid 
walls. Then, the amplitud of oscillation can be represented by means of 
a Fourier series: 
\begin{equation}
\eta \left( x\right) =\sum_{n=0}^\infty q_n\sin \frac{2\pi n\left(
x-x_1\right) }{2L}~.  \label{58}
\end{equation}
Instead of the continuous index $x$ we have the discrete $\eta$. We could use
this representation of $x$ only when $\eta \left(
x\right)$ is a regular function, which happens for many field quantities.

\noindent
We assume that there is only one real field $\eta$ that can be developed in 
a three-dimensional Fourier series 
\begin{equation}
\eta \left( \overrightarrow{r},t\right) =\frac 1{V^{1/2}}\sum_{k=0}q_k\left(
t\right) \exp \left( i\overrightarrow{k}\cdot \overrightarrow{r}\right) 
\label{59}
\end{equation}
Here, $\vec{k}$ is a wave vector that can take only discrete modulus and 
directions, so that in only one lineal dimension there will be an integer
(or sometimes, half-integer) wavelengths.
We say that $\vec{k}$ has a discrete spectrum. The scalar subindex
$k$ represents a certain order of the set of integer subindices that are used 
to enumerate the discrete values of $\vec{k}$; $V$ is the volume of the system,
which appear as a normalization factor.

\noindent
The orthogonality of the exponentials in the entire volume can be stated 
through the relationship
\begin{equation}
\frac 1V\int e^{i\left( \vec{k}-\vec{k}^{^{\prime }}\right) .\vec{r}%
}dV=\delta _{k,k^{^{\prime }}}~.  \label{60}
\end{equation}
As a matter of fact, the allowed values of $k$ are those for which the condition
(\ref{60}) is satisfied, and the coefficients $q_k\left(
t\right)$ are given by
\begin{equation}
q_k\left( t\right) =\frac 1{V^{1/2}}\int e^{-i\overrightarrow{k}\cdot 
\overrightarrow{r}}\eta \left( \overrightarrow{r},t\right) dV~.  \label{61}
\end{equation}
Similarly, for the density of the canonical momentum we have 
\begin{equation}
\pi \left( \overrightarrow{r},t\right) =\frac 1{V^{1/2}}\sum_kp_k\left(
t\right) e^{-i\overrightarrow{k}\cdot \overrightarrow{r}}  \label{62}
\end{equation}
with $p_k\left( t\right) $ defined as
\begin{equation}
p_k\left( t\right) =\frac 1{V^{1/2}}\int e^{-i\overrightarrow{k}\cdot 
\overrightarrow{r}}\pi \left( \overrightarrow{r},t\right) dV~.  \label{63}
\end{equation}
Both
$q_k$ and $p_k$ are integral quantities. Thus, we can look for their Poisson 
brackets. 
Since the exponentials do not contain the fields
we have, according to (\ref{54})
\begin{eqnarray*}
\left[ q_k,p_{k^{^{\prime }}}\right]  &=&\frac 1V\int dVe^{-i\overrightarrow{
k}\cdot \overrightarrow{r}}\left\{ \frac{\delta \eta }{\delta \eta }\frac{
\delta \pi }{\delta \pi }-\frac{\delta \pi }{\delta \eta }\frac{\delta \eta 
}{\delta \pi }\right\}  \\
&=&\frac 1V\int dVe^{-i\overrightarrow{k}\cdot \overrightarrow{r}}
\end{eqnarray*}
that is, by equation (\ref{60}), 
\begin{equation}
\left[ q_k,p_{k^{^{\prime }}}\right] =\delta _{k,k^{^{\prime }}}~.  \label{64}
\end{equation}
From the definition of the Poisson brackets it is obvious that 
\begin{equation}
\left[ q_k,q_{k^{^{\prime }}}\right] =\left[ p_k,p_{k^{^{\prime }}}\right]
=0~.
\label{65}
\end{equation}
The time dependence of $q_k$ is sought starting from
\[
\stackrel{.}{q}_k\left( t\right) =\left[ q_k,H\right] =\frac 1{V^{1/2}}\int
dVe^{-i\overrightarrow{k}\cdot \overrightarrow{r}}\left\{ \frac{\delta \eta 
}{\delta \eta }\frac{\delta \cal{H}}{\delta \pi }-\frac{\delta {\cal H
}}{\delta \eta }\frac{\delta \eta }{\delta \pi }\right\} 
\]
that is 
\begin{equation}
\stackrel{.}{q}_k\left( t\right) =\frac 1{V^{1/2}}\int e^{-i\overrightarrow{k
}\cdot \overrightarrow{r}}\frac{\delta \cal{H}}{\delta \pi}~.  \label{66}
\end{equation}
On the other hand, we have 
\begin{equation}
\frac{\partial H}{\partial p_k}=\int dV\frac{\partial \cal{H}}{\partial
\pi }\frac{\partial \pi }{\partial p_k}  \label{67}
\end{equation}
and therefore we get
\begin{equation}
\frac{\partial \pi }{\partial p_k}=\frac 1{V^{1/2}}e^{-i\overrightarrow{k}
\cdot \overrightarrow{r}}~.  \label{68}
\end{equation}
Comparando las ecuaciones (\ref{67}) y (\ref{66}) tenemos 
\begin{equation}
\stackrel{.}{q}_k\left( t\right) =\frac{\partial H}{\partial p_k}~.  \label{69}
\end{equation}
In a similar way we can obtain the equation of motion for $p_k$ 
\begin{equation}
\stackrel{.}{p}_k=-\frac{\partial H}{\partial q_k}~.  \label{70}
\end{equation}
Thus, $p_k$ y $q_k,$ obey the Hamilton equations of motion.

\section*{11.4 Noether's theorem}

\noindent
We have already seen that the properties of the Lagrangian (or of the
Hamiltonian) imply the existence of conservative quantities. Thus, if the 
Lagrangian does not contain explicitly a particular displacement coordinate,
the corresponding caonical momentum is conserved.
The absence of an explicit dependence on a coordinate means that the 
Lagrangian is not changed by a transformation that alter the value of that
coordinate; we say that it is invariant or symmetric for that transformation.

\noindent
The symmetry under a coordinate transformation refers to the effects of an
infinitesimal transformation as follows 
\begin{equation}
x_\mu \rightarrow x_\mu ^{^{\prime }}=x_\mu +\delta x_\mu ,  \label{71}
\end{equation}
where the variation $\delta x_\mu $ can be a function of all the other
$x_\nu$. Noether's theorem deals with the effect of the transformation of the 
field quantities itselves. Such a transformation can be written
\begin{equation}
\eta \left( x_\mu \right) \rightarrow \eta _\rho ^{^{\prime }}\left( x_\mu
^{^{\prime }}\right) =\eta _{_\rho }\left( x_{_\mu }\right) +\delta \eta
_{_\rho }\left( x_{_\mu }\right) .  \label{72}
\end{equation}
Here $\delta \eta _{_\rho }\left( x_{_\mu }\right) $ is a measure of the effect
of the variations of $x_{_\mu }$ and of $\eta _{_\rho }$. It can be a 
function of all the other fields $\eta _{_\lambda }$. The variation of one
of the field variables in a particular point of space $x_{_\mu}$ is a different
quantity $\overline{\delta }\eta _{_\rho }$:
\begin{equation}
\eta _\rho ^{^{\prime }}\left( x_\mu ^{^{\prime }}\right) =\eta _{_\rho
}\left( x_{_\mu }\right) +\overline{\delta }\eta _{_\rho }\left( x_{_\mu
}\right) .  \label{73}
\end{equation}
The characterization of the transformations by means of infinitesimal 
variations, starting from untransformed quantities means that we consider
only continuous transformations.
Therefore, the symmetry under inversion of the three-dimensional space 
is not a symmetry of the continuous type to which Noether's theorem can be
applied. As a consequence of the transformations both in the coordinates and
the fields, the
Lagrangian will, in general, appear as a different function
of the field and spacetime coordinates:
\begin{equation}
{\cal L}\left( \eta _{_\rho }\left( x_{_\mu }\right) ,\eta _{_{\rho ,\nu
}}\left( x_{_\mu }\right) ,x_{_\mu }\right) \rightarrow {\cal L}^{^{\prime
}}\left( \eta _{_\rho }^{^{\prime }}\left( x_{_\mu }^{^{\prime }}\right)
,\eta _{_{\rho ,\nu }}^{^{\prime }}\left( x_{_\mu }^{^{\prime }}\right)
,x_{_\mu }^{^{\prime }}\right) .  \label{74}
\end{equation}

\noindent
The version of Noether's theorem that we shall present is not of the most
general form possible, but makes easier the proof, without loosing too much
of its generality and the usefulness of the conclusions. We shall suppose 
the following three conditions:

\begin{enumerate}
\item  The spacetime is Euclidean, meaning that the relativity is reduced to
the Minkowski space, which is complex but flat.

\item The Lagrangian density is of the same functional form for the 
transformed quantities as for the original ones, that is 
\begin{equation}
{\cal L}^{^{\prime }}\left( \eta _{_\rho }^{^{\prime }}\left( x_{_\mu
}^{^{\prime }}\right) ,\eta _{_{\rho ,\nu }}^{^{\prime }}\left( x_{_\mu
}^{^{\prime }}\right) ,x_{_\mu }^{^{\prime }}\right) ={\cal L}\left( \eta
_{_\rho }^{^{\prime }}\left( x_{_\mu }^{^{\prime }}\right) ,\eta _{_{\rho
,\nu }}^{^{\prime }}\left( x_{_\mu }^{^{\prime }}\right) ,x_{_\mu
}^{^{\prime }}\right) .  \label{75}
\end{equation}

\item  The value of the action integral is invariant under the transformation 
\begin{equation}
I^{^{\prime }}\equiv \int_{\Omega ^{^{\prime }}}\left( dx_{_\mu }\right) 
{\cal L}^{^{\prime }}\left( \eta _{_\rho }^{^{\prime }}\left( x_{_\mu
}^{^{\prime }}\right) ,\eta _{_{\rho ,\nu }}^{^{\prime }}\left( x_{_\mu
}^{^{\prime }}\right) ,x_{_\mu }^{^{\prime }}\right) =\int_\Omega {\cal L}%
\left( \eta _{_\rho }\left( x_{_\mu }\right) ,\eta _{_{\rho ,\nu }}\left(
x_{_\mu }\right) ,x_{_\mu }\right) .  \label{76}
\end{equation}
\end{enumerate}

\noindent
Combining the equations (\ref{75}) and (\ref{76}) we get the condition
\begin{equation}
\int_{\Omega ^{^{\prime }}}\left( dx_{_\mu }\right) {\cal L}\left( \eta
_{_\rho }^{^{\prime }}\left( x_{_\mu }\right) ,\eta _{_{\rho ,\nu
}}^{^{\prime }}\left( x_{_\mu }\right) ,x_{_\mu }\right) -\int_\Omega {\cal L}
\left( \eta _{_\rho }\left( x_{_\mu }\right) ,\eta _{_{\rho ,\nu }}\left(
x_{_\mu }\right) ,x_{_\mu }\right) =0~.  \label{77}
\end{equation}
From the invariance condition, the equation (\ref{77}) becomes
\begin{eqnarray}
&&\int_{\Omega ^{^{\prime }}}dx_{_\mu }{\cal L}\left( \eta ^{^{\prime
}},x_{_\mu }\right) -\int_\Omega dx_{_\mu }{\cal L}\left( \eta ,x_{_\mu
}\right)   \label{78} \\
&=&\int_\Omega dx_{_\mu }\left[ {\cal L}\left( \eta ^{^{\prime }},x_{_\mu
}\right) -{\cal L}\left( \eta ,x_{_\mu }\right) \right] +\int_s{\cal L}%
\left( \eta \right) \delta x_{_\mu }dS_\mu =0~.  \nonumber
\end{eqnarray}
Here, ${\cal L}\left( \eta ,x_{_\mu }\right)$ is a shorthand notation for the 
total functional dependence,
$S$ is the three-dimensional surface of the region $\Omega$ and
$\delta x_{_\mu }$ is the difference vector between the points of 
$S$ and the corresponding points of the transformed surface $S^{^{\prime }}$.
The last integral can be transformedthrough the theorem of four-dimensional 
divergence. This leads to the following invariance condition
\begin{equation}
0=\int_\Omega dx_{_\mu }\left\{ \left[ {\cal L}\left( \eta ^{^{\prime
}},x_{_\mu }\right) -{\cal L}\left( \eta ,x_{_\mu }\right) \right] +\frac
d{dx_{_\mu }}\left( {\cal L}\left( \eta ,x_{_\mu }\right) \delta x_{_\nu
}\right) \right\} .  \label{79}
\end{equation}
Now, using the equation (\ref{73}), the term in the brackets can be written in 
the first-order of approximation as follows
\[
{\cal L}\left( \eta _{_\rho }^{^{\prime }}\left( x_{_\mu }\right) ,\eta
_{_{\rho ,\nu }}^{^{\prime }}\left( x_{_\mu }\right) ,x_{_\mu }\right) -%
{\cal L}\left( \eta _{_\rho }\left( x_{_\mu }\right) ,\eta _{_{\rho ,\nu
}}\left( x_{_\mu }\right) ,x_{_\mu }\right) =\frac{\partial {\cal L}}{%
\partial \eta _{_\rho }}\overline{\delta }\eta _{_\rho }+\frac{\partial 
{\cal L}}{\partial \eta _{_{\rho ,\nu }}}\overline{\delta }\eta _{_{\rho
,\nu }}.
\]
Using the Lagrange field equations
\[
{\cal L}\left( \eta ^{\prime },x_{_\mu }\right) -{\cal L}\left( \eta
,x_{_\mu }\right) =\frac d{dx_\nu }\left( \frac{\partial {\cal L}}{\partial
\eta _{_{\rho ,\nu }}}\overline{\delta }\eta _{_\rho }\right) .
\]
Then, the invariance condition (\ref{79}) ocurr as
\begin{equation}
\int \left( dx_{_\mu }\right) \frac d{dx_\nu }\left\{ \frac{\partial {\cal L}%
}{\partial \eta _{_{\rho ,\nu }}}\overline{\delta }\eta _{_\rho }-{\cal L}%
\delta x_{_\nu }\right\} =0,  \label{80}
\end{equation}
which already has the form of an equation for the conservation of a current.

\noindent
It is useful to develop more the condition giving the form of the infinitesimal
transformation as a function of the $R$ infinitesimal parameters
$\varepsilon _{r,}r=1,2,...,R$, such that the variations of 
$x_{_\mu }$ y $\eta _{_\rho }$ be lineal in $\varepsilon _r$:
\begin{equation}
\delta x_{_\nu }=\varepsilon _rX_{r\nu },\qquad \delta \eta _{_\rho
}=\epsilon _r\Psi _{_{r\rho }}.  \label{81}
\end{equation}
By substituting these conditions in eq. (\ref{80}) we get
\[
\int \epsilon _r\frac d{dx_\nu }\left\{ \left( \frac{\partial {\cal L}}{%
\partial \eta _{_{\rho ,\nu }}}\eta _{_{\rho ,\sigma }}-{\cal L}\delta
_{_{\nu \sigma }}\right) X_{r\sigma }-\frac{\partial {\cal L}}{\partial \eta
_{_{\rho ,\nu }}}\Psi _{_{r\rho }}\right\} \left( dx_{_\mu }\right) =0~.
\]
Since the $\varepsilon _r$ parameters are arbitrary, there are $r$conservative
currents as solutions of the differential conservation theorems:
\begin{equation}
\frac d{dx_\nu }\left\{ \left( \frac{\partial {\cal L}}{\partial \eta
_{_{\rho ,\nu }}}\eta _{_{\rho ,\sigma }}-{\cal L}\delta _{_{\nu \sigma
}}\right) X_{r\sigma }-\frac{\partial {\cal L}}{\partial \eta _{_{\rho ,\nu
}}}\Psi _{_{r\rho }}\right\} =0~.  \label{82}
\end{equation}
The equations (\ref{82}) are the main conclusion of Noether's theorem, telling
that if the system has symmetry properties fulfilling the conditions (1) and
(2) for transformations of the type given by the equations (\ref{81}),
then there exist $r$ conserved quantities.

\bigskip

\begin{center} Further reading \end{center}

\bigskip

\noindent
R.D. Kamien, {\it Poisson bracket formulation of nematic polymer dynamics},
cond-mat/9906339 (1999)
